\documentclass[aps,pre,twocolumn,longbibliography,nofootinbib]{revtex4-1}

\usepackage{graphicx,xspace}
\usepackage{dcolumn}
\usepackage{bm}

\usepackage[utf8]{inputenc}
\usepackage[T1]{fontenc}
\usepackage{mathptmx}
\usepackage{float}
\usepackage{amsmath}
\usepackage{amssymb}
\usepackage[normalem]{ulem}
\usepackage[dvipsnames]{xcolor}
\usepackage{footnote}
\usepackage{mathtools}
\usepackage{comment} 
\usepackage{hyperref}
\usepackage{tikz}
\usetikzlibrary{positioning}
\input{epsf}

\begin{document}

\preprint{AIP/123-QED}

\title{Effective interactions, structure, and pressure in charge-stabilized colloidal suspensions: \\
Critical assessment of charge renormalization methods}

\author{Mariano E. Brito}
\altaffiliation[Current address: ]{Institute for Computational Physics, University of Stuttgart, D-70569 Stuttgart, Germany, mariano.brito@icp.uni-stuttgart.de}
\affiliation{Institute of Biological Information Processing, IBI-4, Forschungszentrum J\"ulich GmbH, 52425 J\"ulich, Germany}

\author{Gerhard N\"agele}
\email[]{g.naegele@fz-juelich.de}
\affiliation{Institute of Biological Information Processing, IBI-4, Forschungszentrum J\"ulich GmbH, 52425 J\"ulich, Germany}

\author{Alan R. Denton}
\email[]{alan.denton@ndsu.edu}
\affiliation{Department of Physics, North Dakota State University, Fargo, ND 58108-6050 USA}

\date{\today}

\begin{abstract}
Charge-stabilized colloidal suspensions display a rich variety of microstructural and thermodynamic properties, which are determined by electro-steric interactions between all ionic species. The large size asymmetry between molecular-scale microions and colloidal macroions allows the microion degrees of freedom to be integrated out, leading to an effective one-component model of microion-dressed colloidal quasi-particles. For highly-charged colloids with strong macroion-microion correlations, nonlinear effects can be incorporated into effective interactions by means of charge renormalization methods. Here, we compare and partially extend several practical mean-field methods of calculating renormalized colloidal interaction parameters, including effective charges and screening constants, as functions of concentration and ionic strength. Within the one-component description, we compute structural and thermodynamic properties from the effective interactions and assess the accuracy of the different methods by comparing predictions with elaborate primitive-model simulations [P. Linse, J.~Chem.~Phys. {\bf 113}, 4359 (2000)]. We also compare various prescriptions for the osmotic pressure of suspensions in Donnan equilibrium with a salt ion reservoir, and analyze instances where the macroion effective charge becomes larger than the bare one. The methods assessed include single-center cell, jellium, and multi-center mean-field theories. The strengths and weaknesses of the various methods are critically assessed, with the aim of guiding optimal and accurate implementations.
\end{abstract}


\maketitle 

\section{Introduction}

Charge-stabilized colloidal suspensions consist of mesoscopic particles (macroions), typically a few nanometers to microns in size, suspended in a polar, low-molecular-weight solvent (e.g., water) and carrying ionizable chemical groups, which can dissociate, releasing oppositely charged counterions into solution. In the presence of salt in solution, the charged colloids and released counterions coexist with solvent-dissolved counterions and coions (microions). As first explained by the Derjaguin–Landau–Verwey–Overbeek (DLVO\footnote{A list of commonly used acronyms is given at the end in Table~\ref{table1}.}) theory \cite{Derjaguin1941,Verwey1948}, surrounding each colloid is an electric double layer, resulting from the buildup of an inhomogeneous mobile cloud of microions, which carries a charge of opposite sign to that of the colloidal particles and tends to screen their electrostatic potential \cite{HansenLoewen_AnnuRevPhysChem_2000}. The overlap of the double layers of two neighboring colloids generates a repulsive force, which stabilizes the particles against aggregation \cite{Naegele_PhysRep_1996}. Charge-stabilization is important in many practical applications, including water purification and stability of clays, foods, pharmaceuticals, and consumer products.

By taking advantage of the large size and charge asymmetries between macroions and microions, the electrostatic part of the corresponding effective pair potential can be derived from the mean-field Poisson–Boltzmann (PB) theory, where microion-microion correlations are neglected
\cite{Belloni2000,Levin2002_Review,Denton_BookChap_2007}. The nonlinear PB equation for the electrostatic potential can be solved analytically only in certain geometries (planar and cylindrical). Linearizing the PB equation allows for an analytical solution also in spherical geometry. Within the linear Debye–H\"uckel (DH) approximation \cite{DH}, in which the Boltzmann factor for the electrostatic potential is linearized, one obtains an effective macroion-macroion pair potential of Yukawa (screened-Coulomb) form, characterized by the bare valence of the macroions $Z$ and a screening constant $\kappa$ (inverse screening length), which depends on the concentrations of microions. The DH approximation is most accurate for weakly charged species, where the macroion–microion interaction energy is low compared with the thermal energy \cite{Denton_BookChap_2007}.

With increasing charge, when the characteristic Coulomb energy rivals the thermal energy in magnitude, nonlinear screening effects grow and the DH approximation becomes increasingly inaccurate \cite{Belloni1998}. These nonlinear effects arise from a strong electrostatic coupling between oppositely charged colloids and counterions, inducing a significant accumulation of counterions in the vicinity of the macroion surfaces \cite{Belloni1998,Levin1998,delasNieves2000,delasNieves2001}. Even though the nonlinear PB equation for the electrostatic potential cannot be solved analytically in spherical geometry, the concept of {\it charge renormalization} allows for the incorporation of nonlinear effects in a linear DH-like treatment. Numerical simulations also indicate that the electrostatic potential far from colloidal particles saturates as a function of the bare colloidal charge \cite{Alexander1984}, suggesting that the thermodynamics of highly charged colloidal systems can be based on a linear treatment \cite{Levin2002_Review}. 

The basic premise of charge renormalization theories is that a colloidal particle and its cloud of quasi-condensed counterions may be treated as a composite particle that carries a charge much lower than the bare charge \cite{Belloni1998,Belloni2000}, thereby redefining the screening constant of the system. Renormalization methods thus map a highly-charged macroion suspension, with significant nonlinear electrostatic effects, onto an equivalent linearly-behaving system characterized by renormalized effective interaction parameters, namely the renormalized valence $Z_\text{eff}$ and screening constant $\kappa_\text{eff}$. These renormalized parameters are used in the effective DLVO electrostatic pair potential, in order to summarily account for nonlinear effects in the macroion-macroion interactions.

While the concept of renormalized (effective) charge has proven very useful for modeling systems beyond the DH approximation, the absence of a general and precise definition \cite{Belloni1998} has led to a wide variety of proposals. Different charge renormalization methods, being distinguished by their underlying physical assumptions, predict different dependences of the renormalized interaction parameters, $Z_\text{eff}$ and $\kappa_{\text{eff}}$, on the bare parameters, $Z$ and $\kappa$, as well as on colloid and salt concentrations. All methods necessarily rely on approximations (sometimes uncontrolled), which affect their efficacy in describing nonlinear electrostatic screening and their accuracy in predicting system properties \cite{Trizac_PRE_2004}. 

Many charge renormalization methods have been proposed and independently tested within the context of Poisson-Boltzmann theory. Only a few previous studies, however, have aimed to assess the relative strengths and weaknesses of the different methods \cite{Belloni1998,Hallez2017,Hallez_EPJE2018}. The purpose of this work is to thoroughly analyze, and partially extend, a selection of the most commonly used and easily implemented renormalization methods, and to assess their performance and applicability over broad ranges of particle charge and ionic strength. We consider only methods applicable to systems in the weak-coupling limit, where microion-microion correlations are negligible and nonlinear PB theory is a valid starting point. We further restrict our study to methods of the simplest forms -- those that do not require recursive calculation of distribution functions for computing effective interaction parameters \cite{Colla_JCP_2009,Hallez_EPJE2018}. By taking advantage of the effective one-component model of colloidal suspensions, we quantitatively evaluate the performance of the various methods by testing the accuracy of their predictions of structural and thermodynamic properties. 

We focus here on charge renormalization methods based on mean-field PB approximations, including surface charge and extrapolated point charge methods within the spherical cell model with edge and mean potential linearizations, the renormalized jellium model, and the renormalized linear response theory. These methods provide the effective interaction parameters, $Z_\text{eff}$ and $\kappa_\text{eff}$, which we use as input to the effective macroion pair potential for computing pair correlation functions. The latter characterize the structure of colloidal suspensions. We explore and compare the predictions of the different approaches also for thermodynamic quantities, including the suspension pressure and the osmotic compressibility. 

Special attention is devoted to analyzing different prescriptions for calculating the pressure. In particular, we discuss the importance of the different pressure contributions arising from the renormalized, density-dependent, effective macroion pair potential. We also examine concentrated suspensions of weakly coupled macroions, for which some approaches predict an effective macroion valence larger than the bare valence. Our study is restricted to impermeable, rigid, spherical colloidal particles, disregarding methods designed to describe ion-permeable macroions \cite{Colla_JCP_2014,HallezMeireles2016}, colloidal mixtures \cite{GarciaDeSoria2016}, or different colloidal architectures \cite{HallezMeireles2016}. With this work, we aim not only to provide a guide for optimal and accurate implementations of the analyzed methods, but also to bring some clarity to their underlying theoretical foundations.

In the discussed PB mean-field theory methods, the microions are described only by their concentration profiles. The discrete nature of the microions and their correlations, in particular, those induced by their nonzero sizes, are here disregarded. This approach excludes the discussion of macroion overcharging and like-charge attraction effects, induced by strongly correlated polyvalent counterions, and of microion nonzero-size effects relevant to nano-sized colloidal macroion systems and protein solutions. PB-based methods are applicable in the so-called weak-coupling regime of monovalent microions, relatively large macroion-to-microion size ratio, and low to moderately high macroion surface charge densities. For a salt-free suspension of spherical macroions of radius $a$ and, say, negative surface charge $-Ze$, this regime is roughly delineated by low values, $\Gamma_\text{cc}<1$, of the counterion Coulomb coupling parameter, defined as $\Gamma_\text{cc} =\lambda_\text{B} z_{+}^2/l_\text{c}$. Here, $z_{+}e$ is the charge of a counterion (cation) released from the macroion surface, with $e>0$ the proton elementary charge, and $\lambda_\text{B}$ is the Bjerrum length of the polar solvent (typically water). Moreover, $l_\text{c} = a\sqrt{(8\pi/\sqrt{3})z_{+}/|Z|}$ is the characteristic nearest-neighbor distance between counterions quasi-condensed on the macroion surface and taken to reside on the vertices of a triangular lattice \cite{Allahyarov:1998,LevinTrizacBocquet:2003}. For monovalent microions, and macroions of radius $a \gg \lambda_\text{B}$, PB methods work quite well up to very high macroion surface charge densities \cite{LevinTrizacBocquet:2003}. Prominent examples of charge-stabilized colloidal systems amenable to a PB mean-field treatment are polystyrene or silica spheres of radius $a \gtrsim 50$ nm with monovalent, ionizable surface groups that are suspended in an aqueous 1:1 electrolyte solution.

The paper is organized as follows. In Sec.~\ref{Sec:Models}, we first review the primitive and effective one-component models of charge-stabilized colloidal suspensions. In Sec.~\ref{methods}, we then describe several charge renormalization methods, detailing their underlying assumptions and outlining their practical implementations. Next, in Sec.~\ref{sec:StructureThermodynamics}, we discuss predictions of the various methods for structural and thermodynamic properties of suspensions of charged macroions interacting via an effective pair potential in the one-component model. In Sec.~\ref{results}, we assess the performance of the different methods by comparing their predictions against data from simulations of the primitive model. Finally, in Sec.~\ref{conclusions}, we summarize our results and conclude with a discussion of the relative strengths and weaknesses of the various renormalization methods.

\section{Models}\label{Sec:Models}
\subsection{Primitive Model}\label{Sec:PM}
A wide variety of soft materials, including polyelectrolyte solutions, charge-stabilized colloidal suspensions, and globular protein solutions, can be reasonably described by the primitive model (PM). In this idealized model, all ions interact via Coulomb and excluded-volume forces and are immersed in a structureless solvent. Specific properties of the solvent are neglected, except for its static dielectric constant $\epsilon$ and its shear (Newtonian) viscosity $\eta_0$. Here, we neglect also image-charge effects, which can give rise to many-body dielectric interactions, caused by differences in the dielectric properties of the ions and the solvent.
	
Consider a three-component PM system consisting of $N_\text{m}$ negatively charged macroions (species $\alpha=\text{m}$) of negative charge number $z_\text{m}=-Z$, with macroion valence $Z>0$, and $N_\alpha$ microions of valence $z_\alpha$, with $\alpha=\pm$, in a macroscopic suspension of volume $V$ and temperature $T$. The three ion species are modeled as spherical particles of monodisperse radius $a_\alpha$ and mass $m_\alpha$, with $\alpha=\text{m},\pm$. Here, $(+)$ and $(-)$ label the positively charged counterions (cations) and negatively charged coions (anions), respectively, collectively referred to as microions. The counterion species $(+)$ of valence $z_{+}>0$ includes both the salt counterions and counterions released from the macroion surfaces, which are taken as chemically identical for simplicity. The macroion mean number density, $n_\text{m}=N_\text{m}/V$, determines the macroion volume fraction, $\phi=4\pi a_\text{m}^3n_\text{m}/3$. The Hamiltonian of the system can be expressed in the general form,
\begin{equation}
\mathcal{H}=H_\text{mm}+H_{++}+H_{--}+H_{+-}+H_{\text{m}+}+H_{\text{m}-},
\label{GeneralHamiltonian}
\end{equation}
where $H_{\alpha\alpha}$ is the Hamiltonian of the ion species $\alpha$, i.e.,
\begin{equation}
H_{\alpha\alpha}(\{{\bf r}_\alpha^{N_\alpha}\})=K_\alpha+\frac{1}{2}\sum_{i=1}^{N_\alpha}\sum_{j=1}^{N_\alpha}{}^{'} u_{\alpha\alpha}({\bf r}_{ij}),
\label{Halphaalpha}
\end{equation}
and the prime denotes the restriction that $i\neq j$. The first term on the right side of Eq.~(\ref{Halphaalpha}) is the kinetic energy,
\begin{equation}
K_\alpha = \frac{1}{2m_\alpha}\sum_{i=1}^{N_\alpha}\,p_{\alpha\,i}^2,
\end{equation}
where $p_{\alpha\,i}$ is the momentum of the $\alpha$-type ion $i$. The second term is the potential energy of interaction between particles of the same species $\alpha$, described by the bare pair potential $u_{\alpha\alpha}({\bf r}_{ij})$, with ${\bf r}_{ij}={\bf r}_{i}-{\bf r}_{j}$ and ${\bf r}_{i}$ denoting the center position of the $i$th $\alpha$-type particle. The terms $H_{+-}$, $H_{\text{m}+}$, and $H_{\text{m}-}$ in Eq.~(\ref{GeneralHamiltonian}) correspond to the potential energies due to the bare interactions between particles of different species $\alpha\neq\beta$, i.e.,
\begin{equation}
H_{\alpha\beta}=\sum_{i=1}^{N_\alpha}\sum_{j=1}^{N_\beta}u_{\alpha\beta}({\bf r}_{ij}).
\end{equation}
	
The pair-interaction potential between two impermeable ions of species $\alpha$ and $\beta$, with hard-core radii $a_\alpha$ and $a_\beta$, whose center distance is $r$, is of the form $u_{\alpha\beta}(r)=u_{\alpha\beta}^\text{hs}(r)+u_{\alpha\beta}^\text{C}(r)$. Here, excluded-volume interactions are modeled by the hard-sphere pair potential,
\begin{equation}
u_{\alpha\beta}^\text{hs}(r) = \begin{cases} \infty\,, & r<a_\alpha+a_\beta \\ 0\,, & \text{otherwise} \end{cases}
\end{equation}
and electrostatic interactions by the Coulomb pair potential,
\begin{equation}
u_{\alpha\beta}^\text{C}(r)=k_\text{B}T\lambda_\text{B}\frac{z_\alpha z_\beta}{r},\,\,\,r>a_\alpha+a_\beta,
\label{PM_potent}
\end{equation}
where $\lambda_\text{B}=e^2/(\epsilon k_\text{B}T)$ is the Bjerrum length of the solvent, expressed here in Gaussian cgs units, and $k_\text{B}$ is the Boltzmann constant. The Bjerrum length is a length scale defined as the distance between two unit charges at which the Coulombic potential energy equals the thermal energy $k_\text{B}T$. For water at room temperature, $\lambda_\text{B}\approx 7.1$ \AA.
	
The PM system is subject to the global electroneutrality condition
\begin{eqnarray}
\sum_{\alpha=\{m,\pm\}} n_\alpha\;\!z_\alpha =0,
\label{global-electroneutrality}
\end{eqnarray}  
where $n_\alpha=N_\alpha/V$ is the mean number density of species $\alpha$. In equilibrium, the condition of local electroneutrality for the macroion species m can be stated as
\begin{eqnarray}
\sum_{\alpha=\{\pm\}} n_\alpha z_\alpha \int d^3r\, g_{\text{m}\alpha}(r) = - z_\text{m},
\end{eqnarray}  
where $g_{\text{m}\alpha}(r)$ is the partial radial distribution function for ion species m and $\alpha$. 
The partial radial distribution functions $g_{\alpha\beta}(r)$ and partial pair correlation functions $h_{\alpha\beta}(r)\equiv g_{\alpha\beta}(r)-1$ are determined by the partial pair interaction potentials $u_{\alpha\beta}(r)$, which are symmetric, i.e., $u_{\alpha\beta}(r)=u_{\beta\alpha}(r)$. This symmetry is inherited by the partial static (and dynamic) pair correlation functions.

In addition to structural properties, the PM also describes thermodynamic properties, such as the suspension pressure.
From simulations of the PM with its pairwise additive interactions, the suspension pressure $p$ can be obtained from the many-component virial equation of state \cite{Hansen-McDonald}. Using proper regularization of the Coulomb pair potential parts in conjunction with global electroneutrality expressed in Eq.~(\ref{global-electroneutrality}), the PM suspension pressure follows as \cite{Rasaiah-Friedman1968,Hansen-McDonald}
\begin{eqnarray}
\beta\frac{p_\text{PM}}{n}&=&1 + \frac{2\pi n}{3}\!\!\sum_{\alpha,\gamma=\{m,\pm\}}^{ }\!\!x_\alpha x_\gamma \sigma_{\alpha\gamma}^3 
 g_{\alpha\gamma}(\sigma^{+}_{\alpha\gamma}) \nonumber \\
 &-&\frac{\pi n \lambda_\text{B}}{3}\sum_{\alpha,\gamma=\{m,\pm\}}\!\!x_\alpha x_\gamma z_\alpha z_\gamma \sigma_{\alpha\gamma}^2\nonumber\\
 &+& \frac{ 2\pi n \lambda_\text{B}}{3}\sum_{\alpha,\gamma=\{m,\pm\}}\!\!x_\alpha x_\gamma z_\alpha z_\gamma
 \int_{\sigma_{\alpha\gamma}}^\infty\!\!dr\;\!r\;\!h_{\alpha\gamma}(r)\,.
\end{eqnarray}  
Here, $x_\alpha = n_\alpha/n$ is the molar fraction of $\alpha$-type ions and $n=\sum_\gamma n_\gamma = n_\text{m} + n_{+} + n_{-}$ is the total ion number density. Moreover, $\sigma_{\alpha\gamma}=a_\alpha + a_\gamma$. The contact values, $g_{\alpha,\gamma}(\sigma^{+}_{\alpha\gamma})$, of the three partial radial distribution functions differ from those of neutral hard spheres because of the influence of the Coulomb interactions. The pressure contribution involving the square of $\sigma_{\alpha\gamma}$ results from the property, $h_{\alpha\gamma}(r< \sigma_{\alpha\gamma})=-1$.
		
For systems with relatively small size and charge asymmetries between ionic species, bulk properties can be extracted from the PM usually only by computer simulations, except in the dilute limit. With increasing asymmetry and concentration, however, simulations rapidly become  computationally more expensive, especially when dynamic properties are considered. Fortunately, if the asymmetries are sufficiently large, approximations become possible based on exploiting the wide separation of length and times scales of different ionic species. Owing to the large size difference between macroions and microions, the dynamics of the microions is much faster than that of the macroions. If the structure and dynamics of the macroions is of most interest, this asymmetry allows for integrating out the degrees of freedom of the microions, giving rise to structural and dynamic equations for a system of effective (``microion-dressed'') macroions only. (Although this coarse-graining procedure can be formally applied to any species in a classical system, it is of practical utility only when applied to the smaller species.) The price to pay for this reduction is the introduction of three-body and higher-order effective interactions not present in the PM, as well as the occurrence of a one-body contribution to the volume (grand) free energy, associated with microion entropy and macroion-microion interactions. 
		
\subsection{Effective One-Component Model}\label{Sec:OCM}
For simplicity, we now restrict the discussion to monovalent microions ($a_\pm=0, z_\pm=\pm1$) and macroions that are monodisperse in size ($a_\text{m}=a$) and of negative charge $z_\text{m} e = - Z e$ with bare valence $Z>0$. We consider the suspension to be in osmotic (Donnan) equilibrium with a 1:1 electrolyte reservoir of ion concentration $2n_\text{res}$, which is separated from the suspension by an ideal membrane that is permeable only to microions and solvent. For simplicity, we assume the bare macroion valence $Z$ to be constant, independent of ion concentrations and ionic strength. Thus, we neglect possible chemical charge regulation effects. The condition of global electroneutrality can be rewritten as $ZN_\text{m}=\langle N_+\rangle -\langle N_-\rangle$, where $\langle N_{-}\rangle$ denotes the equilibrium number of (monovalent) coions in the system, which equals the number $N_\text{s}$ of salt ion pairs, and $\langle N_{+}\rangle$ is the equilibrium total number of (monovalent) counterions, consisting of $N_s$ salt counterions and $Z N_\text{m}$ counterions released from the macroion (colloid) surfaces. The equilibrium number density of salt ion pairs, $n_\text{s}=N_\text{s}/V$, in the suspension is determined by chemical equilibrium between the suspension and the reservoir, specifically, by the equality, $\mu_\pm=\mu_\text{res}$, of the chemical potentials of cations and anions, $\mu_\pm$, in the suspension and the microion chemical potential, $\mu_\text{res}=k_\text{B}T \ln \left(\Lambda_0^3 n_\text{res}\right)$, in the reservoir, assuming that all microions have the same thermal de Broglie wavelength $\Lambda_0$. Thus, a suspension in Donnan equilibrium has an equilibrium salt concentration that is determined by the salt concentration of the reservoir. A closed suspension, which does not exchange microions with a reservoir, with a given fixed salt concentration, can be directly mapped to an equivalent system in Donnan equilibrium with a reservoir of salt concentration $n_\text{res} \geq n_\text{s}$.
		
In this McMillan-Mayer implicit solvent model, the semi-grand canonical partition function of the suspension can be expressed as
\begin{equation}
\Xi=\langle\langle \text{e}^{-\beta\mathcal{H}}\rangle_\mu\rangle_\text{m}\,,
\end{equation}
where $\beta\equiv 1/(k_\text{B}T)$. The angular brackets denote a canonical trace over macroion (m) center-of-mass coordinates and a grand-canonical trace over microion ($\mu$) coordinates. By tracing out the microion degrees of freedom for a fixed configuration of macroions, one obtains the formally exact relation
\begin{equation}
\Xi= \langle \text{e}^{-\beta H_\text{eff}}\rangle_\text{m}\,,
\end{equation}
with the effective Hamiltonian of pseudo-macroions
\begin{equation}
H_\text{eff} = K_\text{m}+E_\text{vol}(n_\text{m})+U_\text{eff}(\{{\bf r}^{N_\text{m}}\}, n_\text{m})\,,
\label{EffHamilt}
\end{equation}
where $K_\text{m}$ accounts for the translational kinetic energy of the macroions, $E_\text{vol}(n_\text{m})$ is the macroion configuration-independent volume energy, and $U_\text{eff}(\{{\bf r}^{N_\text{m}}\}, n_\text{m})$ is the configuration-dependent effective $N$-particle interaction energy of the macroions in an equivalent one-component model (OCM) of microion-dressed macroions. The latter describes the interactions among macroions mediated by the microions, and can be written as \cite{Denton_PRE_2006,Dijkstra2000}
\begin{equation}
U_\text{eff} = \frac{1}{2}\sum_{i\neq j=1}^{N_\text{m}}u^{(2)}_\text{eff}(r_{ij})+\frac{1}{3!}\sum_{i\neq j\neq k=1}^{N_\text{m}}u^{(3)}_\text{eff}(r_{ij},r_{ik})+\ldots\,,
\label{nbody_Ueff}
\end{equation}
where $u^{(n)}_\text{eff}$, with $2\leq n\leq N_\text{m}$, represent the different n-body effective interactions.
		
Denoting the semi-grand free energy of the suspension in the OCM by $\Omega=-k_\text{B}T \ln \Xi$, the suspension pressure, $p=-(\partial\Omega/\partial V)_{\text{res}}$, is then given by the generalized virial equation in the OCM \cite{Denton_PRE_2006,Denton_BookChap_2007},
\begin{equation}\label{eq:OsmoticPressure}
p = p_\text{vol} + n_\text{m} k_\text{B}T - \frac{1}{3V}\big < \sum_{i=1}^{N_\text{m}} {\bf r}_i\cdot\frac{\partial U_\text{eff}}{\partial {\bf r}_i}\big>_\text{eff} - \big<\frac{\partial U_\text{eff}}{\partial V}\big>_\text{eff}\,,
\end{equation}
including the contribution to the pressure from the volume energy, $p_\text{vol} = -\partial E_\text{vol}/\partial V$, and an extra volume derivative term due to the density dependence of $U_\text{eff}\left(\{{\bf r}^{N_\text{m}}\};n_\text{m}\right)$. Here, the angular brackets $\langle\cdots\rangle_\text{eff}$ denote the canonical ensemble average with respect to the equilibrium distribution function, $P_\text{eq}(\{{\bf r}^{N_\text{m}}\}) \propto\exp[-\beta U_\text{eff}(\{{\bf r}^{N_\text{m}}\})]$, of pseudo-macroions, which should not be confused with the canonical macroion trace $\langle\cdots\rangle_\text{m}$ over macroion positions and momenta. Note that the volume derivative of $U_\text{eff}$ in Eq.~(\ref{eq:OsmoticPressure}) is taken for fixed reservoir ion chemical potentials and, hence, for fixed $n_\text{res}$. It is important to emphasize that the generalized virial equation does not suffer from ambiguities introduced when state-dependent effective pair potentials are combined in an {\em ad hoc} manner with the compressibility and virial equation of state expressions for one-component simple liquids \cite{Louis2002,Hoffmann_JCP_2004}.
	
For sufficiently weakly charged macroions, with electrostatic coupling constant $Z\lambda_\text{B}/a<{\cal O}(10)$, and sufficiently weakly correlated, monovalent microions, the effective potential energy $U_\text{eff}(\{{\bf r}^{N_\text{m}}\};n_\text{m})$ is well-approximated by including only up to two-body contributions in Eq.~(\ref{nbody_Ueff}) with effective macroion pair interaction \cite{Denton_PRE_2000,Denton_BookChap_2007}
\begin{equation}
\beta  u_{\text{eff}}(r)=\lambda_\text{B}Z^2\left(\frac{\exp(\kappa a)}{1+\kappa a}\right)^2\frac{\exp(-\kappa r)}{r},\,\,\,r>\sigma\,,
\label{YukawaPot}
\end{equation}
where $\sigma = 2a$ is the macroion diameter. The screening constant (inverse screening length) is given by
\begin{equation}
\kappa=\sqrt{\frac{4\pi\lambda_\text{B}}{1-\phi}\left(Zn_\text{m}+2n_s\right)}\,,
\label{kappasquare}
\end{equation}
with $n_s\le n_\text{res}$.
Depending on the analytical treatment, the excluded-volume factor $1/(1-\phi)$ may appear \cite{Denton_PRE_2000} or not \cite{vanRoij_PRE_1999}. In the salt dominated regime ($n_\text{res}\gg Zn_\text{m}$),
\begin{equation}
\kappa\simeq\sqrt{8\pi\lambda_\text{B}n_\text{res}}\,.
\end{equation}
We note that the basic approximation underlying the effective pair potential $u_\text{eff}(r)$ [Eq.~(\ref{YukawaPot})] between macroions is a {\it linearized} mean-field Poisson-Boltzmann description of the microions, which neglects all but long-range mutual correlations between free microions, including those due to their nonzero size. This mean-field approximation is most accurate for monovalent microions, a low microion-macroion size ratio, and a relatively low macroion valence. We note further that defining from Eq.~(\ref{kappasquare}) a released-counterion screening constant, $\kappa_\text{c}\equiv\sqrt{4\pi\lambda_\text{B}Zn_\text{m}/(1-\phi)}$, and a salt-ion screening constant, $\kappa_\text{s}\equiv\sqrt{8\pi\lambda_\text{B}n_\text{s}/(1-\phi)}$, identifies the counterion-dominated regime with the condition $\kappa_\text{c}\gg\kappa_\text{s}$ and the salt-dominated regime with $\kappa_\text{c}\ll\kappa_\text{s}$.
		
It should be noted that $Z \lambda_\text{B}/a <{\cal O}(10)$ is a criterion for negligible nonlinear response of {\em monovalent} microions in a mean-field description, i.e., for the validity of linearized PB-based theories without charge renormalization. The criterion has been confirmed \cite{LuDenton2010}, e.g., by comparing with MC data in the PM \cite{Linse2000}. For monovalent microions and $a \gg \lambda_\text{B}$, PB-based mean-field theories are applicable even for strong nonlinear response, where $Z_\text{eff}$ is significantly lower than $Z$, since microion correlations are still very weak \cite{LevinTrizacBocquet:2003}.

When $U_\text{eff}$ may be assumed pairwise additive, the generalized virial equation for the suspension pressure [Eq.~(\ref{eq:OsmoticPressure})] reduces to \cite{Hansen-McDonald}
\begin{equation}\label{eq:PressureTwoBody}
p = p_\text{vol} + p_\text{vir} + p_\text{den},
\end{equation}
where the contribution 
\begin{eqnarray}
p_\text{vir}=n_\text{m}k_\text{B}T-\frac{2\pi}{3}n_\text{m}^2\int_{0}^{\infty}dr\, r^3g(r)\frac{\partial u_\text{eff}(r)}{\partial r},
\label{p-OCM}
\end{eqnarray}
corresponding to the second and third terms on the right side of Eq.~(\ref{eq:OsmoticPressure}), represents the virial pressure of the OCM system {\em without} consideration of the density dependence of $u_\text{eff}(r)$, and the contribution
\begin{eqnarray}
p_\text{den}=2\pi n_\text{m}^3\!
\!\int_{0}^{\infty}\!\!dr\, r^2 g(r)\frac{\partial u_\text{eff}(r)}{\partial n_\text{m}},
\label{p-den}
\end{eqnarray}
corresponding to the last term on the right side of Eq.~(\ref{eq:OsmoticPressure}), accounts for the 
density dependence of $u_\text{eff}(r)$. 

\begin{figure}
\centering
\includegraphics[width=6cm]{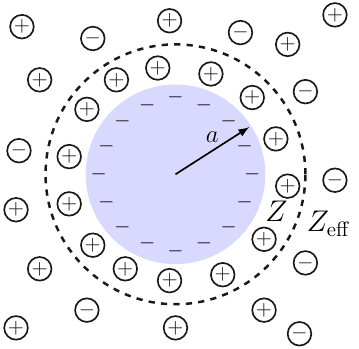}
\caption{Schematic illustration of charge renormalization: Surrounding a colloidal macroion of radius $a=\sigma/2$ and charge $-Z e<0$, some counterions become strongly associated and quasi-condense on the macroion surface, reducing the bare valence $Z>0$ to an effective valence $Z_\text{eff}<Z$.}
\label{fig:sketch}
\end{figure}

Within the mean-field Poisson-Boltzmann-type description of charge-stabilized colloidal suspensions, different theories predict different expressions for $u_\text{eff}$ and $E_\text{vol}$, which are generally accurate only when $Z\lambda_\text{B}/a<{\cal O}(10)$, in which case the microion distributions are relatively weakly perturbed by the macroion charges \cite{vanRoij_PRE_1999,Denton_PRE_2000}. For more strongly coupled suspensions, linearized theories can be extended by renormalizing the bare macroion valence $Z$ and suspension screening constant $\kappa$ to incorporate the nonlinear response of the microions to the strong electric field of the macroions. Essentially, charge renormalization maps the highly-charged nonlinear system onto an equivalent linear one that accounts for the higher-order nonlinear contributions. The latter are subsumed into an effective screening constant $\kappa_\text{eff}$, describing the weakly-associated (linearly responding) microions, and an effective macroion valence $Z_\text{eff}$, describing the mean effect of the colloids and the strongly-associated (nonlinearly responding) microions (see Fig.~\ref{fig:sketch}). This approach leads to an effective pairwise interaction potential between macroions similar to that in Eq.~(\ref{YukawaPot}), but with the replacements $\kappa\rightarrow\kappa_\text{eff}$ and $Z\rightarrow Z_\text{eff}$:
\begin{equation}
\beta  u_{\text{eff}}(r)=\lambda_\text{B}Z_\text{eff}^2\left(\frac{\exp(\kappa_\text{eff} a)}{1+\kappa_\text{eff} a}\right)^2\frac{\exp(-\kappa_\text{eff} r)}{r},\,\,\,r>\sigma.
\label{RenormYukawaPot}
\end{equation}
This newly defined $u_\text{eff}(r)$ allows calculating the macroion-macroion radial distribution function $g(r)$ and static structure factor $S(q)$, which characterize pair correlations in real and Fourier space, respectively, of highly-charged colloidal suspensions.
		
In the next section, we briefly outline the most prominently used charge renormalization methods that we have assessed. We restrict our study to methods of the simplest form, those that do not require recursive calculation of distribution functions for computing effective interaction parameters. We focus on two main classes of method, based on (1) single-center models, where the system is represented by a singled-out colloid, and (2) multi-center models, which include all colloids interacting via effective pair potentials. In the first class, we study {\it cell-model-based} methods and {\it renormalized jellium models}. In the second class, we analyze {\it renormalized linear response theory} and {\it shifted Debye-H\"uckel approximation} methods. We focus especially on the appropriate linearization of the suspension mean-field electrostatic potential $\Phi$.
  
\section{Charge Renormalization Methods}\label{methods}
	
\subsection{Cell Model Approximations} \label{Subsec:CMapp}
Commonly used charge renormalization schemes are based on a simplifying cell model (CM) representation of the suspension. These schemes exploit the observation that for strongly repelling spherical colloids, each colloid is surrounded by a region that is devoid of other colloids \cite{Colla_JCP_2009}. On assuming a crystalline-like structure, a Wigner-Seitz cell tessellation can be applied, where each Wigner-Seitz cell is approximated by a spherical cell containing a single colloid at its center. Each cell is treated identically and independently (no inter-cell correlations) and is considered to be in Donnan equilibrium with a 1:1 strong electrolyte reservoir of monovalent microions of fixed concentration $2n_\text{res}$. The only bulk property of the colloid environment is the volume fraction $\phi$, related to the radius $R$ of the spherical cell via $R=a/\phi^{1/3}$. Considering Boltzmann distributions for the microion number densities, the total electrostatic suspension potential in the cell region $a<r<R$ is determined by the nonlinear PB equation
\begin{equation}
\Phi''(r)+\frac{2}{r}\Phi'(r) = \kappa_{\text{res}}^2 \sinh[\Phi(r)],
\label{non-linPBeq}
\end{equation}
where $\Phi(r)=\beta e\psi(r)$ is the reduced form of the total electrostatic potential $\psi(r)$, and $\kappa_{\text{res}}=\sqrt{8\pi\lambda_{\text{B}}n_\text{res}}$ is the reservoir electrostatic screening constant. The inner and outer boundary conditions for ion-impermeable macroions are given by
\begin{equation}
\Phi'(a) = \frac{Z\lambda_{\text{B}}}{a^2},\quad\Phi'(R) = 0
\label{BC_non-linPBeq}
\end{equation}
at the colloid surface and cell edge, respectively. The inner boundary condition dictates that the radial electric field on the colloid surface obeys Gauss's law, while the outer one ensures that the cell is overall electroneutral. 
The suspension salt ion pair concentration, $n_\text{s}$, is calculated by integrating the coion (anion) number density profile $n_-(r)$ over the volume of the cell, i.e.,
\begin{equation}\label{eq:nsCM}
n_\text{s}=\frac{4\pi}{V_\text{c}}\int_0^R\!\! n_{-}(r)\;\!r^2\;\!dr=\frac{4\pi n_\text{res}}{V_\text{c}^\text{f}}(1-\phi)\int_a^R\!\! e^{\Phi(r)}\;\!r^2\;\!dr,
\end{equation}
thus relating $n_\text{s}$ to the reservoir salt concentration. Here, $V_\text{c}=4\pi R^3/3$ is the cell volume and $V_\text{c}^\text{f}=V_\text{c}(1-\phi)$ is the free volume of the cell (unoccupied by the macroion). Since $\Phi<0$ for negatively-charged macroions, then $n_\text{s}\leq n_\text{res}(1-\phi)$, expressing the Donnan salt-expulsion effect. The CM discussed here is also referred to as the nonlinear Poisson-Boltzmann cell model (PBCM) approximation, since it involves the nonlinear, Boltzmann-distributed microion density profiles. 
				
For defining effective interaction parameters characterizing the effective macroion pair potential with charge renormalization effects included, we expand the nonlinear term in Eq.~(\ref{non-linPBeq}) up to first order around a yet undetermined reference potential $\tilde{\Phi}$, leading to a PB equation for the linearized electrostatic potential $\Phi_l(r)$, i.e.
\begin{equation}
\Phi_{l}''(r)+\frac{2}{r}\Phi_{l}'(r)=\kappa_{\text{eff}}^2[\Phi_l(r)-\tilde{\Phi}+\gamma],
\label{linearPBequation}
\end{equation}
where $\gamma=\tanh(\tilde{\Phi})$ and 
\begin{equation}
\kappa_{\text{eff}}^2 = \kappa_\text{res}^2\cosh(\tilde{\Phi})
\label{EffScren_CM}
\end{equation}
is a new renormalized (effective) screening constant. In order to make further progress, we need to specify the reference potential $\tilde{\Phi}$. Two convenient choices are the (nonlinear) electrostatic potential value at the cell edge, i.e., $\tilde{\Phi} =\Phi(R)$, and the volume-averaged, mean (nonlinear) electrostatic potential, $\tilde{\Phi}=\bar{\Phi}=\frac{4\pi}{V_\text{c}^\text{f}}\int_a^R dr\,r^2\Phi(r)$.
The first choice, which linearizes the potential with respect to its edge value, is referred to as ``edge linearization'' for short. The second choice, which linearizes the potential with respect to its mean value, is called ``mean linearization'' for short. For either choice, one boundary condition that $\Phi_l(r)$ must fulfill is $\Phi_{l}'(R)=0$, which ensures electroneutrality of the cell. The second boundary condition, required to uniquely determine $\Phi_l(r)$, depends on the choice of linearization. For edge linearization, $\Phi_l(R)=\Phi(R)$, equating the linearized and nonlinear potentials at the cell edge. For mean linearization, $\frac{4\pi}{V_\text{c}^\text{f}}\int_a^R dr\,r^2\Phi_l(r)=\bar{\Phi}$, equating the free-volume-averaged mean values of the linear and nonlinear potentials.
The general solution of Eq.~(\ref{linearPBequation}) has the form
\begin{equation}
\Phi_{l}(r) = c_+ \frac{e^{\kappa_\text{eff} r}}{r}+c_- \frac{e^{-\kappa_\text{eff} r}}{r}+\tilde{\Phi}-\gamma,
\label{genLinSol}
\end{equation}
where $c_\pm$ are constants determined by the boundary conditions. Note that, since the cell radius is finite for $\phi>0$, also the positive exponential term must be retained.
				
We now proceed to discuss the renormalized (effective) macroion valence, $Z_\text{eff}$, and screening constant, $\kappa_\text{eff}$, which together determine the renormalized effective pair potential. For a sufficiently weak bare macroion charge, where $Z_\text{eff}=Z$, the linear electrostatic part of the DLVO potential is recovered [Eq.~(\ref{YukawaPot})]. For a specified linearization reference potential $\tilde{\Phi}$, the corresponding renormalized screening constant follows from Eq.~(\ref{EffScren_CM}). For specified $\tilde{\Phi}$ and hence $\Phi_l(r)$, the renormalized valence is then consistently defined by the electrostatic boundary or surface charge (SC) condition in the PBCM,
\begin{equation}
Z_\text{eff}:=\frac{\Phi_l'(a)\,a^2}{\lambda_\text{B}},
\label{AlexZeffdef}
\end{equation}
on the surface of the macroion sphere based on the linearized potential. This definition of $Z_\text{eff}$ was first introduced in the pioneering work of Alexander {\it et al.} using edge linearization~\cite{Alexander1984}. Depending on the choice of $\tilde{\Phi}$, different values for $Z_\text{eff}$ and  $\kappa_\text{eff}$ are obtained. 
				
Following Boon {\it et al.} \cite{Boon_PNAS_2015}, an alternative effective valence can be defined from analytically extrapolating $\Phi_l(r)$ to the center of the cell by assuming a pointlike colloid with effective valence $Q_\text{eff}$ defined by
\begin{equation}
Q_\text{eff}:=\lim_{r\rightarrow 0}\frac{\Phi_l'(r)r^2}{\lambda_\text{B}}.
\label{Qeff_def}
\end{equation}
The quantity $Q_\text{eff}$ is called the extrapolated point charge (EPC) 
and its definition is motivated by noting that $\Phi_l(r\approx0)=Q_\text{eff}\lambda_\text{B}/a$, since no screening is operative at $r=0$. In order to directly compare the two renormalized valence definitions, one needs to account for the geometric factor in the renormalized DLVO potential. Thus, the renormalized valence $Z_\text{eff}$ in the EPC scheme is actually defined as
\begin{equation}
Z_\text{eff}=\frac{1+\kappa_\text{eff}a}{e^{\kappa_\text{eff}a}}Q_\text{eff}.
\label{EPCZeffdef}
\end{equation}
This EPC definition of $Z_\text{eff}$ in Eq.~(\ref{EPCZeffdef}) differs from the SC definition of $Z_\text{eff}$ in Eq.~(\ref{AlexZeffdef}) even though the same $\Phi_l(r)$ is used. The difference can be evaluated fully analytically in the DH regime, where $|\phi(a)|\ll 1$ and counterion quasi-condensation is absent. Here, $\Phi_l(r) = \Phi_\text{DH}(r)$ reduces to the solution of the two-point boundary value problem in Eqs.~(\ref{non-linPBeq}) and (\ref{BC_non-linPBeq}) for $\phi =\phi_\text{DH}(r)$ and $\sinh\left(\phi_\text{DH}\right)=\phi_\text{DH}$. In the DH regime, it holds that $Z^\text{SC}_\text{eff}=Z$ independent of $\kappa_\text{res} a$ whereas \cite{Boon_PNAS_2015}
\begin{equation}
\frac{Z^\text{EPC}_\text{eff}}{Z} =\frac{2 (\kappa a +1) 
\left(\kappa R \cosh(\kappa  R)-\sinh(\kappa  R)\right) e^{\kappa  (R-2 a)} }{(1-\kappa a)(\kappa  R+1)+(\kappa a +1) (\kappa  R-1) e^{2 \kappa  (R-a)}}\,,
\end{equation}
with $\kappa=\kappa_\text{res}$. In the low salinity limit, the EPC effective valence in the DH regime reduces to $Z^\text{EPC}_\text{eff}=Z/(1-\phi) +{\cal O}\left(\left(\kappa_\text{res} a\right)^2\right)$. As further discussed in Sec.~\ref{Zeff>Z}, the EPC definition allows for an effective valence exceeding the bare one (i.e., $Z^\text{EPC}_\text{eff}>Z$). However, this effect results from exclusion of counterions from the macroion core, rather than from nonlinear response. It is not due to counterion quasi-condensation, which tends to decrease $Z_\text{eff}$ below its bare value $Z$.
							
Depending on the choice of $\tilde{\Phi}$ with according boundary conditions, and on the SC or EPC definitions of $Z_\text{eff}$ in Eqs.~(\ref{AlexZeffdef}) and (\ref{EPCZeffdef}), respectively, four different expressions for $Z_\text{eff}$ in terms of the independent system variables $\kappa_\text{res}a$, $Z\lambda_\text{B}/a$, and $\phi$ are obtained:
\begin{equation}
\frac{Z_\text{eff}\lambda_\text{B}}{a}=\gamma\;\!F_i(\kappa_\text{eff}a,\phi^{-1/3}).
\label{Zeff_Def}
\end{equation}
Recall here that $\phi^{-1/3}=R/a$. The four functions $F_i$, with $i\in\{1,...,4\}$, are obtained from $\Phi_l(r)$ in Eq.~(\ref{genLinSol}) with appropriately determined coefficients $c_\pm$. For the SC definition of $Z_\text{eff}$ and linearization with respect to the edge potential (SC edge) in the PBCM \cite{Alexander1984},
\begin{equation}
F_1(x,y)=\frac{1}{x}\{(x^2y-1)\sinh[x(y-1)]+x(y-1)\cosh[x(y-1)]\},
\label{F1_Def}
\end{equation}
while for the SC definition of $Z_\text{eff}$ and linearization with respect to the mean potential (SC mean) in the PBCM,
\begin{equation}
F_2(x,y)=\frac{x^2(y^3-1)}{3}.
\end{equation}
Alternatively, the EPC definition of $Z_\text{eff}$ with edge linearization (EPC edge) yields \cite{Boon_PNAS_2015}
\begin{equation}
F_3(x,y)=\left(\frac{1}{x}+1\right)e^{-x}[xy\cosh(xy)-\sinh(xy)],
\label{F3_Def}
\end{equation}
while the EPC definition of $Z_\text{eff}$ with linearization with respect to the mean potential (EPC mean) gives
\begin{equation}
F_4(x,y)=F_2(x,y)\;\!\xi(x,y),
\label{F4_Def}
\end{equation}
with
\begin{eqnarray}
\xi(x,y)\!=\!\frac{1+x}{e^{x}}\!\!\left[ \frac{e^x\left(1 + xy + e^{2xy}\left(xy-1\right) \right)}
{e^{2xy}\left(1+x\right)\left(xy-1\right) 
+ e^{2x}\left(x-1\right)\left(1+xy\right)}
\right]. \nonumber\\
\label{F4B_Def}
\end{eqnarray}
Notice that, in all four cases, $Z_\text{eff}\to Z$ for $Z\lambda_\text{B}/a\ll1$ and $\phi\to 0$. For larger values of $Z$, nonlinear screening comes into play, triggered by a strong accumulation of counterions near the colloid surface, so that $Z_\text{eff}<Z$. With increasing $Z$, $Z_\text{eff}$ approaches a plateau value $Z_\text{eff}^\text{sat}$. This renormalized charge saturation is due to the invoked mean-field PB approximation, which neglects discreteness and nonzero sizes of microions, allowing for an arbitrarily high surface concentration of counterions \cite{Trizac_JPCM_2002,Aubouy2003}. The approach of $Z_\text{eff}$ to an (apparent) plateau value, however, is a genuine physical effect for large $Z\lambda_\text{B}/a$.
				
Using Eq.~(\ref{non-linPBeq}), one can show that $|\Phi(r)|$ is a strictly monotonically decreasing function with increasing radial distance $r$, implying that $|\bar{\Phi}|>|\Phi(R)|$ and consequently that $\kappa_\text{eff}(\text{mean})>\kappa_\text{eff}(\text{edge})$. Thus, there is stronger effective screening for mean than for edge linearization.
				
For given $\Phi_l$(r), the linear microion number density profiles inside the cell are given by 
\begin{equation}
n_\pm^l(r)=n_\text{res}e^{\mp\tilde{\Phi}}[1\pm\tilde{\Phi}\mp\Phi_l(r)].
\end{equation}
Integrating the linear microion charge density, $n_+^l(r)-n_-^l(r)$, and the total microion density, $n_+^l(r)+n_-^l(r)$, over the free volume $V_\text{c}^\text{f}$ of a cell, and using the definition of $Z_\text{eff}$ in Eq.~(\ref{AlexZeffdef}), we obtain for the effective screening parameter under mean linearization \cite{Trizac_Langmuir:2003}
\begin{equation}
\kappa_\text{eff}^2=\frac{4\pi\lambda_\text{B}}{V_\text{c}^\text{f}}\left(Z_\text{eff}+2N_\text{s}^\text{eff}\right)=\frac{4\pi\lambda_\text{B}}{1-\phi}\left(Z_\text{eff}n_\text{m}+2n_\text{s}^\text{eff}\right).
\label{effkappa_mean}
\end{equation}
Here, $N_s^\text{eff}=4\pi\int_a^R dr\,r^2\,n_-^l(r)$ and $n_\text{s}^\text{eff}=N_\text{s}^\text{eff}/V_\text{c}$ are the mean number and mean concentration, respectively, of free salt ion pairs in the suspension. From Eq.~(\ref{effkappa_mean}), we obtain $n_\text{s}^\text{eff}$ as
\begin{equation}
n_\text{s}^\text{eff}\,a^3=\frac{a}{8\pi\lambda_\text{B}}\left[(1-\phi)(\kappa_\text{eff}\,a)^2-3\phi\frac{Z_\text{eff}\lambda_\text{B}}{a}\right].
\label{nseff_mean}
\end{equation}
Alternatively, for edge linearization, where $\tilde{\Phi}=\Phi(R)$,
\begin{eqnarray}
\kappa_\text{eff}^2&=&\frac{4\pi\lambda_\text{B}}{(1-\gamma)V_\text{c}^\text{f}}\left(Z_\text{eff}+\frac{2N_s^\text{eff}}{1+\gamma}\right)\nonumber\\
&=&\frac{4\pi\lambda_\text{B}}{(1-\phi)(1-\gamma)}\left(Z_\text{eff}n_\text{m}+\frac{2n_\text{s}^\text{eff}}{1+\gamma}\right)\,,
\label{effkappa_edge}
\end{eqnarray}
with the effective suspension salt concentration given by \cite{Trizac_Langmuir:2003}
\begin{equation}
n_\text{s}^\text{eff}=\frac{\kappa_\text{eff}^2}{8\pi\lambda_\text{B}}(1-\phi)(1-\gamma^2)-\frac{1}{2}Z_\text{eff}n_\text{m}(1+\gamma),
\label{nseff_edge}
\end{equation}
where $\gamma = \tanh[\Phi(R)]$. We notice from Eqs.~(\ref{effkappa_mean}) and (\ref{effkappa_edge}) that the square of the effective screening constant consists of two additive contributions, namely a contribution proportional to $Z_\text{eff}n_\text{m}$, arising from the free (i.e., uncondensed) part of the macroion-surface-released counterions, and a contribution due to free salt ion pairs. When implementing the EPC scheme for incorporating nonlinear effects, approximations for the effective suspension salt concentration are obtained by substituting the corresponding renormalized interaction parameters into Eq.~(\ref{nseff_mean}) or (\ref{nseff_edge}) for mean or edge linearization, respectively.
	            
From a physical viewpoint, mean linearization is more consistent than edge linearization, since in the former case the density variations of the nonlinear potential around $\Phi_l$ are only of second order, which is not so in the latter case. Moreover, only in mean linearization is the Donnan expression for $n_\text{s}$ correctly recovered to first order, where $Z_\text{eff}=Z+\mathcal{O}([Z\lambda_\text{B}/a]^2)$, and is the correct expression for $\kappa_\text{eff}^2$ also recovered for $n_\text{res}\rightarrow 0$. Different from Eq.~(\ref{effkappa_edge}), which invokes the factors $1/(1\pm\gamma)$, Eq.~(\ref{effkappa_mean}) naturally splits into free counterion and salt ion contributions. Using a proper pressure definition, mean linearization further guarantees positive pressures \cite{Deserno2002}. Another convenient feature of mean linearization is that $Z_\text{eff}$ is directly obtained from $\bar{\Phi}$, according to 
\begin{equation}
\frac{Z_\text{eff}\lambda_\text{B}}{a}=\left(\frac{1-\phi}{3\phi}\right)(\kappa_\text{res}a)^2\sinh(\bar{\Phi}).
\end{equation}
				
In the salt-free case, where $n_\text{res}=0$, the suspension is a closed system, since no counterions can leave the cell into the microion-empty reservoir because of electroneutrality. For a system in Donnan equilibrium, it holds that $|\Phi(R,n_\text{res})|\to\infty$ for $n_\text{res}\to 0$. To recover the salt-free system as a limiting case, one therefore redefines the salt-free system potential by
\begin{equation}
\Phi(r)=\lim_{n_\text{res}\to 0}\left[\Phi(r;n_\text{res})-\Phi(R;n_\text{res})\right].
\end{equation}
The salt-free potential satisfies the nonlinear PB-type equation for negatively-charged macroions,
\begin{equation}
\Phi''(r)+\frac{2}{r}\Phi'(r) = -k_0^2 e^{-\Phi(r)},\qquad a<r\leq R,
\label{non-linPBeqSaltFree}
\end{equation}
where $k_0=\sqrt{4\pi\lambda_\text{B}n_+^0}$ is the new electrostatic screening constant and $n_+(r)=n_+^0\exp[-\Phi(r)]$ is the nonlinear counterion number density profile. In addition to the boundary conditions on $\Phi(r)$ [Eq.~(\ref{BC_non-linPBeq})], electroneutrality dictates that
\begin{equation}
k_0^2=\frac{Z\lambda_\text{B}}{\int_a^R dr\, r^2\exp[-\Phi(r)]}=-\Phi''(R),
\label{k0-SF}
\end{equation}
since $\Phi(R)=0$. The present nonlinear boundary-value problem can be solved self-consistently. Alternatively, it can be mapped onto a boundary-value problem invoking a third-order differential equation, following ref.~\cite{Ruiz-Reina2008}.
    
Linearization of Eq.~(\ref{non-linPBeqSaltFree}) with respect to an arbitrary reference value $\tilde{\Phi}$ gives
\begin{equation}
\Phi_{l}''(r)+\frac{2}{r}\Phi_{l}'(r)=-k_0^2\exp(-\tilde{\Phi})[1-\Phi_{l}(r)+\tilde{\Phi}],\quad a<r\leq R,
\label{linearPBeqSFedge}
\end{equation}
with boundary conditions similar to systems with salt, and $k_0$ determined from the nonlinear boundary-value problem. The renormalized screening parameter in the salt-free case follows then explicitly as 
\begin{equation}
\kappa_{\text{eff}}^2 = k_0^2\times\begin{cases}e^{-\bar{\Phi}},& \text{mean}\\e^{-\Phi(R)},& \text{edge}\end{cases}.
\end{equation}
Notice that $e^{-\Phi(R)}=1$ under salt-free conditions. Therefore, as in the Donnan equilibrium case, $\kappa_\text{eff}(\text{mean})>\kappa_\text{eff}(\text{edge})$. The renormalized macroion valence in the four considered cases follows from Eq.~(\ref{Zeff_Def}) with $\gamma=1$,
\begin{equation}
\frac{Z_\text{eff}\lambda_\text{B}}{a}=F_i(\kappa_\text{eff}a,\phi^{-1/3}),
\label{Zeff_DefSF}
\end{equation}
for $i\in\{1,...,4\}$ with $F_i(x,y)$ still given by Eqs.~(\ref{F1_Def})-(\ref{F4B_Def}). For $Z_\text{eff}$ in the SC mean model, it follows explicitly that
\begin{equation}
\frac{Z_\text{eff}\lambda_\text{B}}{a}(n_\text{res}=0)=\left(\frac{1-\phi}{3\phi}\right)(k_0 a)^2e^{-\tilde{\Phi}}.
\label{ZeffSCmodel}
\end{equation}
Since $\tilde{\Phi}=\Phi(R)=0$ in edge linearization, and $\tilde{\Phi}=\bar{\Phi}$ in mean linearization, with $\bar{\Phi}<0$ for negatively-charged macroions, it follows that $Z_\text{eff}(\text{mean})>Z_\text{eff}(\text{edge})$ under salt-free conditions. This inequality holds empirically also for salty systems. 
				
In the salt-free case, and in the dilute limit $n_\text{m}\to 0$ ($R\to \infty$) when $Z_\text{eff}=Z$ and $\kappa_\text{eff}=0$, we notice that $\Phi(r)\to Z\lambda_\text{B}/r$. This limit expresses the domination of the counterion entropy over the electrostatic energy of counterion-macroion attraction in an unbounded three-dimensional space ($a<r<R\to\infty$), in which case the $Z$ counterions around the central macroion are randomly distributed throughout space and thus do not contribute to screening.

\subsection{Renormalized Jellium Model (RJM)}\label{Subsec:RJM}
				In contrast to the spherical CM, which is motivated by a crystalline-like structure of the suspension, the renormalized jellium model (RJM) presumes a fluid suspension. Based on the jellium approximation (JA) \cite{Beresford-Smith1985}, this charge renormalization scheme provides renormalized parameters, $Z_\text{eff}$ and $\kappa_\text{eff}$, as input to the linear electrostatic part of the DLVO potential. In the JA, a spherical macroion of radius $a$ is singled out and placed at the origin of the coordinate system. The remaining  $N_\text{m}-1$ macroions and their quasi-condensed counterions are assumed to form a uniform neutralizing background, called the jellium, smeared out in the space $r>a$, where the uncondensed counterions and coions can freely move \cite{Pianegonda2007}. The RJM has also been generalized to polydisperse suspensions of macroions \cite{Bareigts_JCP_2018}. Assuming a Boltzmann distribution for the monovalent free microions and a uniform, structureless jellium for the remaining macroions, i.e., $g_\text{mm}(r)=1$ for $r>a$, the resulting nonlinear PB equation for the total (reduced) electrostatic potential $\Phi$ in Donnan equilibrium with a salt reservoir takes the form \cite{Trizac_PRE_2004}
\begin{equation}
\Phi''(r)+\frac{2}{r}\Phi'(r) = \kappa_{\text{res}}^2 \sinh[\Phi(r)]+3\phi \frac{Z_\text{back}\lambda_{\text{B}}}{a^3},\quad r>a,
\label{NonlinPBeqJell}
\end{equation}
where $\kappa_{\text{res}}^2=8\pi\lambda_\text{B}n_\text{res}$. The first term on the right side is the (reduced) microion charge density contribution, while the second term is the contribution from the (reduced) uniform jellium charge density outside the central macroion, with $Z_\text{back}$ denoting the background macroion valence. The latter is taken to be equal to $Z_\text{eff}$ and is self-consistently determined. The boundary conditions guaranteeing a unique solution are $\Phi'(a) = Z\lambda_\text{B}/a^2$, with $Z$ the bare macroion valence, and $\Phi'(r\rightarrow\infty) = 0$, accounting for electroneutrality of the suspension with an asymptotically decaying electric field. The electrostatic (Donnan) potential at infinity ($r\rightarrow\infty$), $\Phi_\infty$, is nonzero and related to $Z_\text{back}$ by
\begin{equation}
\Phi_\infty = \text{arsinh}\left[- \frac{3\phi}{(\kappa_\text{res}a)^2}\frac{Z_\text{back}\lambda_\text{B}}{a}\right].
\end{equation}
This boundary-value problem can be solved numerically only. Linearizing Eq.~(\ref{NonlinPBeqJell}) with respect to $\Phi_\infty$, we obtain the linearized electrostatic potential, $\Phi_l(r)$, satisfying
\begin{equation}
\Phi_{l}''(r)+\frac{2}{r}\Phi_{l}'(r)=\kappa_{\text{eff}}^2[\Phi_{l}(r)-\Phi_\infty], \quad r>a,
\label{linPBeqJell}
\end{equation}
with effective screening constant $\kappa_{\text{eff}}^2 = \kappa_{\text{res}}^2\cosh(\Phi_\infty)$, which can be alternatively written as
\begin{equation}
(\kappa_\text{eff}a)^4 = (\kappa_\text{res}a)^4+\left(3\phi \frac{Z_\text{back}\lambda_{\text{B}}}{a}\right)^4.
\label{screenparamjell}
\end{equation}
We see that $(\kappa_\text{eff}a)^4$ has two clearly distinguishable contributions: the first due to the salt microions, involving the reservoir salt concentration, and the second due to microion-dressed macroions. The boundary conditions that uniquely determine $\Phi_l$ are $\Phi_l'(r\rightarrow\infty)= 0$, expressing overall electroneutrality of the infinite jellium system, and $\Phi_l(r) \sim \Phi(r)$ for $r\rightarrow\infty$, demanding asymptotic matching of the linear and nonlinear solutions. As in the CM, the effective valence, $Z_\text{eff}$, is obtained from $\Phi_l$ using Eq.~(\ref{AlexZeffdef}) \cite{Trizac_PRE_2004,Pianegonda2007}.
The solution of Eq.~(\ref{linPBeqJell}) can be analytically expressed in terms of $Z_\text{eff}$ as
\begin{equation}
\Phi_l(r)-\Phi_\infty = \lambda_\text{B}Z_\text{eff} \, \frac{e^{\kappa_\text{eff}a}}{1+\kappa_\text{eff}a} \, \frac{e^{-\kappa_\text{eff} r}}{r}, \qquad r>a.
\label{sol2}
\end{equation}
Finally, from comparing Eq.~(\ref{sol2}) against the similarly exponential asymptotic form of the nonlinear solution $\Phi(r)$, we obtain $Z_\text{eff}$ in terms of $Z_\text{back}$. In order to find $Z_\text{eff}$, we demand self consistency by requiring that $Z_\text{eff}=Z_\text{back}$ \cite{Trizac_PRE_2004}, from which $Z_\text{eff}$ is obtained iteratively using a selected starting value $Z_\text{back}<Z$.
				
Similarly to the CM, there is a relation between semi-open and closed systems. This relation follows from the mean concentration of the free co- and counterions inside the suspension \cite{Pianegonda2007},
\begin{equation}
n_\pm=n_\text{res}\exp(\mp\Phi_\infty).
\end{equation}
The effective suspension salt concentration, $n_\text{s}^\text{eff}$, is thus obtained in terms of $n_\text{res}$ and $\Phi_\infty$ as
\begin{equation}
n_\text{s}^\text{eff}=n_-=n_\text{res}\exp(\Phi_\infty)\leq n_\text{res},
\label{nseff_RJM}
\end{equation}
since $\Phi_\infty<0$ holds for negatively charged colloids. Just as in the CM, the renormalized valence, $Z_\text{eff}$, in the RJM asymptotes to a saturation value when $Z\to \infty$. Likewise, typically $Z_\text{eff}<Z$ and it is found that $Z_\text{eff}\to Z$ and $\kappa_\text{eff}\to \kappa_\text{res}$ in the dilute limit $\phi\to 0$ and the salt-free limit $n_\text{res}\to 0$. 

In the salt-free case, suspension electroneutrality requires that $n_+^0 \exp(-\Phi_\infty)=Z_\text{back} n_\text{m}$ and $\Phi(r)\rightarrow 0$ as $r\rightarrow\infty$. The nonlinear PB equation [Eq.~(\ref{NonlinPBeqJell})] then acquires the form
\begin{equation}
\Phi''(r)+\frac{2}{r}\Phi'(r) = -3\phi \frac{Z_\text{back}\lambda_{\text{B}}}{a^3} \{\exp[-\Phi(r)]-1\},\qquad r>a,
\label{non-linPBeqSaltFreeJell}
\end{equation}
with boundary conditions $\Phi'(a) = Z\lambda_\text{B}/a^2$, $\Phi(r\rightarrow\infty) = 0$. Linearizing the right side of Eq.~(\ref{non-linPBeqSaltFreeJell}) around $\Phi_\infty=0$ yields
\begin{equation}
\Phi_l''(r)+\frac{2}{r}\Phi_l'(r) =  \kappa_\text{eff}^2\Phi_l(r),\qquad r>a,
\label{linPBeqSaltFreeJell}
\end{equation}
where 
\begin{equation}
\kappa_\text{eff}^2=4\pi \lambda_\text{B}Z_\text{back}n_\text{m},
\label{Screen_RJMSF}
\end{equation}
and the boundary conditions are $\Phi_l'(r\rightarrow\infty) = 0$ and $\Phi_l(r) \sim \Phi(r)$ for $r\rightarrow\infty$. With the effective valence defined by Eq.~(\ref{AlexZeffdef}), the solution of Eq.~(\ref{linPBeqSaltFreeJell}) has the form
\begin{equation}
\Phi_l(r) = \lambda_\text{B}Z_\text{eff}\frac{e^{\kappa_\text{eff}a}}{1+\kappa_\text{eff}a} \frac{e^{-\kappa_\text{eff} r}}{r},\qquad r>a.
\label{sol2b}
\end{equation}
As for a system with added salt, $Z_\text{eff}$ is determined by asymptotically matching the linearized potential $\Phi_l(r)$ to the nonlinear potential for a selected $Z_\text{back}$ and self-consistently assuming $Z_\text{back}=Z_\text{eff}$.
				
A variant of the discussed non-penetrating RJM is the penetrating RJM, where the neutralizing jellium is not only smeared out across the volume $r>a$, but also penetrates the volume of the central colloid while the microions are still expelled. This leads to the same PB equation [Eq.~(\ref{NonlinPBeqJell})] for $r>a$, but now with boundary conditions
\begin{equation}
\Phi'(a)=\frac{\lambda_{\text{B}}}{a^2}(Z+Z_\text{back}\phi),\qquad
\Phi'(r\rightarrow\infty)=0.
\label{BC_RJMpenetrat}
\end{equation}
The inner boundary condition states now that the electric field on the surface of the central colloid is due to the bare valence, $Z$, of the central colloid plus an additional contribution, $Z_\text{back}\phi$, arising from the penetrating jellium charge inside the volume of the central colloid. Charge renormalization is introduced identically to the non-penetrating jellium case by linearizing with respect to $\Phi_\infty$. The inner boundary condition in Eq.~(\ref{BC_RJMpenetrat}) involves now the modified self-consistency condition $Z_\text{eff}=Z+Z_\text{back}\phi$, which implies that $Z_\text{eff}=Z(1+\phi)+\mathcal{O}(Z^2)$ for small $Z$. This is the same effective charge as obtained from the high-temperature-limiting mean spherical approximation (MSA) solution for the PM \cite{Belloni_JCP_1986,Ruiz-Estrada1990}.

\subsection{Renormalized Linear Response Theory (RLRT)}\label{Sec:RLRT}
When applied to charge-stabilized colloidal suspensions, linear response theory (LRT) provides analytic expressions for the one-body volume energy $E_\text{vol}$ and the effective pair potential between the microion-dressed macroions $u_\text{eff}(r)$ [Eq.~(\ref{YukawaPot})] in the effective one-component Hamiltonian [Eq.~(\ref{EffHamilt})]. The screening constant [Eq.~(\ref{kappasquare})] accounts for the excluded volume of the macroion hard cores through the factor $1/(1-\phi)$.
				
For coupling parameters $Z\lambda_\text{B}/a\gtrsim{\cal O}(10)$, the LRT fails when nonlinear screening effects associated with macroion-microion correlations lead to a strong accumulation of counterions near the macroion surface \cite{Denton_JPCM_2008,LuDenton2010}. However, the theory can be extended to more strongly coupled macroion suspensions by explicitly distinguishing between surface-associated (bound) microions and dissociated (free) microions in the volume energy \cite{Denton_JPCM_2008,LuDenton2010,Denton_JPCM_2010}, according to
\begin{equation}
E_\text{vol} = \Omega_\text{free}+F_\text{bound}.
\end{equation}
Here, $\Omega_\text{free}$ is the grand free energy of the free microions and $F_\text{bound}$ is the free energy of the bound microions. This approach yields the same form of effective pair potential as in Eq.~(\ref{YukawaPot}), but with $Z$ and $\kappa$ replaced by a renormalized effective valence, $Z_\text{eff}\leq Z$, and a renormalized effective screening constant, $\kappa_\text{eff}$, both depending on the state-dependent concentration of free microions. The fraction of strongly-associated (quasi-condensed) counterions is related to an association shell of thickness $\delta$ surrounding a macroion ($a<r<a+\delta$), which is defined as the distance from the macroion surface at which the electrostatic energy of attraction of a counterion is comparable to its thermal energy, i.e.,
\begin{equation}
e|\psi(a+\delta)-\bar{\psi}|=Ck_\text{B}T,
\end{equation}
or in reduced form,
\begin{equation}
|\Phi(a+\delta)-\bar{\Phi}|=C,
\label{assoceq}
\end{equation}
where $\Phi(r)$ is the (LRT-orbital) reduced electrostatic potential at distance $r$ from a macroion center, with mean (spatially averaged) value
\begin{equation}
\bar{\Phi}=\beta e\bar{\psi}=-(\tilde{n}_+-\tilde{n}_-)/(\tilde{n}_++\tilde{n}_-)\,,
\label{Phibar}
\end{equation}
and $C$ is a constant of order unity. Here
\begin{equation}
\tilde{n}_\pm=\frac{\tilde{N}_\pm}{V(1-\tilde{\phi})} 
\label{tilden}
\end{equation}
are mean number densities of free microions, defined as the numbers of free (uncondensed) microions $\tilde{N}_\pm$ in the effective free volume, $V(1-\tilde{\phi})$, where $\tilde{\phi}=\phi(1+\delta/a)^3$ is the effective volume fraction of the macroions including their (quasi-condensed) counterion-association shells.

Combining the LRT with this approximate scheme for incorporating nonlinear microion response yields the renormalized electrostatic potential around a dressed macroion,
\begin{equation}
\Phi(r) = -\lambda_\text{B}Z_\text{eff}\frac{e^{\kappa_\text{eff} (a+\delta)}}{1+\kappa_\text{eff} (a+\delta)}\frac{e^{-\kappa_\text{eff} r}}{r},\,\,\,\, r\geq a+\delta,
\label{dresselectpot}
\end{equation}
with $\kappa_\text{eff}=\sqrt{4\pi\lambda_\text{B}(\tilde{n}_++\tilde{n}_-)}$ being the effective (renormalized) screening constant. Notice that $\kappa_\text{eff}$ can be rewritten as
\begin{equation}
\kappa_\text{eff}=\sqrt{4\pi\lambda_\text{B}\left(\frac{Z_\text{eff}\,n_\text{m}}{1-\tilde\phi}+2\tilde{n}_\text{s}^\text{eff}\right)}\,,
\label{screenparamRLRT}
\end{equation}
using electroneutrality and $\tilde{n}_-=\tilde{n}_\text{s}^\text{eff}$, with $\tilde{n}_\text{s}^\text{eff}$ the renormalized salt concentration of free microion pairs. Substituting Eq.~(\ref{dresselectpot}) into Eq.~(\ref{assoceq}), the association shell thickness is determined by 
\begin{equation}
\left|\frac{Z_\text{eff}\lambda_\text{B}}{[1+\kappa_\text{eff} (a+\delta)](a+\delta)}+\bar{\Phi}\right|=C
\label{assoceq2}
\end{equation}
for given $Z_\text{eff}$, $\phi$, and $C$, on noting that $\kappa_\text{eff}$ depends self-consistently on $\delta$. The physical requirement that the coion density be non-negative dictates that $C=1$, although numerical results prove to be not strongly sensitive to variations of $C$ within the range $1\le C\le 2$ \cite{Denton_JPCM_2008,LuDenton2010,Denton_JPCM_2010}.
				
The distinction between free and bound microions implies a corresponding separation of the total grand free energy $\Omega$. The free microions are only weakly correlated with the macroions, and thus well described by linear-response theory. The volume energy per macroion, $\varepsilon_\text{vol}=E_\text{vol}/N_\text{m}$, has the form
\begin{eqnarray}
\varepsilon_\text{vol}&=&\sum_{i=\pm}\frac{\tilde{N}_i}{N_\text{m}}\left[\ln\left(\frac{\tilde{n}_i}{n_\text{res}}\right)-1\right]-\frac{Z_\text{eff}^2}{2}\frac{\kappa_\text{eff}\lambda_\text{B}}{1+\kappa_\text{eff}(a+\delta)}\nonumber\\&-&\frac{Z_\text{eff}}{2}\frac{\tilde{n}_+-\tilde{n}_-}{\tilde{n}_++\tilde{n}_-}+f_\text{bound},
\end{eqnarray}
with the bound counterion free energy per macroion being approximated by
\begin{equation}
f_\text{bound}\approx(Z-Z_\text{eff})\left[\ln\left(\frac{Z-Z_\text{eff}}{v_s}\Lambda_0^3\right)-1\right]+\frac{Z_\text{eff}^2\lambda_\text{B}}{2a}.
\end{equation}
The first term on the right side is the ideal-gas free energy of the bound counterions in the association shell of volume $v_\text{s}=(4\pi/3)[(a+\delta)^3-a^3]$ and the second term accounts for the self-energy of a dressed macroion of valence $Z_\text{eff}$, assuming the bound counterions to be localized near the macroion surface ($r=a$). 
				
The effective macroion valence $Z_\text{eff}$, and hence the association shell thickness $\delta$, can be determined then by equating the chemical potentials of microions in the free and bound phases, which is equivalent to minimizing the volume energy at fixed temperature and mean microion densities \cite{Denton_JPCM_2008}. For given bare valence $Z$, the effective valence is then obtained from the variational condition,
\begin{equation}
\left(\frac{\partial\varepsilon_\text{vol}}{\partial Z_\text{eff}}\right)_{T,\tilde{n}_\pm}=0\,.
\end{equation}
Notice that $Z_\text{eff}$ and $\delta$ are inter-related by Eq.~(\ref{assoceq2}) and together determine the effective screening constant $\kappa_{\text{eff}}$.

\subsection{Shifted Debye-H\"uckel Approximation (SDHA)}\label{Subsec:SDHA}
	
The SDHA method \cite{Boon_PNAS_2015}, like the RLRT method, is based on a multi-colloid-center model and combines density-functional theory (DFT) with PB-type approximations. Following ref.~\cite{Boon_PNAS_2015}, we present its essential features for suspensions of impermeable colloids.
	
Assuming for the moment pointlike macroions, we formally expand the DFT-PB grand-free energy functional $\hat{\Omega}_\mu({\bf X})$ in the presence of $N_\text{m}$ (pointlike) macroions at positions ${\bf X}$ up to quadratic order in the microion trial densities $\rho_\pm({\bf r};{\bf X})$, measured relative to yet-unknown constant densities $\bar{n}_\pm$ \cite{Boon_PNAS_2015}. In Donnan equilibrium, the densities $\bar{n}_\pm$ are not independent, but are related by
\begin{equation}
\bar{n}_\pm=n_\text{res}e^{\mp\tilde{\Phi}},
\end{equation}
for a yet unspecified potential value, $\tilde{\Phi}$, so that $\bar{n}_+\bar{n}_-=n_\text{res}^2$. On minimizing $\hat{\Omega}_\mu({\bf X})$, quadratically expanded with respect to the trial microion densities $\rho_\pm({\bf r};{\bf X})$, the linearized equilibrium microion profiles
\begin{equation}
n_\pm({\bf r};{\bf X})=n_\text{res}\,e^{\mp \tilde{\Phi}}\left\{1\mp\left[\Phi_l({\bf r};{\bf X})-\tilde{
\Phi}\right]\right\} 
\label{equildensprofiles}
\end{equation}
are obtained, with the linearized suspension potential $\Phi_l({\bf r};{\bf X})$. The so-called shifted linearized potential, $\Phi_l^\text{s}({\bf r};{\bf X})$, defined by
\begin{equation}
\Phi_l^\text{s}({\bf r};{\bf X})=\Phi_l({\bf r};{\bf X})-\tilde{\Phi}+\gamma,
\end{equation}
for $\gamma=\tanh(\tilde{\Phi})$, fulfills the multi-colloid-center linearized PB equation (shifted Debye-H\"uckel equation)
\begin{equation}
\nabla^2\Phi_l^\text{s}({\bf r};{\bf X})=\kappa_\text{eff}^2\,\Phi_l^\text{s}({\bf r};{\bf X})-4\pi\lambda_\text{B}q({\bf r};{\bf X}). 
\end{equation}
Here,
\begin{equation}
q({\bf r};{\bf X})=\sum_{j=1}^{N_\text{m}}\delta({\bf r}-{\bf R}_j)Q_\text{eff}
\end{equation}
is the charge number density of pointlike macroions at positions ${\bf X}=\{{\bf R}_1,\ldots,{\bf R}_{N_\text{m}}\}$, where charge renormalization (discussed below) is accounted for in the effective macroion valence $Q_\text{eff}$. Moreover, $\kappa_\text{eff}^2=\kappa_\text{res}^2\,\cosh(\tilde{\Phi})$ is taken as the renormalized screening parameter linked to $Q_\text{eff}$. Fourier transformation straightforwardly yields the solution
\begin{equation}
\Phi_l^\text{s}({\bf r};{\bf X})=\lambda_\text{B}Q_\text{eff}\sum_{j=1}^{N_\text{m}}\frac{e^{-\kappa_\text{eff}|{\bf r}-{\bf R}_j|}}{|{\bf r}-{\bf R}_j|}
\end{equation}
for the shifted linearized potential, which is a superposition of $N_\text{m}$ Yukawa-type orbitals. Substitution of $n_\pm({\bf r};{\bf X})$, according to Eq.~(\ref{equildensprofiles}), into the quadratic-order expanded microion grand free energy results in
\begin{equation}
\beta \Omega_\mu({\bf X})=\beta E_\text{vol}+\lambda_\text{B}Q_\text{eff}^2\sum_{i<j}^{N_\text{m}}\frac{e^{-\kappa_\text{eff}|{\bf R}_i-{\bf R}_j|}}{|{\bf R}_i-{\bf R}_j|},
\end{equation}
with volume energy per macroion
\begin{equation}
\beta \varepsilon_\text{vol}=-\frac{\kappa_\text{eff}^2}{8\pi\lambda_\text{B}n_\text{m}}\left(\frac{\kappa_\text{res}^4}{\kappa_\text{eff}^4}+1\right)+Q_\text{eff}(\tilde{\Phi}-\gamma).
\label{Evol_SDHA}
\end{equation}
	
So far, we have not specified the value of $\tilde{\Phi}$, and hence the resulting value for $\kappa_\text{eff}$ and $Q_\text{eff}$. This is done now using, for simplicity, a spherical CM with $\tilde{\Phi}$ identified as $\Phi(R)$ or $\bar{\Phi}$ \cite{Boon_PNAS_2015}. The macroion effective valence $Q_\text{eff}$, defined using Eq.~(\ref{Qeff_def}), is related to the effective valence $Z_\text{eff}$ by Eq.~(\ref{EPCZeffdef}), i.e., $Z_\text{eff}$ is obtained from $Q_\text{eff}$ by multiplying the latter with a geometric factor due to the actually nonzero radius of the macroions. The hard-core of the macroions is reintroduced {\it a posteriori} by enforcing electroneutrality of the individual orbitals, leading to the geometric factor in the relation between $Z_\text{eff}^\text{EPC}$ and $Q_\text{eff}$.
	
Interestingly, in contrast to the CM with SC definition of the renormalized valence in Eq.~(\ref{AlexZeffdef}), $Z_\text{eff}^\text{EPC}$ in the EPC definition can exceed the bare colloid valence $Z$ for sufficiently high colloid concentration and small coupling parameter $Z\lambda_\text{B}/a$, where nonlinear renormalization effects are negligible. 
	
In the salt-dominated regime, where $2n_\text{res}\gg Zn_\text{m}$, one obtains $Z_\text{eff}^\text{EPC}\to Z$, $\kappa_\text{eff}\to\kappa_\text{res}$, and $\tilde{\Phi}-\gamma \to 0$. The volume energy reduces then to minus the kinetic energy density, $\varepsilon_\text{vol} \to -2k_\text{B}T\;\!n_\text{res}$, of the reservior ions, and the effective pair potential to $\beta u_{\text{eff}}(r) \to \lambda_{\text{B}} Z^2\exp(2\kappa_\text{res})/(1+\kappa_\text{res}a)^2\exp(-\kappa_{\text{res}}r)/r$ for $r>2a$ \cite{NBoon_diss}.
	
In the present paper, we analyze the SDHA method for the volume energy and effective pair potential, in combination with the EPC charge renormalization scheme originally introduced, for edge linearization only, by Boon {\it et al.} \cite{Boon_PNAS_2015}. Although the SDHA may be combined also with SC, for conciseness we do not consider this combination here. A study of the combined SDHA-SC method using edge linearization is presented in ref.~\cite{Hallez_EPJE2018}.

\section{Thermodynamics and Structure}\label{sec:StructureThermodynamics}
Before comparing predictions from the various renormalization methods for thermodynamic and structural properties of charge-stabilized colloidal suspensions, we first recall that the CM approximates the bulk pressure $p$ using the contact (cell) theorem (CT) as \cite{Wennerstrom1982},
\begin{equation}
\beta p\approx \beta p_\text{CT}=n_+(R) + n_-(R) + n_\text{m}\,,
\label{contacttheocell}
\end{equation}
where $n_\pm(R)$ are the microion densities at the cell edge and the ideal-gas contribution $n_\text{m}$ of the macroions is included. Since the CM neglects contributions to the pressure due to macroion correlations, in concentrated suspensions $p_\text{CT}$ can differ significantly from the actual suspension pressure $p$, except at low salt reservoir concentrations, where the backbone-released counterions overwhelm salt ions ($ZN_\text{m} \gg N_\text{s}$) and make the dominant contribution \cite{Dobnikar2006}.
 
Similar to the CM, in RJM the bulk pressure is determined from the microion concentrations in the electric field free region at the edge of the system, reached here in the asymptotic limit $r\rightarrow\infty$ \cite{Trizac_PRE_2004}, according to
\begin{eqnarray}\label{jelliumPress_saltadded}
\beta p_\text{jell} &=& n_+(\infty)+n_-(\infty)+n_\text{m}\,\nonumber\\
&=& \sqrt{(2n_\text{res})^2+(n_\text{m}Z_\text{eff})^2}+n_\text{m}\,.
\end{eqnarray}
Here, $n_\pm(\infty)$ are the microion densities infinitely far from the central macroion, and again the ideal-gas macroion contribution is included. The first term on the right-hand side of Eq.~(\ref{jelliumPress_saltadded}) is the microion contribution, resulting from salt ions of concentration $n_\text{res}$ and non-condensed surface-released counterions. In the salt-free case, Eq.~(\ref{jelliumPress_saltadded}) reduces to
\begin{equation}
\beta p_\text{jell}=n_\text{m}(1+Z_\text{eff}),
\label{jelliumPress_saltffree}
\end{equation}
which is of the same form as the ideal-gas pressure $p_\text{id}$ of macroions and counterions,
\begin{equation}
\beta p_\text{id}=n_\text{m}(1+Z),
\end{equation}
but with $Z$ replaced by $Z_\text{eff}<Z$, expressing that only free counterions contribute directly to the pressure.

In the one-component model of a suspension of charged colloids interacting via the effective pair potential in Eq.~(\ref{RenormYukawaPot}), thermodynamic and structural properties can be computed once the renormalized interaction parameters, $\kappa_\text{eff}$ and $Z_\text{eff}$, are determined for given system parameters $Z$, $n_\text{res}$, and $\phi$. Comparing predictions of these properties by the various renormalization methods against data from simulations of the PM gauges the performance of the different methods.

Since $u_\text{eff}(r)$ is purely repulsive, we can use the thermodynamically self-consistent Rogers-Young (RY) integral-equation scheme \cite{Hansen-McDonald} for calculating the static structure factor, $S(q)$, of colloids and the associated radial distribution function, $g(r)$. This hybrid  integral-equation scheme, based on a mixing function that interpolates between the hypernetted chain (HNC) and Percus-Yevick (PY) closures \cite{Hansen-McDonald}, is known from comparisons with computer simulation data to accurately predict structural properties for a variety of repulsive interaction potentials, including the screened-Coulomb potential \cite{Banchio_JCP_2008,Banchio_JCP_2018}.
The mixing parameter, $\alpha$, in the RY mixing function is determined self-consistently by enforcing equality of the suspension osmotic compressibility derived from the one-component compressibility and virial equation of states, respectively, i.e., by demanding \cite{Hoffmann_JCP_2004,Dobnikar2006}
\begin{equation}
\lim_{q\to 0}\;\!\frac{1}{S(q;\alpha)} = \beta \left(\frac{\partial p(\alpha)}{\partial n_\text{m}}\right)_{u_\text{eff}},
\label{Sq0}
\end{equation}
where $S(q)$ and $p$ are the static structure factor and pressure, respectively,
of the OCM system, and the density derivative of $p$ is taken at fixed $u_\text{eff}(r)$, 
i.e., disregarding any $n_\text{m}$-dependence of the effective macroion pair potential.

In general, the suspension pressure $p$ can be calculated by several methods. For instance, $p$ can be determined from the thermodynamic relation
\begin{equation}
p= -\left(\frac{\partial \Omega}{\partial V}\right)_{\text{res}}=n_\text{m}^2\left(\frac{\partial\omega}{\partial n_\text{m}}\right)_{\text{res}},
\label{susppress_derivdef}
\end{equation}
provided the semi-grand suspension free energy per macroion, $\omega=\Omega/N_\text{m}$, is known, including its volume energy contribution. In practice, the renormalized interaction parameters ($Z_\text{eff}$, $\delta$, $\kappa_{\text{eff}}$) are held constant when taking thermodynamic derivatives \cite{Denton_JPCM_2010}. Within RLRT and SDHA, $\omega$ is obtained for impermeable, rigid macroions as
\begin{equation}
\omega=\varepsilon_\text{vol}+f_\text{m},
\label{omega}
\end{equation}
where $\varepsilon_\text{vol}=E_\text{vol}/N_\text{m}$ is the volume energy per macroion and $f_\text{m}$ is the macroion free energy per macroion. From Eqs.~(\ref{susppress_derivdef}) and (\ref{omega}), the total pressure correspondingly separates into
\begin{equation}
p=p_\text{vol}+p_\text{m},
\label{pressure1}
\end{equation}
where the contributions $p_\text{vol}$ and $p_\text{m}$ are associated with the volume energy and the
effective macroion-macroion interactions, respectively. 

With renormalized interaction parameters, $\kappa_\text{eff}$ and $Z_\text{eff}$, as input, the RLRT method predicts the pressure contribution arising from the renormalized volume energy as \cite{Denton_JPCM_2008,LuDenton2010,Denton_JPCM_2010}:
\begin{equation}
\beta p_\text{vol}=n_\text{m}^2\left(\frac{\partial\beta\varepsilon_\text{vol}}{\partial n_\text{m}}\right)_{\text{res}}=\tilde{n}_++\tilde{n}_--\frac{Z_\text{eff}(\tilde{n}_+-\tilde{n}_-)\kappa_\text{eff}\lambda_\text{B}}{4[1+\kappa_\text{eff}(a+\delta)]^2},
\label{pfree_RLRT}
\end{equation}
where $\tilde{n}_{\pm}$ are number densities of free (uncondensed) microions, corrected for the effective excluded volume of the macroions plus their (quasi-condensed) counterion-association shells [Eq.~(\ref{tilden})].
In taking the macroion density derivative in Eq.~(\ref{pfree_RLRT}), the temperature and reservoir salt concentration, $n_\text{res}$, are held constant and system electroneutrality is maintained for given $Z$. The bound counterions make no direct pressure contribution, since the effective interaction parameters are kept fixed in the concentration derivative. 

In the salt-dominated regime, $2 n_\text{res} \gg Z n_\text{m}$, where $\tilde{n}_\pm \to n_\text{res}$ and $\kappa_\text{eff}a \to \kappa_\text{res} a \gg 1$, the RLRT volume pressure reduces to the reservoir pressure, i.e., $p_\text{vol} \to p_\text{res}= 2 n_\text{res}$. In the salt-free case of vanishing coion density ($\tilde{n}_{-}=0$), and in the DH regime ($Z\lambda_B/a \ll 1$) of electrostatically weakly interacting macroions, the volume energy-related pressure reduces in the RLRT to
\begin{equation}
\label{eq:pvolRLRT}
\beta p_\text{vol}= n_\text{m} Z - \frac{\kappa^3}{16\pi}\;\!Z +{\cal O}\left(n_\text{m}^2\right)\,,
\end{equation}
where $\kappa^2 = 4\pi\lambda_\text{B} n_\text{m} Z$ here involves only the counterions.
Thus, the correct ideal-gas contribution to the volume pressure, $k_\text{B}T n_\text{m} Z$, is recovered. 
The exact leading order non-ideal (DH limiting law) volume pressure contribution, proportional to $n_\text{m}^{3/2}$, as obtained from the DH limiting law (i.e., macroion size-independent) volume energy $\epsilon_\text{vol}$ for a salt-free system \cite{Chan-PRE2001}, differs from the RLRT expression in Eq.~(\ref{eq:pvolRLRT}) only in that the former has a factor $(Z+2/3)$ instead of $Z$. The difference, however, is negligible for the commonly encountered case of $Z \gg 1$.     
Appendix \ref{AppendixRLRT} describes how the pressure contribution, $p_\text{m}$, due to effective macroion interactions is determined in the RLRT method.

In the SDHA method, $p_\text{vol}=n_\text{m}^2(\partial\varepsilon_\text{vol}/\partial n_\text{m})$ is obtained analytically from  Eq.~(\ref{Evol_SDHA}) by keeping constant the effective interaction parameters, $Z_\text{eff}$ and $\kappa_\text{eff}$, and the linearization points, $\bar{n}_\pm$, of the microion concentrations in the DFT-PB grand free energy functional. The result is \cite{Boon_PNAS_2015}
\begin{equation}
\beta p_\text{vol}=-\left(\frac{\partial \beta E_\text{vol}}{\partial V}\right)_{N,T,\bar{n}_\pm}=\frac{\kappa_{\text{eff}}^2}{8\pi\lambda_\text{B}}\left[ \left(\frac{\kappa_{\text{res}}}{\kappa_{\text{eff}}}\right)^4+1\right]
\label{pvol_SDHA}
\end{equation}
for a suspension in osmotic equilibrium with a reservoir of microion pair concentration $2n_\text{res}$. We have followed here Boon {\it et al.} \cite{Boon_PNAS_2015} in holding fixed the effective pair potential and the linearization points while taking the volume derivative in Eq.~(\ref{pvol_SDHA}). A thorough theoretical study \cite{Deserno2002} based on the linearized cell model has come to the conclusion that, while treating the linearization points of the microion concentrations as volume-dependent in calculating $p$ is admissible, there are advantages in treating them as independent variables. In particular, in a linear approximation the system pressure remains always positive for a proper choice of the linearization point.

In the salt-dominated regime, where $\kappa_\text{eff}=\kappa_\text{res}$, the reservoir pressure is recovered from the SDHA volume pressure, i.e., $p_\text{vol}=p_\text{res}$.
In the salt-free case, where $\kappa_\text{res}=0$, the volume pressure in SDHA reduces to 
\begin{equation}
\beta p_\text{vol}= \frac{\kappa_{\text{eff}}^2}{8\pi\lambda_\text{B}}\,.
\label{pvol_BoonSF}
\end{equation}
For weakly interacting macroions, where $Z\lambda_\text{B}/a \ll 1$ and $\kappa _\text{eff}^2=4\pi \lambda_\text{B} n_\text{m} Z$, the SDHA volume pressure reduces further to $\beta p_\text{vol}= n_\text{m}Z/2$, which is only one-half of its exact ideal-gas value. Moreover, no leading-order non-ideal pressure contribution is predicted. Interestingly enough, there is a pressure contribution $n_\text{m}\left(1+ Z/2\right)$ to first order in macroion concentration arising from the macroion-related virial pressure contribution $\beta p_\text{vir}$ in the no-salt DH regime, such that overall the correct suspension ideal-gas pressure limit, $\beta p_\text{id}= n_\text{m}\left(1+Z\right)$, is recovered in SDHA, where the pressure contribution $p_\text{den}$ is not considered. Actually, the latter gives, to linear order in concentration, the negative pressure contribution, $\beta p_\text{den}=-n_\text{m} Z/2+{\cal O}\left(n_\text{m}^2\right)$, in the zero-salt DH regime, whose inclusion in the SDHA would spoil again the recovery of the exact ideal-gas suspension pressure. While interesting from a principal viewpoint, the peculiarities of the SDHA method in the DH regime are of minor concern in the present study, which focuses mainly (with the exception of Sec.~\ref{sec:VB}) on systems in which $Z\lambda_\text{B}/a \gtrsim {\cal O}(10)$, where counterion quasi-condensation is operative.

Since three-particle or higher-order effective interaction contributions are neglected for the considered mean-field methods, the suspension pressure $p$ can be computed from the generalized virial equation for pairwise-interacting systems [Eq.~(\ref{eq:PressureTwoBody})], from which the macroion pressure contribution can be expressed as
\begin{eqnarray}
p_\text{m} = p_\text{vir} + p_\text{den},
\label{eq:PressureTwoBodyII}
\end{eqnarray}
with $p_\text{vir}$ and $p_\text{den}$ given by Eqs.~(\ref{p-OCM}) and (\ref{p-den}), respectively.
Note that, in taking the density derivative of the effective pair potential, $\partial u_\text{eff}/\partial n_\text{m}$, in Eq.~(\ref{p-den}) for $p_\text{den}$, only the density dependence of $u_\text{eff}(r;n_\text{m})$ arising from tracing out the microions should be considered, and not that due to the extra imposed charge renormalization. Only the first density dependence is thermodynamically relevant, and accordingly accounted for in the RLRT calculation of $p_\text{den}$. In contrast, in the SDHA calculation, the pressure contribution $p_\text{den}$ is disregarded. This treatment is consistent with Eq.~(\ref{pvol_SDHA}) for $p_\text{vol}$ derived from the volume derivative of Eq.~(\ref{Evol_SDHA}), keeping $Z_\text{eff}$ and $\kappa_\text{eff}$ fixed, and the recovery of the correct ideal-gas suspension pressure limit in the zero-salt DH regime.

Assuming a semi-open system of macroions in Donnan (osmotic) equilibrium with a microion reservoir, the osmotic compressibility $\chi_\text{osm}$ of the suspension of macroions plus dispersed microions is given by \cite{Belloni2000}
\begin{equation}
\chi_\text{osm}=\frac{1}{n_\text{m}}\left(\frac{\partial n_\text{m}}{\partial \pi_\text{osm}}\right)_{\text{res}}
= \frac{1}{n_\text{m}}\left(\frac{\partial n_\text{m}}{\partial p}\right)_{\text{res}}\,,
\label{OsmCompress_def}
\end{equation}
where $\pi_\text{osm}=p-p_\text{res}$ is the osmotic pressure of the suspension, relative to the reservoir pressure $p_\text{res}=2 k_\text{B}T n_\text{res}$. The leading order non-ideal (limiting-law) contribution to the reservoir pressure is negative and scales with the power $3/2$ of the reservoir salt concentration, according to $-k_\text{B}T\kappa_\text{res}^3/(24\pi)$ \cite{Roa_SoftMatter_2016}. Therefore, non-ideality contributions to the reservoir pressure are negligible for the considered reservoir ionic strengths of monovalent electrolyte ions, which is consistent with the implemented PB description of microions.

The macroion density derivative in Eq.~(\ref{eq:OsmoticPressure}) is taken for fixed reservoir properties, i.e., for fixed $T$ and $\mu_\text{res}$ (or, equivalently, $n_\text{res}$), summarily denoted by the subscript $\text{res}$.  Quite remarkably, in Donnan equilibrium the osmotic compressibility $\chi_\text{osm}$ divided by its ideal-gas value, $\chi_\text{osm}^\text{id} = 1/(n_\text{m} k_\text{B} T)$, can be expressed via the Kirkwood-Buff relation \cite{Kirkwood_JCP_1951,Belloni2000,Dobnikar2006},
\begin{equation}
\frac{\chi_\text{osm}}{\chi_\text{osm}^\text{id}} = 1 + n_\text{m} \int d^3r\left[g_\text{mm}(r)-1\right] = \lim_{q\to 0}S_\text{mm}(q) \,,
\end{equation}
solely in terms of the solvent-averaged macroion-macroion static structure factor $S_\text{mm}(q)$ and its associated radial distribution function, $g_\text{mm}(r)$, as defined in the PM for colloidal macroions and microions. The Kirkwood-Buff relation remains valid in the zero-salt limit ($n_s\to 0$) of a binary PM mixture, where $\pi_\text{osm}$ reduces, due to global electroneutrality, to the system (osmotic) pressure relative to a pure-solvent reservoir.

One can exploit here a theorem by Henderson \cite{Henderson_PhysLettA_1974} asserting that for an (effective) one-component model system with only pairwise interactions, for each thermodynamic state (density $n$ and temperature $T$) there is a one-to-one correspondence between the associated pair distribution function $g(r;n)$ and an associated pair potential $u_\text{eff}(r)$, up to an irrelevant additive constant for the latter. As discussed in \cite{Henderson_PhysLettA_1974}, at given macroion density $n_\text{m}$ and temperature, the osmotic compressibility in Donnan equilibrium can thus be obtained also from the density derivative of the suspension  pressure $p$ of the fictitious OCM system with state-independent pair potential $u_\text{eff}(r;n_\text{m},T)$, i.e., \cite{Dobnikar2006},
\begin{equation}\label{eq:OCM_vS_Mixture}
 \left(\frac{\partial p}{\partial n_\text{m}}\right)_\text{res} =  \left(\frac{\partial p}{\partial n_\text{m}}\right)_{u_\text{eff}} = \frac{k_\text{B}T}{S(q\to 0)}\,,
\end{equation}
where $S(q)$ is the static structure factor of the OCM system with radial distribution function $g(r)$.  
The second equality expresses the OCM compressibility equation in which, when taking the density derivative of $p$, any $n_\text{m}$-dependence of the effective pair potential $u_\text{eff}$ is disregarded. For a direct application, consider a salt-free suspension at low macroion concentration, where $p \approx p_\text{jell}$ can be used as an  analytic approximation for $p$, with $p_\text{jell}$ given in Eq. (\ref{jelliumPress_saltffree}). The OCM compressibility equation, where $Z_\text{eff}$ is kept constant in the density derivative, leads here to $S(q\to 0) \approx 1/\left(1+Z_\text{eff}\right)$. This expression is consistent with the DH limiting law result $S_\text{mm}(q\to 0) = 1/(1+Z) + {\cal O}\left(\sqrt{n_\text{m}}\right)$ for a binary $Z:1$ PM system with $Z\lambda_\text{B}/a \ll 1$.

For the exact validity of the OCM compressibility equation, it is understood that $u_\text{eff}$ is determined such that $g(r)=g_\text{mm}(r)$ and hence $S(q)=S_\text{mm}(q)$ is exactly valid, with no approximations involved in obtaining $u_\text{eff}$. Of course, these premises are not fulfilled, in particular, by the discussed PBCM approaches, which {\em per se} do not provide an effective macroion pair potential expression, rendering the substitution of the PBCM-generated effective interaction parameters into the Yukawa form in Eq. (\ref{p-OCM}) arbitrary to some extent. An additional ambiguity arises from the different PB-based charge renormalization schemes, which are not uniquely defined. Based on the PM, a unique effective macroion pair potential with uniquely defined interaction parameters can be obtained, in principle, from the dressed ion theory (DIT) of Kjellander and Mitchell \cite{Kjellander-Mitchell1994,Kjellander-Mitchell1997}. DIT, however, requires as input the partial pair distribution functions of the PM, which can be obtained only from elaborate simulations or accurate Ornstein-Zernike integral-equation theory calculations, as performed, e.g., in \cite{Gonzalez-Mozuelos2013}. Keeping in mind these facts, Eq. (\ref{eq:OCM_vS_Mixture}) is useful as a  means of checking the degree of internal consistency of an OCM charge renormalization scheme, by comparing its (approximate) prediction for the suspension osmotic compressibility in Donnan equilibrium with the (approximate) prediction of the zero-wavenumber limit OCM static structure factor $S(q)$ based on the (approximate) $u_\text{eff}(r)$ and calculated using an (approximate) Ornstein-Zernike integral equation scheme, such as RY.

\section{Results}\label{results}

For suspensions of spherical colloidal macroions in osmotic equilibrium with a 1:1 electrolyte (salt) reservoir, the effective one-component description in the weak-coupling limit is based on three reduced parameters that uniquely determine the system: the colloid concentration, quantified by the macroion volume fraction $\phi$; the reduced reservoir screening parameter $\kappa_\text{res}a$, related to the salt reservoir concentration; and the electrostatic coupling parameter $Z\lambda_{\text{B}}/a$, with $Z$ denoting the bare macroion valence.

To assess the performance of the various renormalization methods and directly compare their predictions for thermodynamic and structural properties with PM computer simulation data \cite{Linse2000}, we choose the following system parameters: solvent Bjerrum length $\lambda_\text{B} = 0.714\;\!\text{nm}$ (corresponding to water at $T = 293\;\!K$), bare macroion valence $Z =40$, and macroion radius $a$ corresponding to $\lambda_{\text{B}}/a = 0.0222, 0.0445, 0.0889, 0.1779, 0.3558, 0.7115$, and coupling parameter $Z\lambda_{\text{B}}/a = 0.89, 1.78, 3.56, 7.12, 14.23, 28.5$.
The reservoir salt pair concentration, $n_\text{res}$, is chosen such that $\kappa_\text{res}a$ is in the range of $0-18$, which connects the counterion-dominated regime at low $\kappa_\text{res}$ with the salt-dominated regime at high $\kappa_\text{res}$. The selected values of the volume fraction $\phi=(4\pi/3)\,n_\text{m}\, a^3$ are in the range of $1\times10^{-4}-3.75\times 10^{-1}$. 
We numerically solve the nonlinear PB equation using the MATLAB routine bvp4c \cite{Kierzenka_2001}, and the RY integral-equation scheme using the same code as in ref.~\cite{Roa_SoftMatter_2016}.

\subsection{Renormalized Interaction Parameters}
	
We employ the charge renormalization methods discussed in Sec.~\ref{methods} to compute the renormalized interaction parameters, $\kappa_\text{eff}$ and $Z_\text{eff}$, and use them as input to the effective macroion-macroion pair potential, $u_\text{eff}(r)$ [Eq.~(\ref{RenormYukawaPot})]. In the RJM and RLRT methods, the renormalized interaction parameters are directly connected to $u_\text{eff}(r)$, while in the CM-based methods, such as SDHA \cite{Boon_PNAS_2015}, only an {\it ad hoc} connection can be established.
\begin{figure}
\centering
\includegraphics[width=8cm]{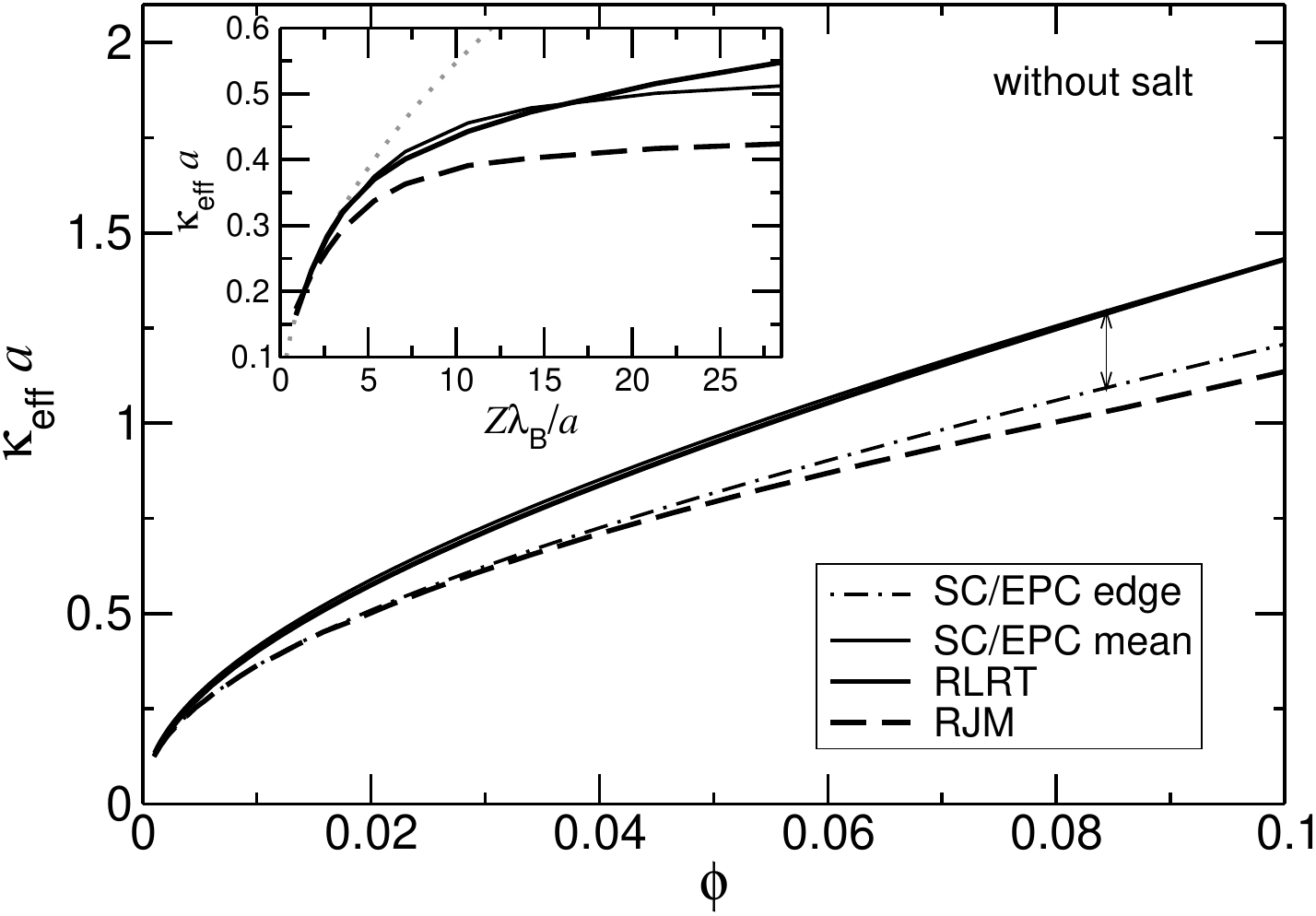}
\caption{Reduced renormalized screening constant, $\kappa_\text{eff}a$, versus colloid volume fraction, $\phi$, for different renormalization methods as indicated. A salt-free suspension ($n_\text{res} = 0$) is considered with bare macroion valence $Z=40$ and $Z\lambda_\text{B}/a=7.12$ for $\lambda_\text{B} = 0.714\;\!\text{nm}$. Inset: $\kappa_\text{eff}a$ versus coupling parameter, $Z\lambda_\text{B}/a$, for the different methods at $\phi=0.01$. In the inset, the SC/EPC results with edge linearization are indistinguishable from the RJM result on the scale of the figure. Dotted grey line represents the screening constant $\kappa a= \sqrt{3\phi Z\lambda_\text{B}/a}$ without accounting for charge renormalization.}
\label{effect_screening_SF}
\end{figure}
		
To explore predictions for the effective interaction parameters, we first consider the salt-free case. In this limiting case, the suspension is closed, since electroneutrality prevents counterions from leaving the suspension into the salt-free reservoir. The parameter space is here two-dimensional only and spanned by $\phi$ and $Z\lambda_{\text{B}}/a$. 
	
Figure~\ref{effect_screening_SF} shows predictions of the renormalization methods for $\kappa_\text{eff}$ as a function of $\phi$ and $Z\lambda_{\text{B}}/a$ (inset) in the salt-free case. For all considered methods, $\kappa_\text{eff}$ increases with increasing $\phi$ and $Z\lambda_{\text{B}}/a$. Notice that, for the CM-based SC and EPC methods, $\kappa_\text{eff}$ differs only by the invoked linearization. In the assessed mean-field methods, $\kappa_\text{eff}$ depends only on the concentration of free (uncondensed) microions. Therefore, $\kappa_\text{eff}$ increases with increasing concentration of (surface-released) free counterions, triggered by increasing colloid volume fraction $\phi$ or bare valence $Z$. This trend follows clearly from Eqs.~(\ref{effkappa_mean}) and (\ref{effkappa_edge}) for the CM-based methods with edge and mean linearization, respectively, from Eq.~(\ref{Screen_RJMSF}) for RJM, and from Eq.~(\ref{screenparamRLRT}) for RLRT. The potential linearization in RLRT is equivalent to the potential linearization with respect to the (volume-averaged) mean electrostatic potential. The screening constant comparison in Fig.~\ref{effect_screening_SF} allows for a distinction between methods that linearize with respect to the mean electrostatic potential and those that linearize with respect to the minimum absolute value of the potential at the cell edge. The methods based on mean potential linearization predict a larger $\kappa_\text{eff}$, a distinguishing feature that becomes more pronounced with increasing $\phi$ and $Z\lambda_{\text{B}}/a$.
\begin{figure}
\centering
\includegraphics[width=8cm]{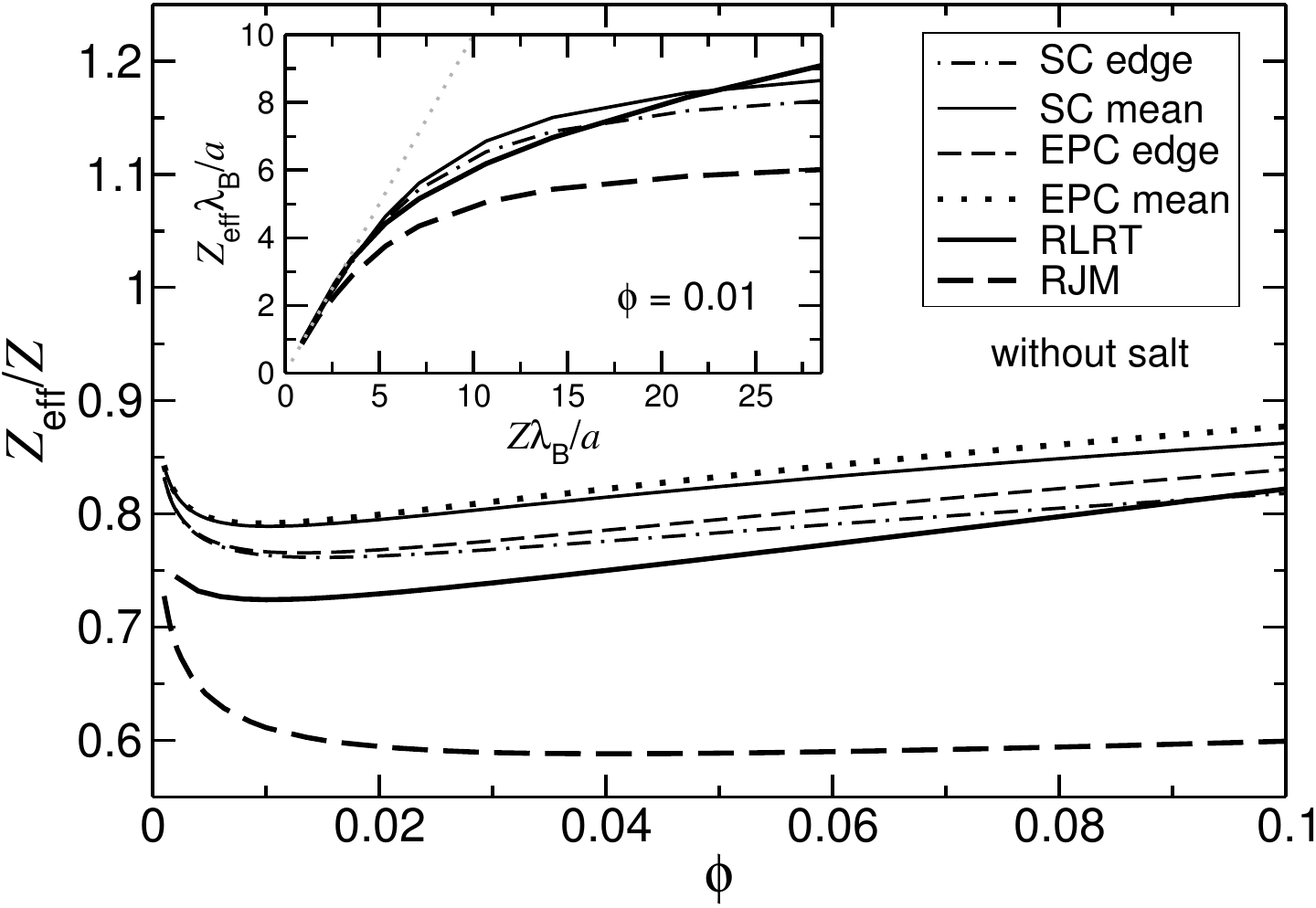}
\caption{Renormalized valence, $Z_\text{eff}/Z$, in units of the bare valence, as function of volume fraction $\phi$ for the indicated renormalization methods and a salt-free suspension ($n_\text{res} = 0$).  The bare macroion valence is $Z=40$ with coupling parameter $Z\lambda_\text{B}/a=7.12$ for $\lambda_\text{B} = 0.714\;\!\text{nm}$. Inset: $Z_\text{eff}\lambda_\text{B}/a$ versus coupling parameter, $Z\lambda_\text{B}/a$, at $\phi=0.01$. In the inset, EPC edge and SC edge curves coincide, and likewise so do the EPC and SC mean potential linearization curves.}
\label{effect_charge_SF}
\end{figure}

Figure~\ref{effect_charge_SF} shows predictions of the different methods for the effective valence $Z_\text{eff}$ as a function of $\phi$ and $Z\lambda_{\text{B}}/a$ (inset). Renormalization arises from the interaction of the charged colloidal surfaces with strongly (nonlinearly) associated counterions \cite{Alexander1984,Belloni1998,Levin2002_Review}, and from the competition between a reduction in the electrostatic energy of quasi-condensed counterions and the gain in entropy of free counterions spread across the system. 
In the considered zero-salt case and for spherical macroions, $Z_\text{eff} =Z$ at $\phi = 0$, since the gain in entropy by dispersing the finite number of counterions over an infinite (cell) volume outweighs here the electrostatic attraction energy tending to confine the counterions near to the macroion surface. 
With $\phi$ increasing above $0$, the entropy per counterion decreases, since the accessible volume decreases, and the initially zero fraction of counterions subject to quasi-condensation (with electrostatic attraction energy $> k_B T$) increases. This effect results in the steep initial decline of $Z_\text{eff}$ at very small $\phi$. With further increasing $\phi$, electrostatic screening becomes operative, roughly when the effective Debye length $1/\kappa_\text{eff}$ decreases below the geometric mean macroion distance $n_\text{m}^{-1/3}$. The electrostatic potentential difference, $|\Phi(a) - \Phi(R)|$, decreases then at such a rate that the fraction of quasi-condensed counterions decreases. As a result, $Z_\text{eff}$ has a minimum, roughly here at $\phi \sim 0.01$, and it  increases subsequently with increasing $\phi$.  Notice further that $Z_\text{eff}$ tends asymptotically to $Z$ for large values of $\phi$. All considered renormalization methods share this qualitative behavior, except for the RJM, where $Z_\text{eff}$ increases only weakly with increasing $\phi$. The non-monotonic behavior of $Z_\text{eff}(\phi)$ becomes less pronounced with increasing salt content, where the minimum turns more shallow and is shifted to larger volume fractions.
 
The inset of Fig.~\ref{effect_charge_SF} shows how $Z_\text{eff}$ depends on the bare valence $Z$. For low values of $Z$, the potential difference between a colloid surface and the bulk region of the suspension is small enough that $Z\approx Z_\text{eff}$, implying no significant counterion condensation. As $Z$ increases, condensation sets in, leading to $Z_\text{eff}<Z$. Notice that all renormalization methods predict $Z_\text{eff}\approx Z$ and $\kappa_\text{eff}\approx \kappa_0=\sqrt{4\pi\lambda_{\text{B}}n_+}$ for $Z\lambda_{\text{B}}/a\lesssim5$ (c.f. insets in  Figs.~\ref{effect_screening_SF} and \ref{effect_charge_SF}). Charge renormalization first becomes relevant for $Z\lambda_{\text{B}}/a\gtrsim 5$, beyond which the predictions for $\kappa_\text{eff}$ and $Z_\text{eff}$ progressively differ from one other as the coupling strengthens. For sufficiently high $Z$, most methods predict that $Z_\text{eff}$ saturates to a value dependent on the invoked method. The exception is RLRT, for which $Z_\text{eff}$ continues to grow gradually with increasing $Z$.
	
All considered methods make comparable, though quantitatively different, predictions for $Z_\text{eff}$ and $\kappa_\text{eff}$. The only exception is RJM, which predicts distinctly stronger counterion condensation, with accordingly lower $Z_\text{eff}$. Comparing the CM-based methods only shows that $Z_\text{eff}$ is higher for mean than for edge linearization. The fact that $Z_\text{eff}(\text{mean})>Z_\text{eff}(\text{edge})$ was discussed above in relation to Eq.~(\ref{ZeffSCmodel}). It is important to realize, however, that $Z_\text{eff}$ and $\kappa_\text{eff}$ have physical significance only insofar as they affect predictions of thermodynamic and structural properties. 

We analyze next how salt concentration variation affects the renormalized interaction parameters. Since we assume Donnan equilibrium, results are presented as functions of the reservoir screening constant, $\kappa_\text{res}\propto \sqrt{n_\text{res}}$, which is proportional to the square-root of the reservoir salt concentration.
\begin{figure}
\centering
\includegraphics[width=8cm]{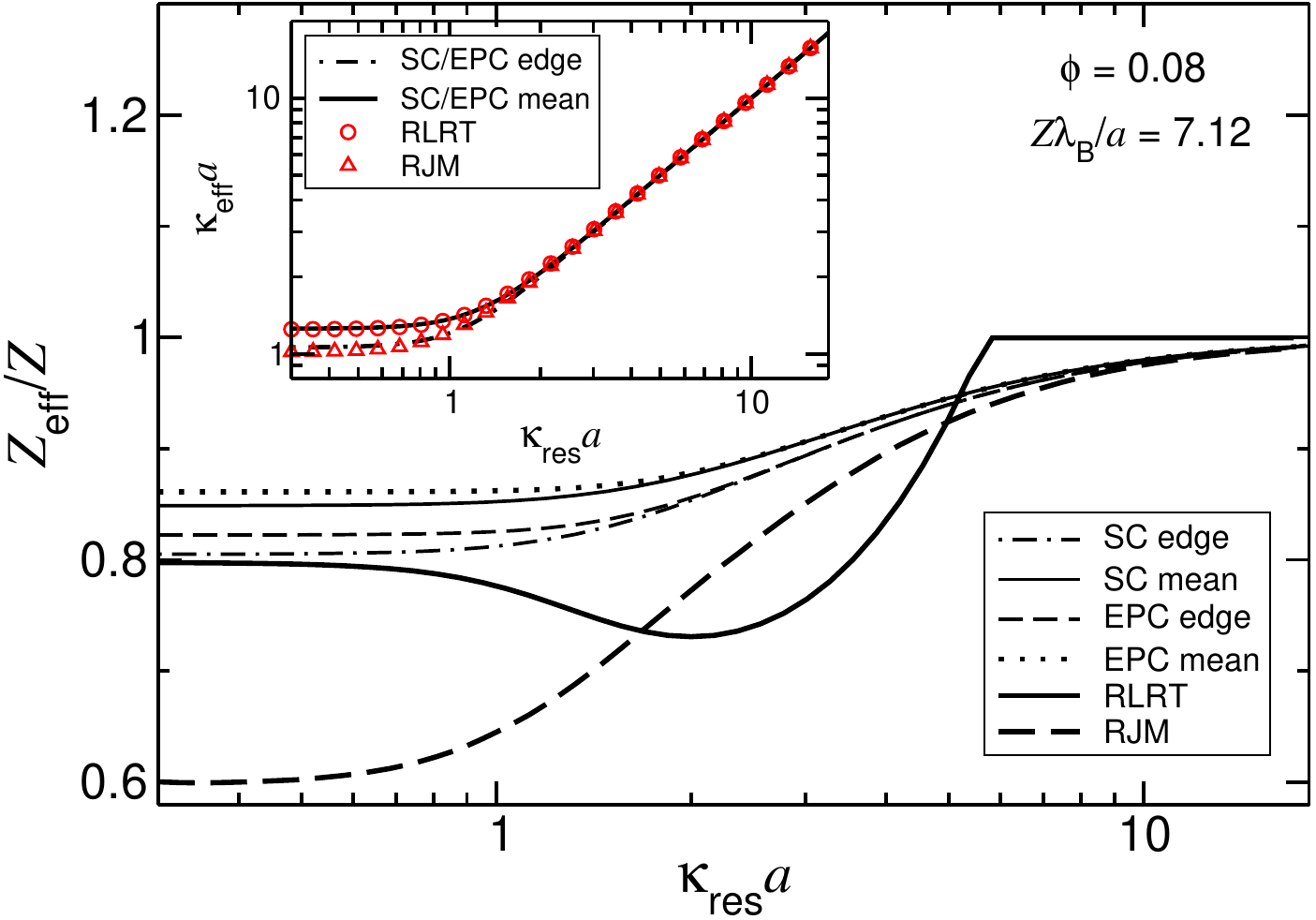}
\caption{Predictions of different renormalization methods for ratio of renormalized (effective) and bare macroion valence, $Z_\text{eff}/Z$, as function of reduced reservoir screening constant $\kappa_\text{res}a$ ($\kappa_\text{res}\propto \sqrt{n_\text{res}}$), for $\phi=0.08$, $Z=40$, $Z\lambda_\text{B}/a=7.12$, and $\lambda_\text{B}=0.714$ nm. Inset: reduced effective screening constant, $\kappa_\text{eff}a$, as function of $\kappa_\text{res}a$.}
\label{effect_param_salt}
\end{figure}
Figure \ref{effect_param_salt} depicts how $Z_\text{eff}$ and $\kappa_\text{eff}$ (inset) vary with reservoir salt concentration. In the limit of high salt concentration, all considered renormalization methods predict that the $Z_\text{eff}$ curve flattens out and approaches a limiting value of $Z_\text{eff}\rightarrow Z$. Most methods predict $Z_\text{eff}$ to grow monotonically with increasing salt concentration. The exception is RLRT, which, despite showing the correct limiting behavior for low and high salt concentrations, is nonmonotonic for intermediate concentrations and saturates at $Z_\text{eff}=Z$ above a critical value of $\kappa_\text{res}$. This unusual behavior can be attributed to the property of the RLRT, described in Sec.~\ref{Sec:RLRT}, by which counterion quasi-condensation ceases abruptly when the counterion-macroion electrostatic energy of attraction is comparable to the thermal energy [Eq.~(\ref{assoceq})]. In concert with $Z_\text{eff}$, the renormalized screening constant $\kappa_\text{eff}$ (inset of Fig. \ref{effect_param_salt}) is constant in the counterion-dominated regime and monotonically grows with increasing salt concentration, tending to the reservoir screening constant $\kappa_\text{res}$ in the salt-dominated regime. Again, while the various renormalization methods define the effective interaction parameters differently, only the measurable thermodynamic and structural properties are physically meaningful.
	
According to Eq.~(\ref{kappasquare}), $\kappa_{\text{eff}}^2$ from the different mean-field approaches has two additive contributions -- one associated with free (uncondensed) counterions, $(\kappa^\text{eff}_\text{c})^2$, and the other with salt ions, $(\kappa^\text{eff}_\text{s})^2$, i.e.,
\begin{equation}\label{eq:KappaEffTwo}
\kappa_\text{eff}^2=(\kappa^\text{eff}_\text{c})^2+(\kappa^\text{eff}_\text{s})^2.
\end{equation}
From Eqs.~(\ref{effkappa_mean}), (\ref{effkappa_edge}), (\ref{screenparamjell}), and (\ref{screenparamRLRT}), it follows that $\kappa^\text{eff}_\text{c}\propto \sqrt{\phi Z_\text{eff}}$ for the CM-based methods and for RLRT and RJM, whereas $\kappa^\text{eff}_\text{s}\propto \sqrt{n_\text{res}}$ for RJM. Equation~(\ref{eq:KappaEffTwo}) allows to specify two regimes: a counterion-dominated regime for $\kappa^\text{eff}_\text{c} \gg \kappa^\text{eff}_\text{s}$, where $\kappa_\text{eff} \approx\kappa^\text{eff}_\text{c}$, and a salt-dominated regime for $\kappa^\text{eff}_\text{c} \ll \kappa^\text{eff}_\text{s}$, where $\kappa_\text{eff} \approx\kappa^\text{eff}_\text{s}$.
The major differences in the predictions for $Z_\text{eff}$ and $\kappa_\text{eff}$ between the various renormalization methods are in the counterion-dominated regime, where these parameters are insensitive to changes in salt concentration. 
In contrast, in the salt-dominated regime, all methods converge to $\kappa_\text{eff}\rightarrow\kappa_\text{res}$ and $Z_\text{eff}\rightarrow Z$. Therefore, differences between the methods tend to emerge at low salt concentrations.
\begin{figure}
\centering
\includegraphics[width=8cm]{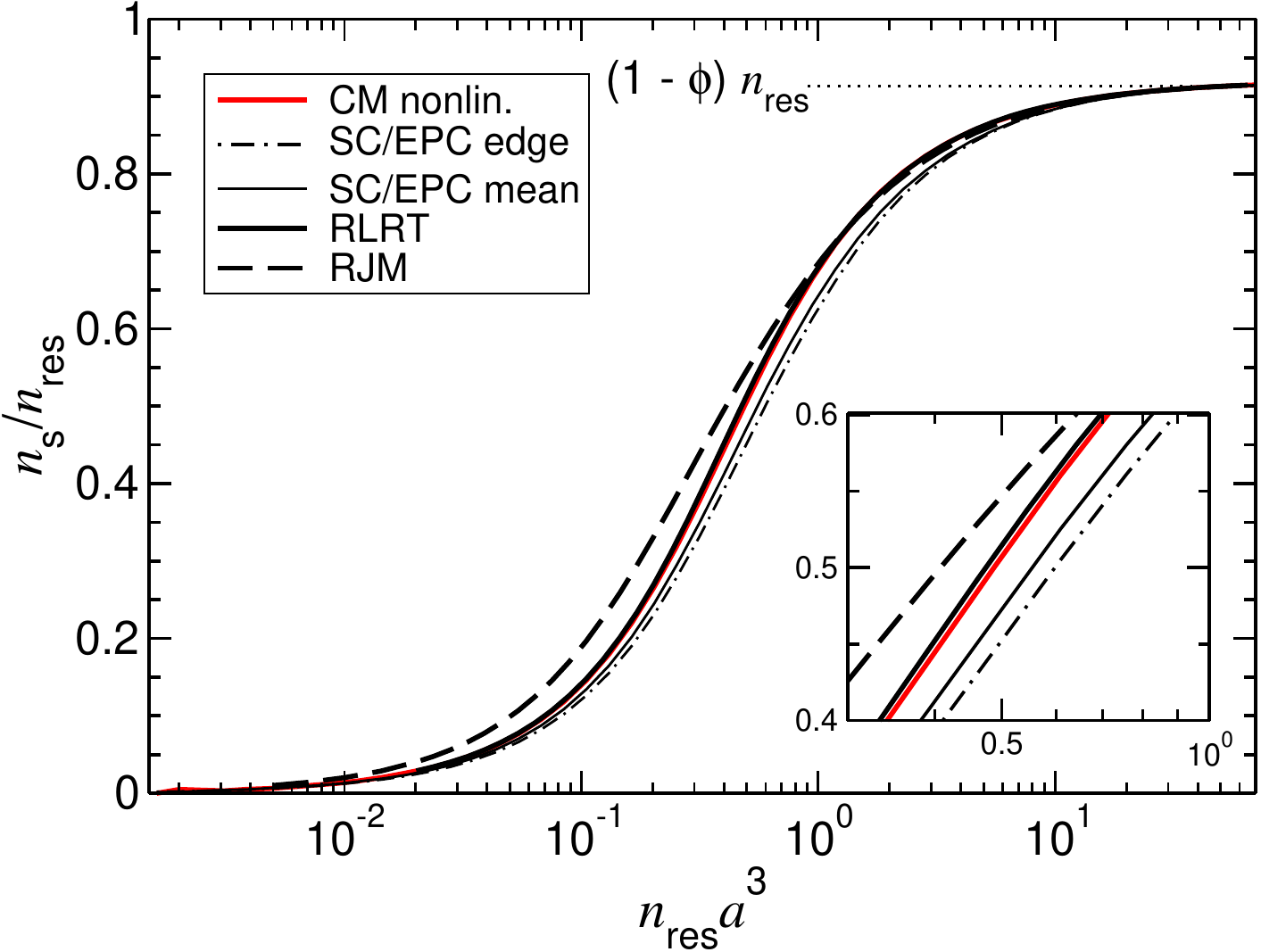}
\caption{Predictions of different renormalization methods for effective suspension salt concentration in units of reservoir salt concentration, $n_\text{s}^\text{eff}/n_\text{res}$, as function of reduced reservoir concentration, $n_\text{res}a^3$, for $\phi=0.08$, $Z=40$, $Z\lambda_\text{B}/a=7.12$, and $\lambda_\text{B}=0.714$ nm. Results for the nonlinear (unrenormalized) cell model, according to Eq.~(\ref{eq:nsCM}), are included for comparison (solid red curve). Inset: Magnification of the transition region from counterion- to salt-dominated regime.}
\label{syst_salt_conc}
\end{figure}
	
In Donnan equilibrium, the suspension salt concentration, $n_\text{s}$, is determined by equating the microion chemical potentials in the reservoir and suspension. In the nonlinear CM approximations, $n_\text{s}$ is calculated from Eq.~(\ref{eq:nsCM}). When charge renormalization is operative, the renormalized suspension salt concentration, $n_\text{s}^\text{eff}$, follows from the renormalized interaction parameters and depends on the concentration of free salt ion pairs. Explicit expressions for $n_\text{s}^\text{eff}$ predicted by the various methods are given in Eqs.~(\ref{nseff_mean}) and (\ref{nseff_edge}) for SC (and EPC) with mean and edge linearization, respectively, in Eq.~(\ref{screenparamRLRT}) for RLRT, and in Eq.~(\ref{nseff_RJM}) for RJM. 
Figure \ref{syst_salt_conc} illustrates how the suspension salt concentration, $n_\text{s}^\text{eff}$, varies with reservoir salt concentration $n_\text{res}$. The fact that $n_\text{s}^\text{eff}<n_\text{res}$ reflects the salt expulsion effect for a colloidal suspension in Donnan equilibrium. Two limiting plateau regions are observed: one in the counterion-dominated regime, where $n_\text{s}^\text{eff}\approx 0$, and the other in the salt-dominated regime, where $n_\text{s}^\text{eff}\rightarrow (1-\phi)n_\text{res}$. The predictions for $n_\text{s}^\text{eff}$ by the different methods tend to coincide in these limits. In the transition regime at intermediate salt concentrations, the relative differences are $20\%$ at most. It is interesting to compare the various predictions for $n_\text{s}^\text{eff}$ with $n_\text{s}$ obtained [from Eq.~(\ref{eq:nsCM})] by integrating the coion density profile in the nonlinear CM approximation (red curve). As seen, $n_\text{s}$ is approximately equal to the RLRT prediction of $n_\text{s}^\text{eff}$.
	
As previously noted, salt ions are expelled from the suspension into the reservoir at low to intermediate reservoir salt concentrations, leading to $n_\text{s}^\text{eff}< (1-\phi)n_\text{res}$, since the counterions make here the largest contribution to the suspension chemical potentials. According to Fig.~\ref{syst_salt_conc}, RJM predicts the weakest salt-expulsion effect, while CM-based methods predict the strongest effect.

\subsection{Case of Effective Valence Exceeding Bare Valence}\label{Zeff>Z}
\label{sec:VB}
On comparing predictions by the various charge renormalization methods in the $(\phi,n_\text{res}a^3)$ parameter space, one notices that the EPC methods predict an effective valence higher than the bare valence ($Z_\text{eff}>Z$) for sufficiently high colloid concentrations and low $Z$. Since quasi-condensation of counterions at higher $Z$ always results in $Z_\text{eff}<Z$, this observation is not a nonlinear renormalization effect. Instead, it is due to a reduction in the linear microion screening of a macroion caused by other nearby macroions. In the context of the EPC method, Boon \cite{DiscussionBoon} attributes this effect to the creation of a spherical hole in the one-component plasma (counterions plus neutralizing background) by the hard core of a central macroion. Since the hole acts effectively as a region of charge density opposite to that of the counterions, the pointlike effective macroion particle in EPC attains an increased effective charge. The effective coupling parameter, $Z_\text{eff}\lambda_\text{B}/a$, increases here with increasing volume fraction.
	
\begin{figure}
\centering
\includegraphics[width=8cm]{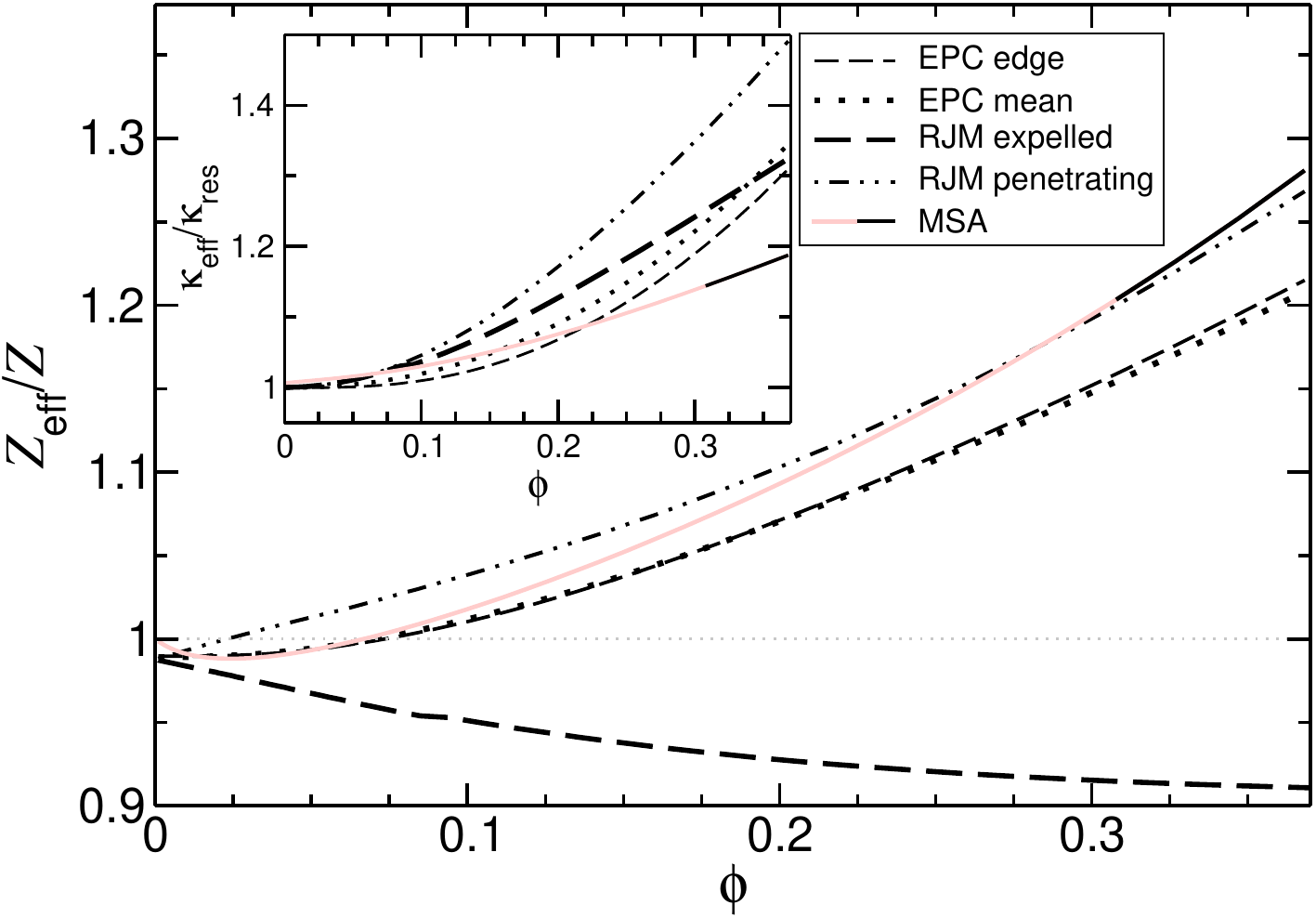}
\caption{Effective macroion valence in units of the bare valence, $Z_\text{eff}/Z$, as function of volume fraction $\phi$ for the indicated methods. Note that EPC and penetrating RJM allow for $Z_\text{eff}>Z$ at sufficiently large $\phi$ and low coupling. Inset: Reduced effective screening constant $\kappa_\text{eff}/\kappa_\text{res}$ versus $\phi$ for the different methods. The solid black part of the PM-MSA curve is for $g_\text{MSA}(\sigma^+\geq 0)$, as physically demanded, while the red part violates this condition. Other system parameters are $Z=80$, $Z\lambda_{\text{B}}/a=1.52$, $\lambda_\text{B}=0.714$ nm and $\kappa_\text{res}a=1.327$.}
\label{effect_param_specialcase}
\end{figure}

In Fig.~\ref{effect_param_specialcase}, $Z_\text{eff}$ and $\kappa_\text{eff}$ are plotted for an exemplary, weakly-coupled system with $Z\lambda_{\text{B}}/a=1.52$. Notice that $Z_\text{eff}>Z$ in EPC for $\phi\geq0.075$, with no noticeable difference between edge and mean linearization. There is a remnant of quasi-condensation of counterions visible for $\phi\leq 0.075$, where $Z_\text{eff}<Z$. The inset displays the expected monotonic increase of $\kappa_\text{eff}$, with the results for edge and mean linearizations being qualitatively similar.

The EPC method is not the only method predicting $Z_\text{eff}>Z$ for concentrated suspensions of weakly charged colloids. Another method is the penetrating RJM, described in its simplest version in Sec.~\ref{Subsec:RJM}, where, different from the non-penetrating (expelled) RJM, the homogeneous neutralizing jellium also penetrates the macroion core, as quantified by the boundary conditions in Eq.~(\ref{BC_RJMpenetrat}). 
	
As noted above in Sec.~\ref{Subsec:RJM}, the RJM is based on the jellium approximation (JA), widely used in solid state physics and originally applied to charge-stabilized colloidal suspensions by Beresford-Smith {\it et al.} \cite{Beresford-Smith1985}. The JA approximates the macroion radial distribution function by $g_\text{mm}(r)=1$ for all $r>0$. By combining this approximation with the linear MSA closure for the microion-microion direct correlation functions of pointlike microions, plus the MSA closure for the macroion-microion direct correlation functions at $r>a$, the effective macroion pair potential for weak macroion coupling and high dilution is obtained in the linear penetrating JA as \cite{Beresford-Smith1985}
\begin{equation}
\beta u_\text{eff}(r)=\lambda_{\text{B}}Z^2(1+\phi)^2\left(\frac{e^{\kappa a}}{1+\kappa a}\right)^2\frac{e^{-\kappa r}}{r},
\label{EffPot_Jellium}
\end{equation}
with $\kappa^2 = 4\pi\lambda_{\text{B}}(n_++n_-)$ in the case of monovalent microions. Contrasting this pair potential with the one in Eq.~(\ref{YukawaPot}), one notices that they are identical apart from the factor $(1+\phi)^2$, which is due to the macroion core penetration by the jellium. By combining the (penetrating) JA with the nonlinear HNC closure for the macroion-microion direct correlation functions, nonlinear effects can be included, leading to a description equivalent to the nonlinear PB equation in Eq.~(\ref{NonlinPBeqJell}), with associated boundary conditions for the penetrating jellium in Eq.~(\ref{BC_RJMpenetrat}). Thus, one can apply a charge renormalization procedure, similar to the one for the nonpenentrating case, resulting in renormalized valence and screening parameter used as inputs to Eq.~(\ref{EffPot_Jellium}) in order to incorporate nonlinear screening effects. The effective pair potential derived within the nonlinear penetrating JA approximation has proven to accurately describe the pair structure of highly-coupled suspensions for $Z\lambda_\text{B}/a\approx 13$ up to macroion concentrations of $\phi=0.13$ \cite{Beresford-Smith1985}.
	
Another model predicting $Z_\text{eff}>Z$ at high concentration and small $Z\lambda_\text{B}/a$ is the PM-MSA scheme, which gives rise to the effective pair potential \cite{Belloni_JCP_1986,Ruiz-Estrada1990}
\begin{equation}
\beta u_{\text{eff}}(r)=\lambda_\text{B}Z^2X_\text{MSA}^2\frac{\exp(-\kappa r)}{r},\,\,\, r>2a\,,
\label{EffectPotMSA1}
\end{equation}
with prefactor \cite{Belloni_JCP_1986}
\begin{equation}
X_\text{MSA} = \cosh(\kappa a)+U\,[\kappa a\cosh(\kappa a)-\sinh(\kappa a)]
\label{XMSA}
\end{equation}
and screening constant $\kappa=\sqrt{4\pi\lambda_\text{B}\left(Zn_\text{m}+2 n_\text{s}\right)}$ for monovalent counterions and added 1:1 electrolyte of pair concentration $n_\text{s}$. Notice that $\kappa$ here does not include the excluded-volume factor $1/(1-\phi)$. 
		
The PM-MSA effective pair potential in Eq.~(\ref{EffectPotMSA1}) is of DLVO-type with effective macroion valence $Z_\text{eff}=X_\text{MSA}Z$. 
The parameter $U$ is determined by $U = c/(\kappa a)^3-\gamma/(\kappa a)$, where $c=3\phi/(1-\phi)$ and $\gamma=(c+\Gamma_\text{MSA} a)/(1+c+\Gamma_\text{MSA} a)$.
The MSA screening parameter $\Gamma_\text{MSA}$ is the unique positive solution of the biquadratic equation
\begin{equation}
(\Gamma_\text{MSA} a)^2=(\kappa a)^2+\frac{(q_0a)^2}{(1+c+\Gamma_\text{MSA} a)^2}\,,
\label{equation_GammaMSA}
\end{equation}
fulfilling $\Gamma_\text{MSA} > \kappa$, where $(q_0a)^2=3\lambda_\text{B}\phi Z^2/a$. In the infinite dilution limit ($n_\text{m}\rightarrow0$), one obtains $X_\text{MSA}\rightarrow e^{\kappa a}/(1+\kappa a)$, recovering hereby the standard DLVO potential. Within the considered Donnan equilibrium, $n_s$ is given in PM-MSA by
\begin{align}
\frac{n_\text{s}}{n_\text{res}} &=-\frac{n_\text{m} Z}{2 n_\text{res}}\nonumber\\
&+ \left[(1-\phi)^2 \exp\left[\left(\Gamma_\text{MSA}-\kappa_\text{res}\right)\lambda_\text{B}\right] + \left( \frac{n_\text{m} Z}{2 n_\text{res}}\right)^2\right]^{1/2},
\label{equation_MSA_Donnan}
\end{align}
which needs to be solved in conjunction with Eq. (\ref{equation_GammaMSA}) for $\Gamma_\text{MSA}$. The above expression for $n_\text{s}$ reduces, in the limit $a\to 0$, to the standard Debye-H\"uckel limiting law result for the salt concentration in Donnan equilibrium. In the infinite dilution limit of macroions ($n_\text{m}\to 0$), $X_\text{MSA}$ tends to the geometric factor $\exp(\kappa a)/\left(1+\kappa a \right)$ of spheres.
	
Furthermore, in the high-temperature limit, where $\kappa a\ll 1$, it follows that
\begin{equation}
Z_\text{eff}^\text{MSA} = Z\frac{1+\kappa a}{\exp(\kappa a)}X_\text{MSA} 
~~\to~~\frac{Z}{1-\phi}\approx (1+\phi)\;\!Z +{\cal O}\left(\kappa a\right)\,.
\end{equation}  
Notice that $Z_\text{eff}^\text{MSA}$ differs from the linear effective valence $Z_\text{eff}=Z(1+\phi)$ appearing in the penetrating JA pair potential of Eq.~(\ref{EffPot_Jellium}) only by small corrections of quadratic order in $\phi$ and linear order in $\kappa$. An explicit calculation shows indeed that $Z_\text{eff}^\text{MSA}>Z$.
	
Figure~\ref{effect_param_specialcase} depicts $Z_\text{eff}$ and $\kappa_\text{eff}$, predicted by the EPC methods and by the penetrating and non-penetrating (expelled) RJM, for weakly coupled macroion suspensions with $Z\lambda_{\text{B}}/a=1.52$ and system parameters where the EPC-based $g(r)$ calculated using RY is of higher accuracy than that based on the SC methods \cite{Boon_PNAS_2015}. The curves of $\kappa_\text{eff}/\kappa_\text{res}$, depicted in the inset, show the expected monotonic increase with increasing $\phi$ for all cases. However, it is only in EPC and penetrating RJM that $Z_\text{eff}>Z$ for $\phi\gtrsim0.1$. The non-penetrating (expelled) RJM predicts $Z_\text{eff}\le Z$ for all concentrations. 
	
The solid black part of the PM-MSA curve in Fig.~\ref{effect_param_specialcase} represents $Z_\text{eff}^\text{MSA}$ and $\kappa$ obtained in linear PM-MSA for $\phi>0.3$, where the MSA contact value $g(\sigma^+)$ is non-negative. As noted, $Z_\text{eff}^\text{MSA}$ is larger than $Z$ and approximately equal to $Z_\text{eff}$ as predicted by the penetrating RJM in its simplest version described in Sec.~\ref{Subsec:RJM}. According to ref.~\cite{Belloni_JCP_1986}, $Z_\text{eff}^\text{MSA}>Z$ or equivalently, $X_\text{MSA}>e^{\kappa a}/(1+\kappa a)$, can be attributed to a decreased screening ability of the microions around a given macroion owing to a steric constriction by neighboring macroions, leading to stronger effective macroion-macroion repulsion. Such reduced screening occurs when neighboring electric double layers overlap.
\begin{figure}
\centering
\includegraphics[width=8.2cm]{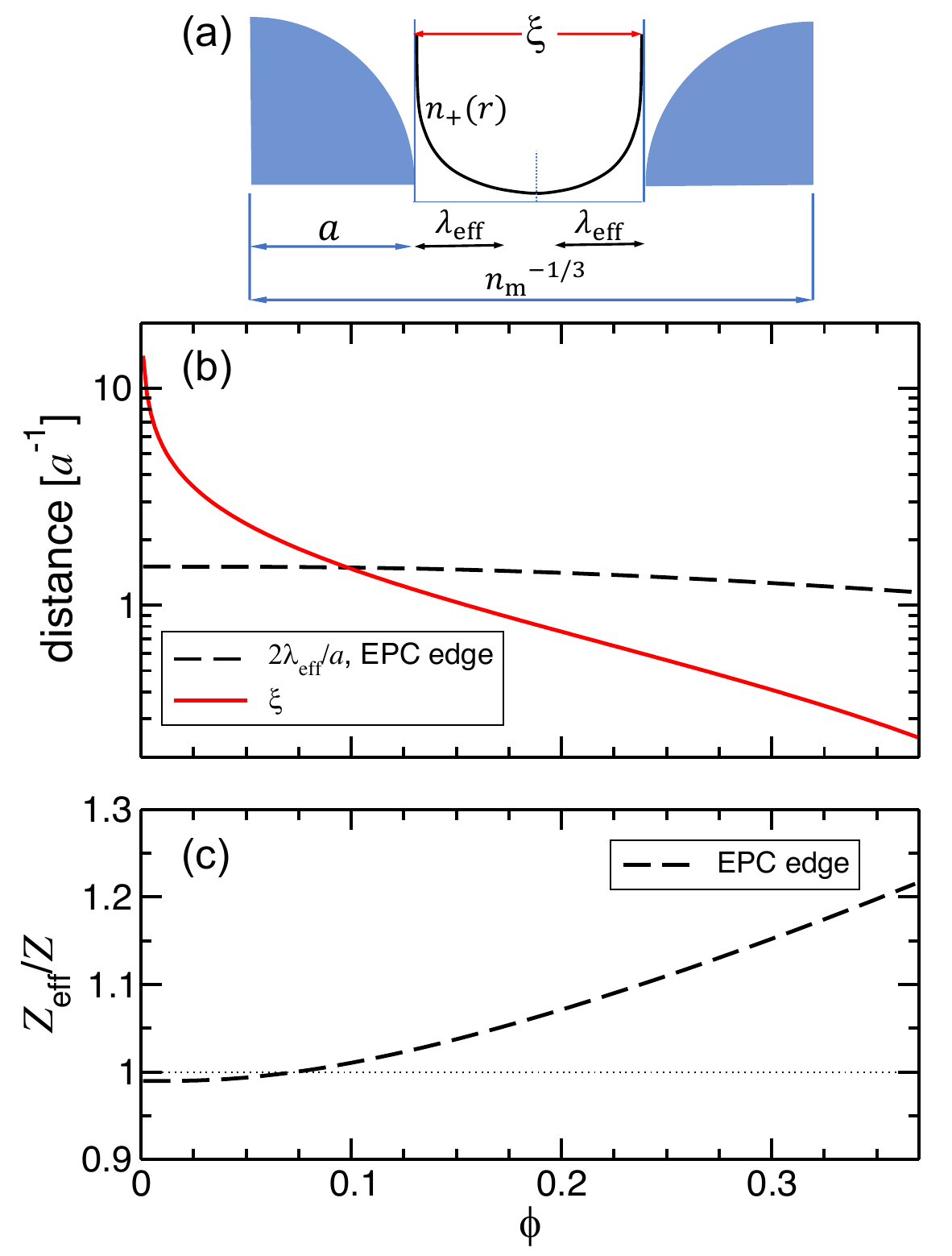}
\caption{(a) Sketch of two spherical macroions (blue), labeled with relevant parameters. (b) Reduced effective screening length, $\lambda_\text{eff}/a$, with red curve depicting reduced mean distance, $\xi=n_\text{m}^{-1/3}/a-2$, between two macroion surfaces. (c) Reduced effective valence, $Z_\text{eff}/Z$, plotted versus $\phi$ as predicted by EPC edge method with system parameters from Fig.~\ref{effect_param_specialcase}. Note $Z_\text{eff}>Z$ for sufficiently large $\phi$.}
\label{colloidalmeandistance}
\end{figure}

To examine how electric double layer overlap relates to the EPC prediction of $Z_\text{eff}>Z$, in Fig.~\ref{colloidalmeandistance}(a) we compare the EPC effective electrostatic screening length, $\lambda_\text{eff}=\kappa_\text{eff}^{-1}$, with the (reduced) mean surface-to-surface distance of neighboring macroions, $\xi=n_\text{m}^{-1/3}/a-2$. This comparison reveals the volume fraction $\phi$ at which overlap of double layers sets in. As seen in Fig.~\ref{colloidalmeandistance}(c), $Z_\text{eff}$ becomes {\it larger} than $Z$ at about the same concentration where neighboring double layers begin to overlap, suggesting that $Z_\text{eff}>Z$ is not a single-macroion effect. This linear screening effect should play a role in concentrated solutions of weakly charged proteins. At sufficiently large $\phi$ and strong electrostatic coupling, however, $Z_\text{eff}<Z$ due to significant counterion quasi-condensation (charge renormalization). Remnant charge renormalization is predicted by all considered schemes (except MSA), and noticeable in Fig.~\ref{colloidalmeandistance}(c) for $\phi < 0.05$ when the electric double layers are practically non-overlapping. The linear MSA does not account for counterion quasi-condensation. Actually, the red part of the MSA curve in Fig.~\ref{effect_param_specialcase} is in the regime $\phi<0.3$ where the MSA predicts unphysical negative values of the macroion radial distribution function $g_\text{mm}(r)$ at contact distance $r= \sigma=2a$.  

The prediction $Z_\text{eff}>Z$ by the EPC and penetrating RJM mean-field methods for concentrated macroion dispersions, where nonlinear microion response (counterion quasi-condensation) is absent or weak, should be compared with the so-called macroion charge amplification effect described by Gonz{\'a}lez-Mozuelos, Guerrero-Garcia {\it et al.} \cite{Gonzalez-Mozuelos2013,Guerrero-Garcia2013}. The latter effect, predicted on the basis of elaborate molecular dynamics and dressed ion theory calculations, where microion correlations are included, is found for dilute dispersions of charged nano-sized particles at larger concentrations of the suspending 1:1 electrolyte solution, and for a small macroion-counterion size ratio of 5:1. This effect likely originates from microion correlations, in particular, from the non-negligible size of the counterions and (to a lesser extent) coions close to the macroion surface attenuating the electrostic shielding of the macroion charge \cite{Gonzalez-Mozuelos2013}. Microion correlation effects are outside the scope of the discussed PB mean-field methods, and $a \gg \lambda_\text{B}$ is implicitly assumed for their validity. In the PM-MSA method, microion correlations are accounted for to reasonable accuracy only for concentrated dispersions of weakly charged macroions, where nonlinear microion response is negligible. The microions in PM-MSA are taken here as pointlike for simplicity, allowing for an analytic expression for the effective macroion pair potential in Eq.~(\ref{EffectPotMSA1}) of single-Yukawa form. While microion nonzero size effects are excluded hereby, remaining electrostatic correlations between the pointlike microions contribute in PM-MSA to an effective macroion valence larger than the bare one.

\subsection{Pair Structure}
Having discussed the renormalized interaction parameters, $\kappa_\text{eff}$ and $Z_\text{eff}$, predicted by the considered renormalization methods, we next compute structural suspension properties within the OCM governed by the effective pair potential, $u_\text{eff}(r)$. Explicitly, we compute the macroion-macroion radial distribution function, $g(r)$, and static structure factor, $S(q)$, using the thermodynamically self-consistent Rogers-Young (RY) integral-equation scheme, which provides accurate results for dispersions with Yukawa-type repulsive interactions \cite{Gapinski_JCP2012,Gapinski_JCP_2014,Banchio_JCP_2018}. We first consider salt-free systems, for which differences in the renormalized parameters are largest (cf.~Fig.~\ref{effect_param_salt}), and for which Monte Carlo (MC) simulation data based on the PM are available from Linse \cite{Linse2000}.

\begin{figure}
\centering
\includegraphics[width=7.5cm]{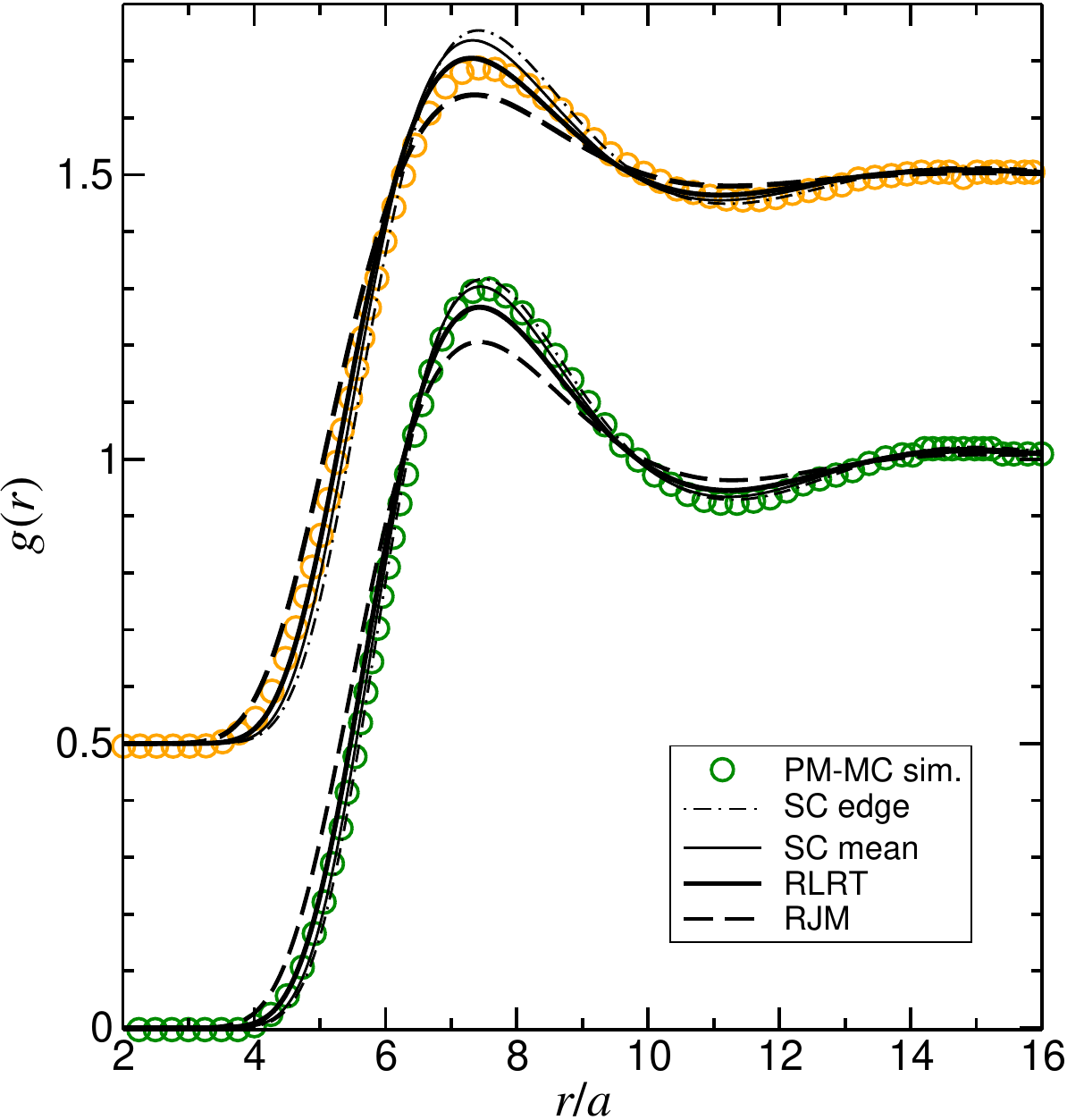}
\caption{Macroion-macroion radial distribution function, $g(r)$, of a salt-free suspension ($n_\text{res} = 0$) at volume fraction $\phi=0.01$, for different coupling parameters, $Z\lambda_\text{B}/a$. The curves are computed using the RY method with renormalized interaction parameters, $Z_\text{eff}$ and $\kappa_\text{eff}$, from the considered renormalization methods inserted into the OCM $u_\text{eff}(r)$. Results from the EPC edge and mean methods are omitted, since these are close to the SC edge and mean results, respectively. Green and yellow open symbols are PM-MC simulation data \cite{Linse2000} for $Z\lambda_\text{B}/a=7.12$ and $14.23$, respectively. The latter are vertically upshifted by 0.5 for better visibility. Other system parameters are $Z=40$ and $\lambda_\text{B}=0.714$ nm.
}
\label{g_of_r_comparison_0-01}
\end{figure}

Figure~\ref{g_of_r_comparison_0-01} compares OCM-based theoretical predictions for $g(r)$ with PM-MC simulation data of Linse \cite{Linse2000} for $\phi=0.01$ for the two highest considered coupling parameters, $Z\lambda_\text{B}/a$, where the renormalization methods show the greatest differences. Qualitatively, all considered methods provide good results for $g(r)$ in the explored parameter range. With increasing $Z\lambda_\text{B}/a$, there is a gradual decrease in the principal peak height, $g(r_\text{m})$, and increase in the peak position, $r_\text{m}$, as seen in the simulations. The analyzed renormalization methods reproduce this behavior, but slightly overestimate $g(r_\text{m})$, increasingly so with increasing coupling, with the exception of RJM. In the simulation study \cite{Linse2000}, it is shown that stronger coupling promotes macroion aggregation.

\begin{figure}
\centering
\includegraphics[width=8cm]{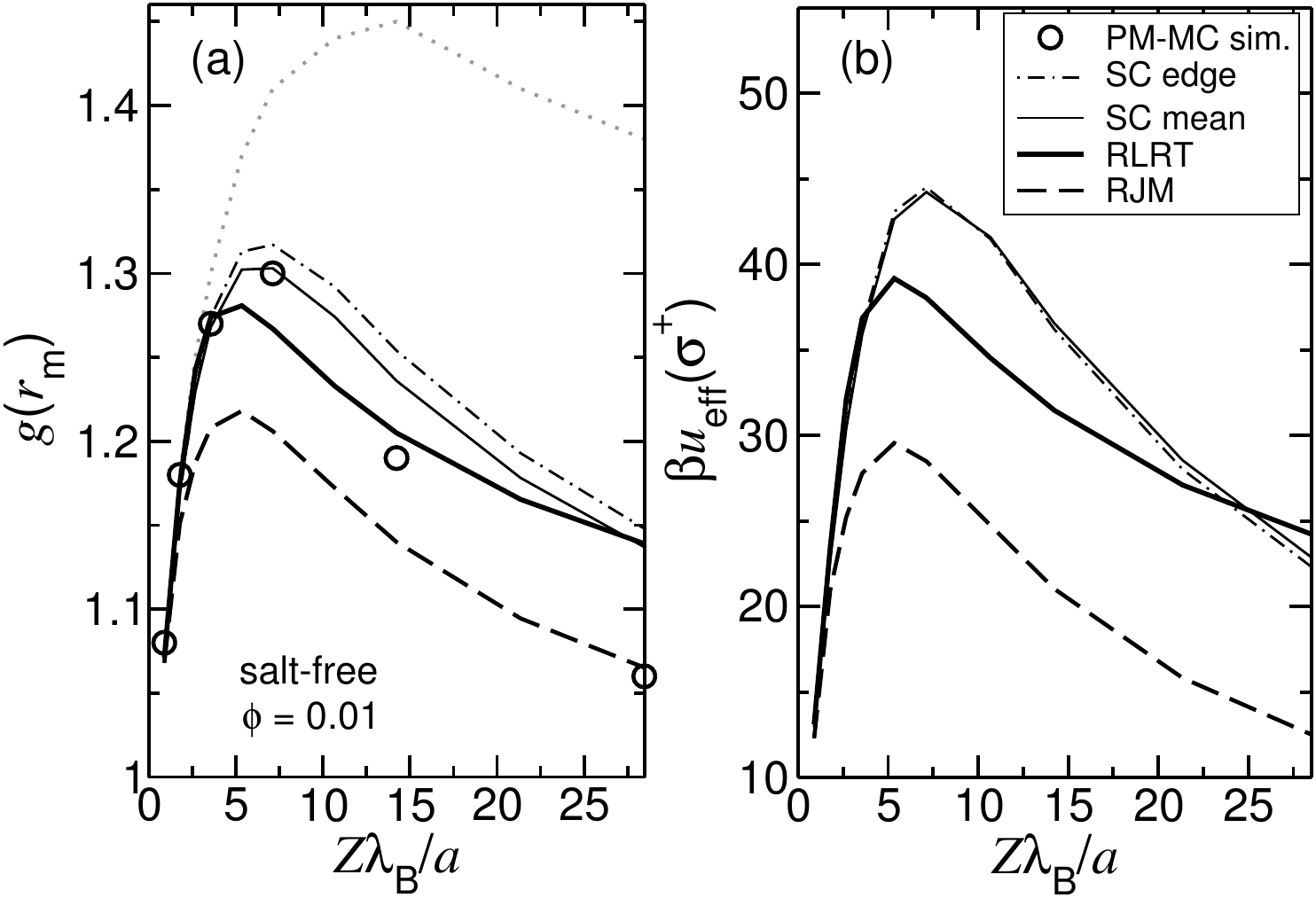}
\caption{(a) Principal peak height, $g(r_\text{m})$, of the macroion-macroion radial distribution function, and (b) reduced effective pair potential at contact, $\beta u_\text{eff}(\sigma^+)$, versus bare coupling, $Z\lambda_\text{B}/a$. The indicated charge renormalization schemes describe a salt-free suspension ($n_\text{res} = 0$) of bare macroion valence $Z=40$ at $\phi=0.01$ and for $\lambda_\text{B}=0.714$ nm. SC edge and SC mean results are very close to the EPC edge and EPC mean results, respectively (the latter, therefore, not shown). The dotted grey line in panel (a) is a prediction without charge renormalization. Curves in panel (a) are OCM-RY results, while symbols are PM-MC data \cite{Linse2000}.}
\label{summary_structure}
\end{figure}

Figure~\ref{summary_structure}(a) shows the principal peak height, $g(r_\text{m})$, as a function of $Z\lambda_\text{B}/a$ as predicted by the various renormalization methods. Renormalization becomes relevant for $Z\lambda_{\text{B}}/a\gtrsim 5$ (see insets of Figs.~\ref{effect_screening_SF} and \ref{effect_charge_SF}). At weak coupling below the onset of renormalization, all methods accurately reproduce the initial increase of $g(r_\text{m})$ visible in the PM-MC simulation data. This agreement simply reflects the accuracy of the RY method for suspensions with repulsive Yukawa-type interactions. The dotted curve in Fig.~\ref{summary_structure}(a) illustrates that $g(r_\text{m})$ is significantly overestimated at stronger coupling, if charge renormalization is neglected.
It is obtained from using the non-renormalized pair potential in Eq. (\ref{YukawaPot}) in the RY calculation of $g(r)$, for $\kappa$ determined by Eq. (\ref{kappasquare}), but without invoking the $1/(1-\phi)$ free volume correction factor.
Provided renormalization is incorporated, the nonmonotonic behavior of the MC data for $g(r_\text{m})$ in panel (a) is qualitatively captured by all renormalization methods. Consistent with its comparatively strong renormalization of $Z_\text{eff}$ (cf. Fig.~\ref{effect_charge_SF}), the RJM strongly underestimates $g(r_\text{m})$. In contrast, RLRT is more accurate, overestimating $g(r_\text{m})$ only for the highest considered coupling. The CM-based methods for mean and edge linearization reproduce the PM-MC $g(r_\text{m})$ most accurately at moderate coupling, with mean linearization giving slightly better results than edge linearization. Except for RJM, all methods overestimate $g(r_\text{m})$ at the highest coupling. The EPC and SC results for peak height are hardly distinguishable from each other when either edge or mean linearization is used.

For salt-free suspensions, $g(r_\text{m})$ has a dependence on the coupling parameter similar to that of the contact value, $u_\text{eff}(\sigma^+)$, of the effective pair potential. This similarity is evident from Fig.~\ref{summary_structure}~(b), where $\beta u_\text{eff}(\sigma^+)$ is plotted as a function of $Z\lambda_\text{B}/a$. The curves for the contact value of the effective pair potential predicted by the different renormalization methods qualitatively reflect the curves for $g(r_\text{m})$ in panel (a), with peaks located at the same respective coupling parameter values. Notice that $\beta u_\text{eff}(\sigma^+)\propto Z_\text{eff}^2(1+\kappa_\text{eff} a)^{-2}$, with the geometric factor $(1+\kappa_\text{eff} a)^{-2}$ arising from the impermeability of the macroion cores.
\begin{figure}
\centering
\includegraphics[width=8.5cm]{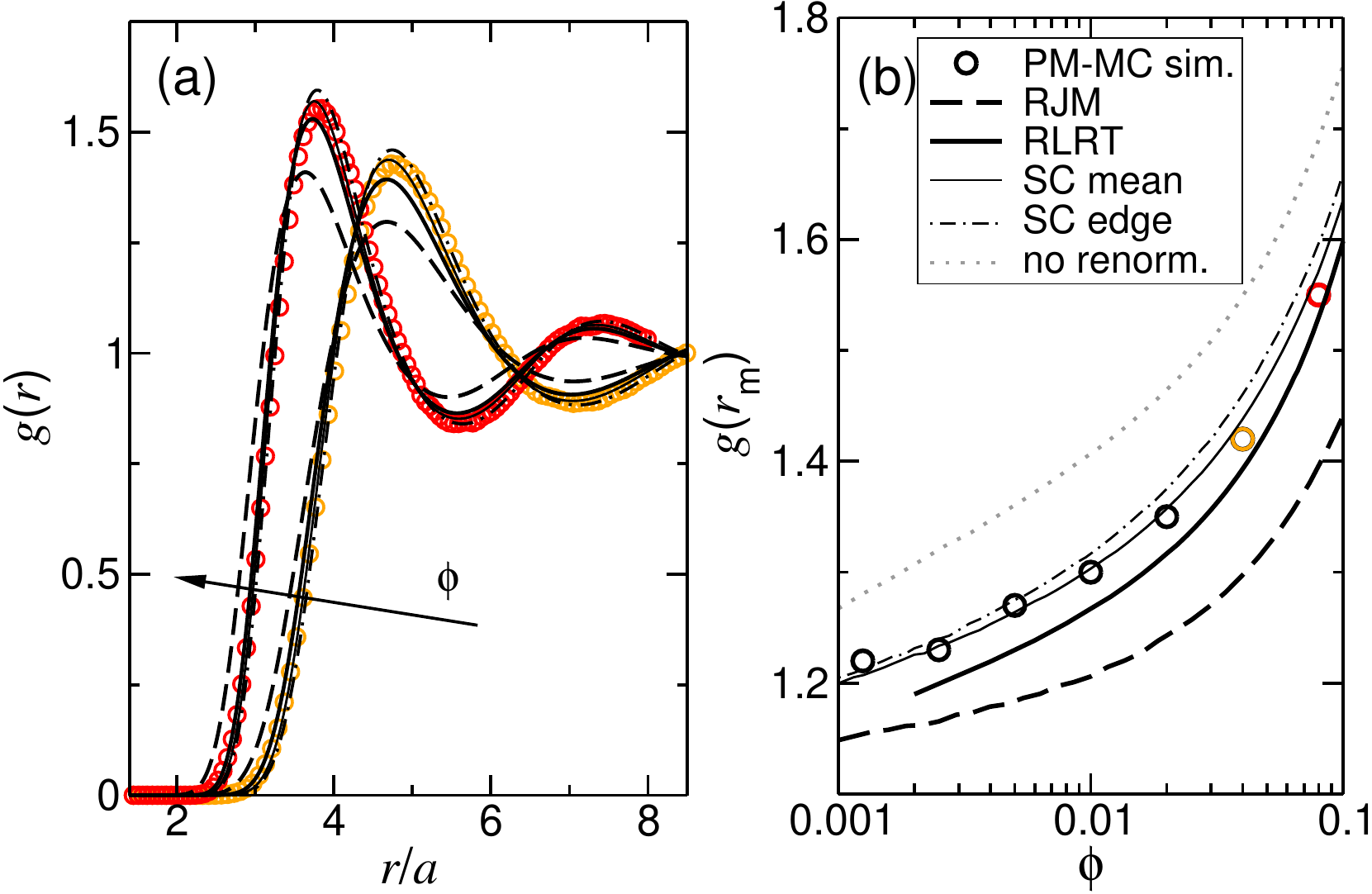}
\caption{(a) Macroion-macroion radial distribution function, $g(r)$, for $\phi=0.04$ and $0.08$. (b) Principal peak height, $g(r_\text{m})$, versus $\phi$ for salt-free suspensions ($n_\text{res} = 0$) with $Z\lambda_{\text{B}}/a=7.12$. Curves in panel (a) are obtained using the RY method with renormalized interaction parameters from the indicated renormalization methods. Open circles are PM-MC simulation data \cite{Linse2000}. In panels (a) and (b), SC edge and mean results for $g(r_\text{m})$ are very close to those for EPC edge and mean, respectively (the latter therefore not shown). The dotted grey line in panel (b) is the result without charge renormalization. Other system parameters are $Z=40$ and $\lambda_\text{B}=0.714$ nm.}
\label{summary_structure2}
\end{figure}

Figure~\ref{summary_structure2} displays the concentration dependence of $g(r)$ and $g(r_\text{m})$ for salt-free suspensions with $Z\lambda_{\text{B}}/a=7.12$, for which $g(r_\text{m})$ in Fig.~\ref{summary_structure} is most structured. In Fig.~\ref{summary_structure2}(a), predictions by different renormalization methods are compared with PM-MC simulation data for $\phi=0.04$ and $0.08$. With increasing $\phi$, the suspension becomes more structured, reflected in a sharpening of the principal peak of $g(r)$, whose position $r_\text{m}$ shifts to smaller inter-particle distances. Moreover, the secondary peak becomes more pronounced, reflecting the build-up of the second-neighbor shell. The position of the principal peak is approximately equal to the macroion next-neighbor distance, $r_\text{m}\approx n_\text{m}^{-1/3}=[3\phi/(4\pi a^3)]^{-1/3}$, typical of suspensions whose structure is determined by long-range repulsive interactions \cite{Linse2000}, for which $g(\sigma^+)=0$. In Fig.~\ref{summary_structure2}(b), predictions of $g(r_\text{m})$ by the different methods are plotted as functions of $\phi$. As seen, $g(r_\text{m})$ is underestimated by RJM and, to a lesser degree, also by RLRT, while the CM-based methods give quite accurate results in the full concentration range. Unlike the coupling variation depicted in Fig.~\ref{summary_structure}, the performance of the renormalization methods is relatively insensitive to variations in colloid concentration. The red and yellow open circles in Fig.~\ref{summary_structure2}(b) relate to the according $g(r)$ depicted in Fig.~\ref{summary_structure2}(a).
\begin{figure}
\centering
\includegraphics[width=8cm]{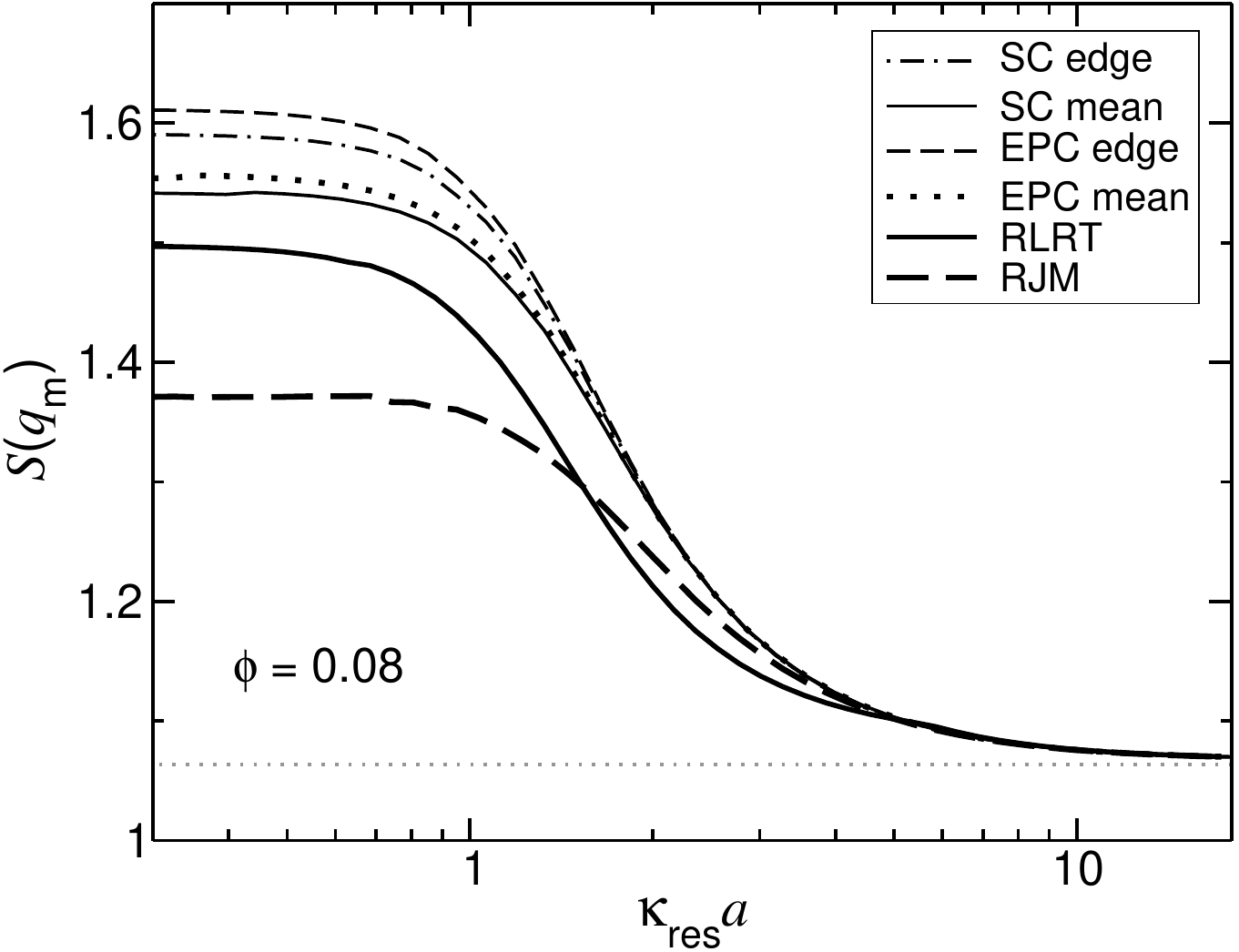}
\caption{Static structure factor peak height, $S(q_\text{m})$, versus reduced reservoir screening constant, $\kappa_\text{res}a$ ($\kappa_\text{res}^2\sim n_\text{res}$), for different renormalization methods as indicated, with $\phi=0.08$ and $Z\lambda_\text{B}/a=7.12$. The horizontal dotted line is $S(q_\text{m})$ of a hard-sphere fluid from Eq.~(\ref{CS-Sqm}). Other system parameters are $Z=40$ and $\lambda_\text{B}=0.714$ nm.}
\label{structure_vs_salinity}
\end{figure}
	
The static structure factor, $S(q)$, characterizes pair correlations in Fourier space. Its principal peak value, $S(q_\text{m})$, at wavenumber $q_\text{m}\approx 2\pi/r_\text{m}$, allows to roughly identify the freezing transition of suspensions of spherical particles with Yukawa-type repulsion, while $S(0)$ provides the osmotic compressibility factor of a monodisperse suspension. We investigate next the effect of added salt on $S(q_\text{m})$ as predicted by the considered renormalization methods, focusing on the largest concentration, $\phi=0.08$, to highlight structural differences. Figure~\ref{structure_vs_salinity} shows the RY-generated principal peak height, $S(q_\text{m})$, as a function of the reduced reservoir screening constant, $\kappa_\text{res}a=\sqrt{8\pi\lambda_{\text{B}}a^2n_\text{res}}$, for salt concentrations (reservoir screening parameter values) spanning the range from the counterion- to the salt-dominated regimes. Two plateaus are visible at low and high salt concentrations, reflecting the counterion- and salt-dominated regimes, respectively. The largest differences in the $S(q_\text{m})$ predictions occur in the counterion-dominated regime, consistent with the observation regarding the renormalized parameters in Fig.~\ref{effect_param_salt}. RJM predicts the smallest $S(q_\text{m})$ value at low salt content, consistent with its overestimate of the renormalized charge (cf.~Fig.~\ref{effect_param_salt}), followed by RLRT and CM-based methods. Regarding the SC and EPC methods, $S(q_\text{m})$ obtained for mean linearization is smaller than for edge linearization, and this difference is more pronounced for $S(q_\text{m})$ than for $g(r_\text{m})$. These distinguishing features are of key relevance in estimating freezing transition concentrations based on the Hansen-Verlet rule. This empirical rule states that $S(q_\text{m})\approx 3.1$ at freezing of a charge-stabilized system with $g(\sigma^+) \approx 0$ \cite{Roa_SoftMatter_2016}, as it applies to salt-free systems. According to Figs.~\ref{summary_structure} and \ref{summary_structure2}, $Z_\text{eff}$ and $\kappa_\text{eff}$ obtained from SC with mean potential linearization give the most accurate prediction of the freezing concentration based on the Hansen-Verlet rule.

In the transition region between counterion- and salt-dominated regimes, $S(q_\text{m})$ decreases with increasing $n_\text{res}$, reflecting loss in structure. The different renormalization methods predict the same $S(q_\text{m})$ in the salt-dominated regime, where $S(q_\text{m})$ approaches the hard-sphere value, $S^\text{HS}(q_\text{m})$, for the considered $\phi$. The horizontal, dotted line in Fig.~\ref{structure_vs_salinity} indicates the hard-sphere fluid peak value $S^\text{HS}(q_\text{m})\approx 1.06$, obtained from the expression
\begin{equation}
S^\text{HS}(q_\text{m})= 1+0.644\,\phi\frac{1-\phi/2}{(1-\phi)^3}
\label{CS-Sqm}
\end{equation}
provided by Banchio {\it et al.} \cite{Banchio_JCP_2000}, which quantitatively reproduces the simulation data and Verlet-Weis corrected PY structure factor peak height values for the hard-sphere fluid.

\subsection{Osmotic Pressure and Compressibility}
Having assessed the implications of different charge renormalization methods for the structural properties, $g(r)$ and $S(q)$, we address next their effect on various contributions to the pressure $p$ and osmotic compressibility $\chi_\text{osm}$ of a suspension. We first consider the salt-free case. 

In the CM, $p$ can be computed by solving the nonlinear PB equation and using the contact theorem [Eq.~(\ref{contacttheocell})]. The results are in reasonable agreement with PM-MC simulations for salt-free suspensions \cite{Denton_JPCM_2010}. In the RJM, we compute $p$ from Eq.~(\ref{jelliumPress_saltffree}), and in RLRT from the generalized virial equation [Eq.~(\ref{eq:PressureTwoBodyII})] with $p_\text{vol}$ determined according to Eq.~(\ref{pfree_RLRT}). Details of the RLRT pressure calculation are given in Appendix \ref{AppendixRLRT}). The latter method yields essentially the same results for the pressure as the variational method in refs.~\cite{Denton_JPCM_2008,LuDenton2010}. In the SDHA-EPC method, we compute $p$ using Eq.~(\ref{eq:PressureTwoBodyII}), with $p_\text{vol}$ according to Eq.~(\ref{pvol_BoonSF}) in the salt-free case, by holding the effective pair potential and the potential linearization point fixed in taking the macroion density derivative according to ref.~\cite{Boon_PNAS_2015}. 
We stress that if $p$ is determined in this way using the SDHA with EPC macroion charge renormalization, the correct ideal-gas limit $\beta p = n_\text{m}\left(1+Z\right)$ is recovered to first order in $n_\text{m}$ \cite{DiscussionBoon}.
This procedure is equivalent to computing the pressure of a suspension of pointlike macroions with fixed effective charge, $Q_\text{eff}$ [Eq.~(\ref{Qeff_def})], on neglecting the macroion density-dependence of $\kappa_\text{eff}$ in $u_\text{eff}(r)$.
\begin{figure}
\includegraphics[width=8cm]{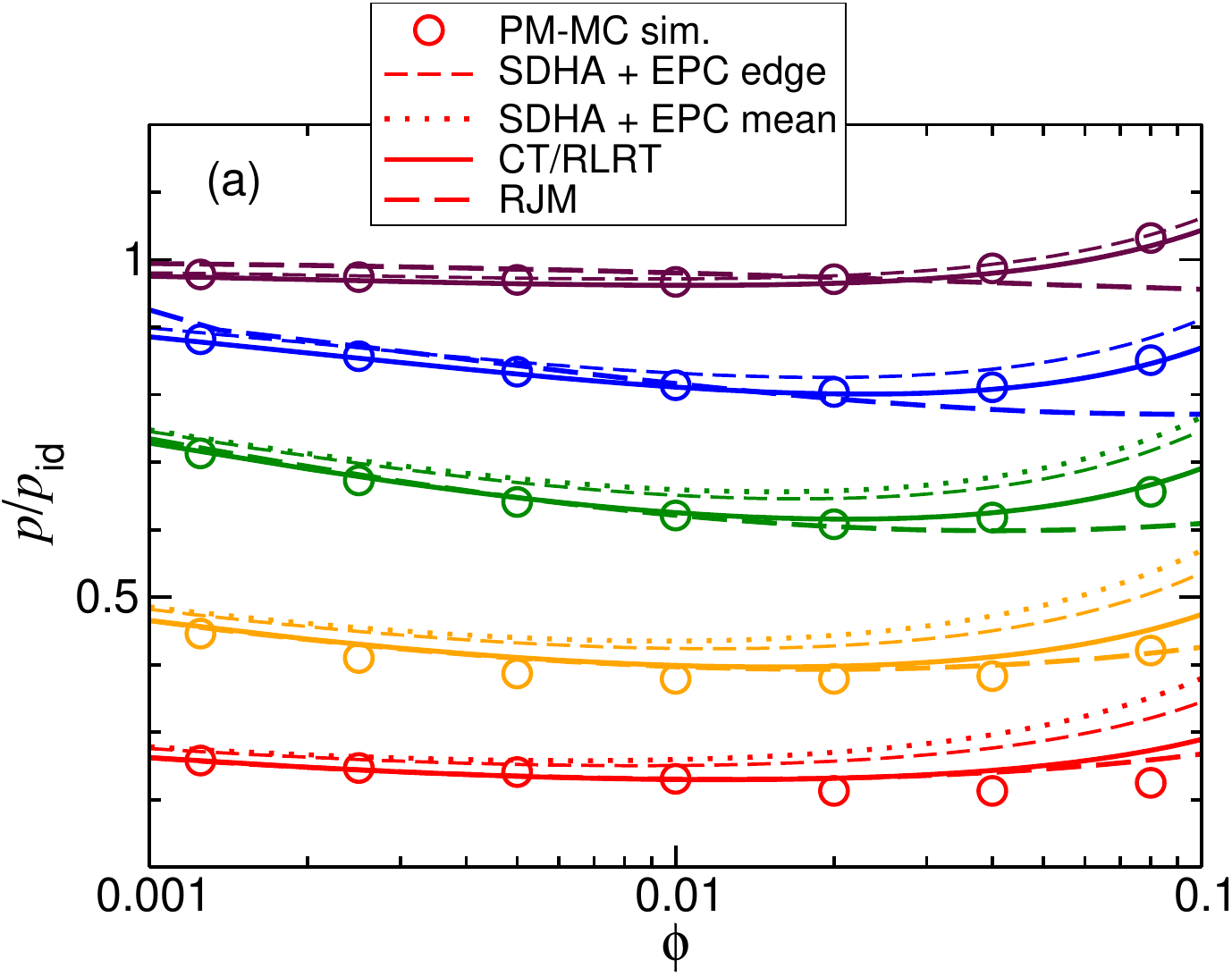} 
\includegraphics[width=8cm]{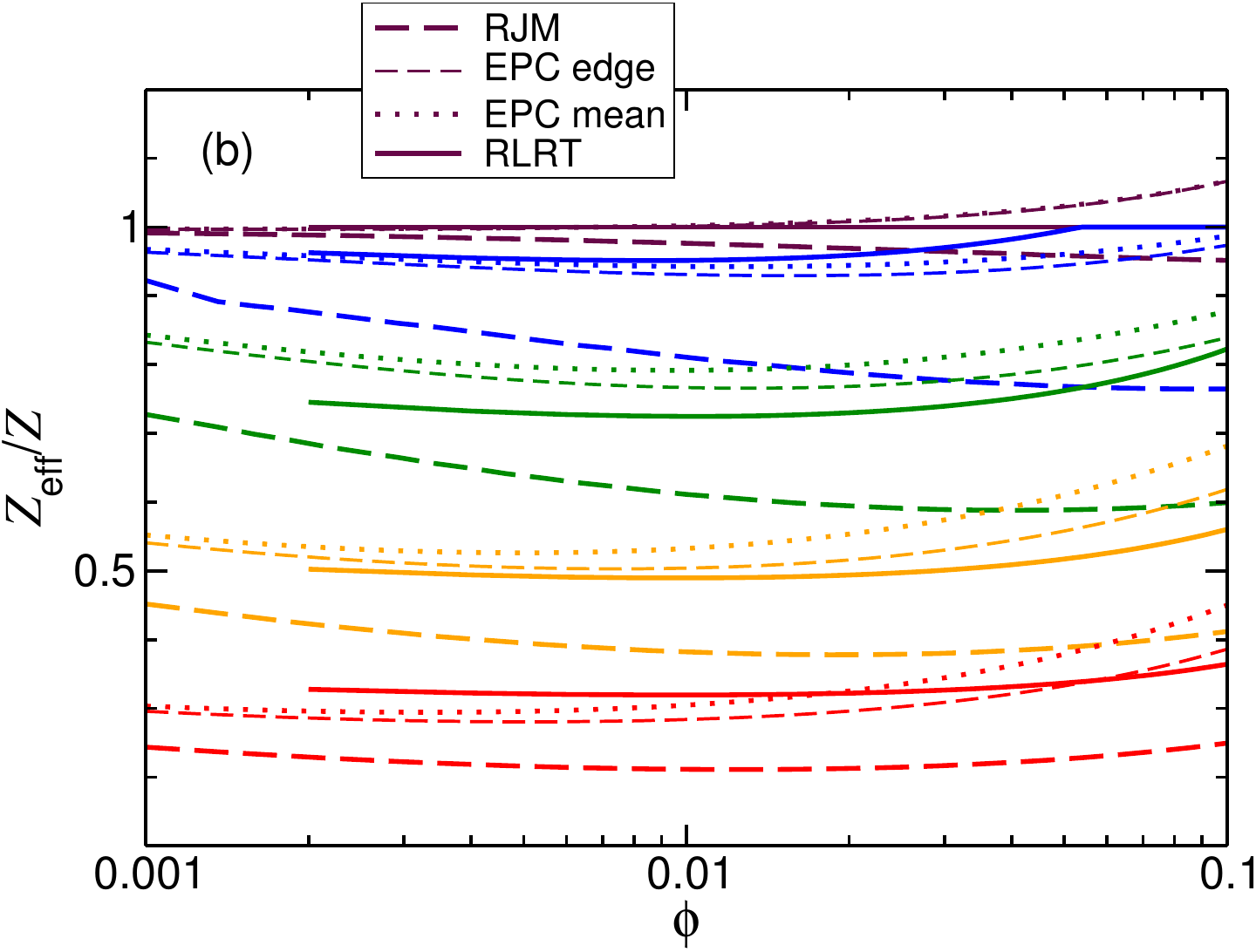}
\caption{(a) Reduced pressure, $p/p_\text{id}$, with $p_\text{id}=n_\text{m}k_\text{B}T(1+Z)$ and (b) reduced renormalized valence, $Z_\text{eff}/Z$, as functions of volume fraction $\phi$ for salt-free suspensions ($n_\text{res} = 0$) and coupling strengths $Z\lambda_\text{B}/a=0.89,\, 3.56,\, 7.12,\, 14.23,\, 28.46$ (top to bottom), distinguished by different colors. The curves correspond to different charge renormalization methods, while the symbols are PM-MC simulation data \cite{Linse2000}. In panel (a), for $Z\lambda_\text{B}/a=0.89$ and 3.56, the EPC edge and mean curves practically overlap. The nonlinear CT and RLRT pressure curves are also practically indistinguishable and, therefore, are represented by a single curve.} Other system parameters are $Z=40$, $\lambda_\text{B}=0.714$ nm.
\label{pressure_linse}
\end{figure}

In Fig.~\ref{pressure_linse}(a), we compare predictions for $p$ versus $\phi$ by the renormalization methods with one another and with PM-MC simulation data \cite{Linse2000} over a range of coupling parameter values $Z\lambda_\text{B}/a$. We normalize $p$ by $p_\text{id}$, to reveal deviations from $p_\text{id}$ due to pair interactions. Notice that $p<p_\text{id}$, except for sufficiently weak couplings and high volume fractions where charge renormalization ceases and excluded-volume interactions play a role. At fixed $Z\lambda_\text{B}/a$, the reduced pressure has a weakly nonmonotonic $\phi$-dependence with a shallow minimum, while at fixed $\phi$, $p$ decreases with increasing $Z\lambda_\text{B}/a$. Analysis of the PM-MC data for $p$ attributes deviations from $p_\text{id}$ and its decrease with increasing coupling to the strong accumulation of counterions near the macroion surfaces \cite{Linse2000}. This accumulation reduces the number of free counterions that contribute to the pressure. As seen, $p/p_\text{id}$ displays a minimum at $\phi\approx 0.03$, where the decrease in reduced pressure due to strengthening electrostatic interactions with increasing $\phi$ is balanced by an increase in the excluded-volume pressure contribution \cite{Linse2000}.

All considered effective one-component methods in Fig.~\ref{pressure_linse} agree closely with the PM-MC data for $p/p_\text{id}$, except for the RJM pressure, $p_\text{jell}$, at larger $\phi$. Also shown in Fig.~\ref{pressure_linse} [panel (b)] is the effective valence, $Z_\text{eff}$, divided by $Z$ to reveal the influence of charge renormalization. The similarity in shape between the reduced pressure and effective valence curves shows that, without salt, $p/p_\text{id}\approx Z_\text{eff}/Z$ is valid for relatively low values of $\phi$, as suggested by Eq.~(\ref{jelliumPress_saltffree}). In particular, the minima of $p/p_\text{id}$ and $Z_\text{eff}/Z$ are located at roughly the same volume fraction.

We recall that the considered charge renormalization methods are all based on linear screening and mean-field approximations. The overall good agreement between the OCM-based pressure predictions and the PM-MC simulation data suggests that effective many-body interactions and microion correlations have negligible impact on the thermodynamic properties of the considered suspensions with monovalent microions. A more detailed comparison of the different methods reveals that the PB contact theorem and RLRT accurately predict the pressure, yielding nearly identical curves on the scale of Fig.~\ref{pressure_linse}(a). SDHA with EPC methods (edge and mean linearization) tend to overestimate $p$, particularly at large $\phi$ and for strong coupling. Since SDHA-EPC determines $p$ using the generalized virial equation [Eq.~(\ref{eq:PressureTwoBodyII})], which overestimates the pair structure for strong coupling [cf. Fig.~\ref{summary_structure}(a)], it is not surprising that the predicted $p$ is less accurate. On contrasting the different linearizations within SDHA-EPC, one notices that $p$ is larger in mean than in edge linearization. This difference is understood from the effective valence predictions in Fig.~\ref{pressure_linse}(b), showing that $Z_\text{eff}^\text{mean}>Z_\text{eff}^\text{edge}$, expressing that counterion condensation is less pronounced for mean linearization. The strong effective coupling predicted in mean linearization implies that $p^\text{mean}>p^\text{edge}$. Even though the RJM performs rather poorly in predicting structure, it accurately predicts the pressure for strong coupling and underestimates $p$ only mildly for weak coupling and high concentrations. According to Eq.~(\ref{jelliumPress_saltffree}), this behavior is consistent with the RJM prediction of strong counterion condensation.
\begin{figure}
\includegraphics[width=8cm]{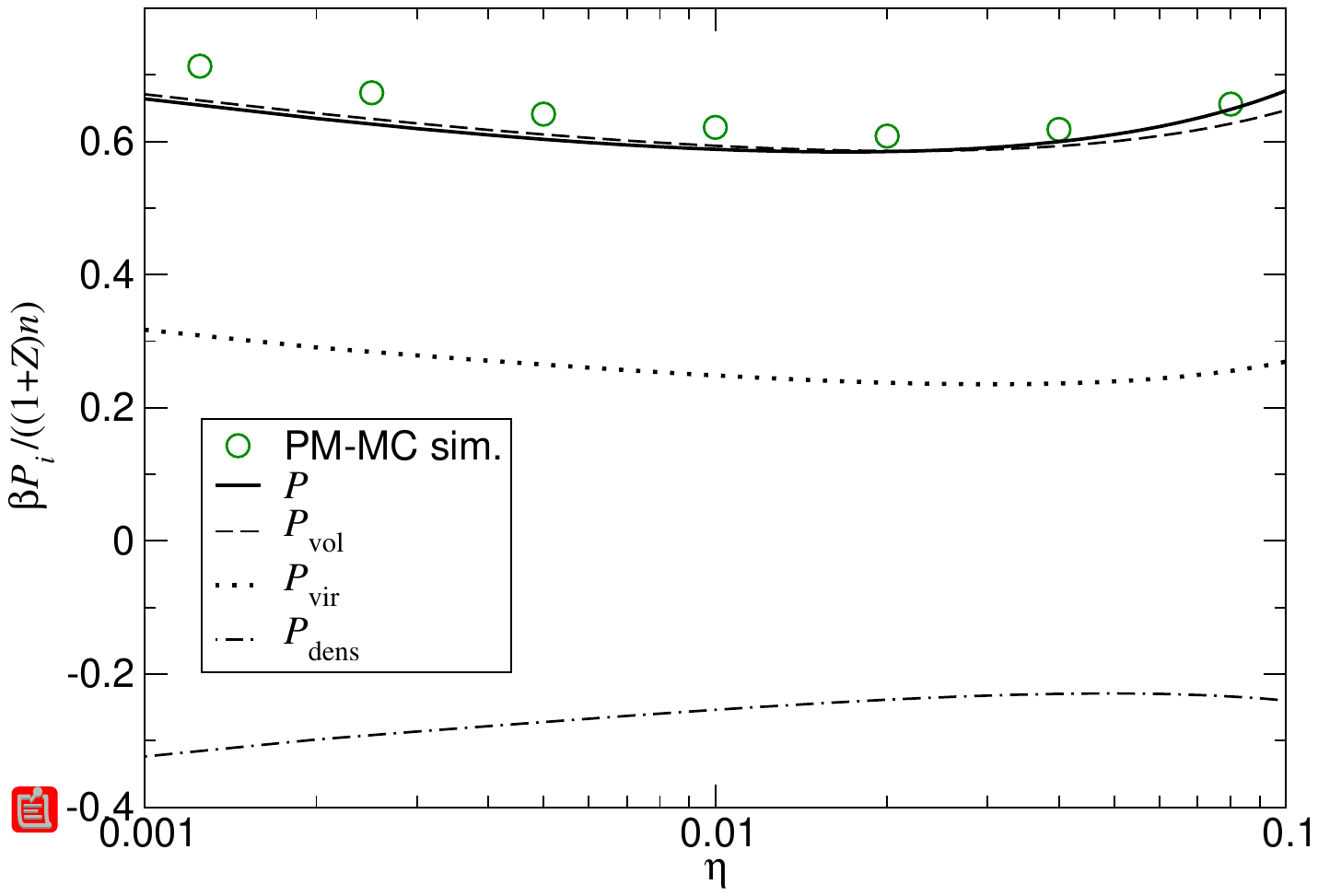}
\includegraphics[width=8cm]{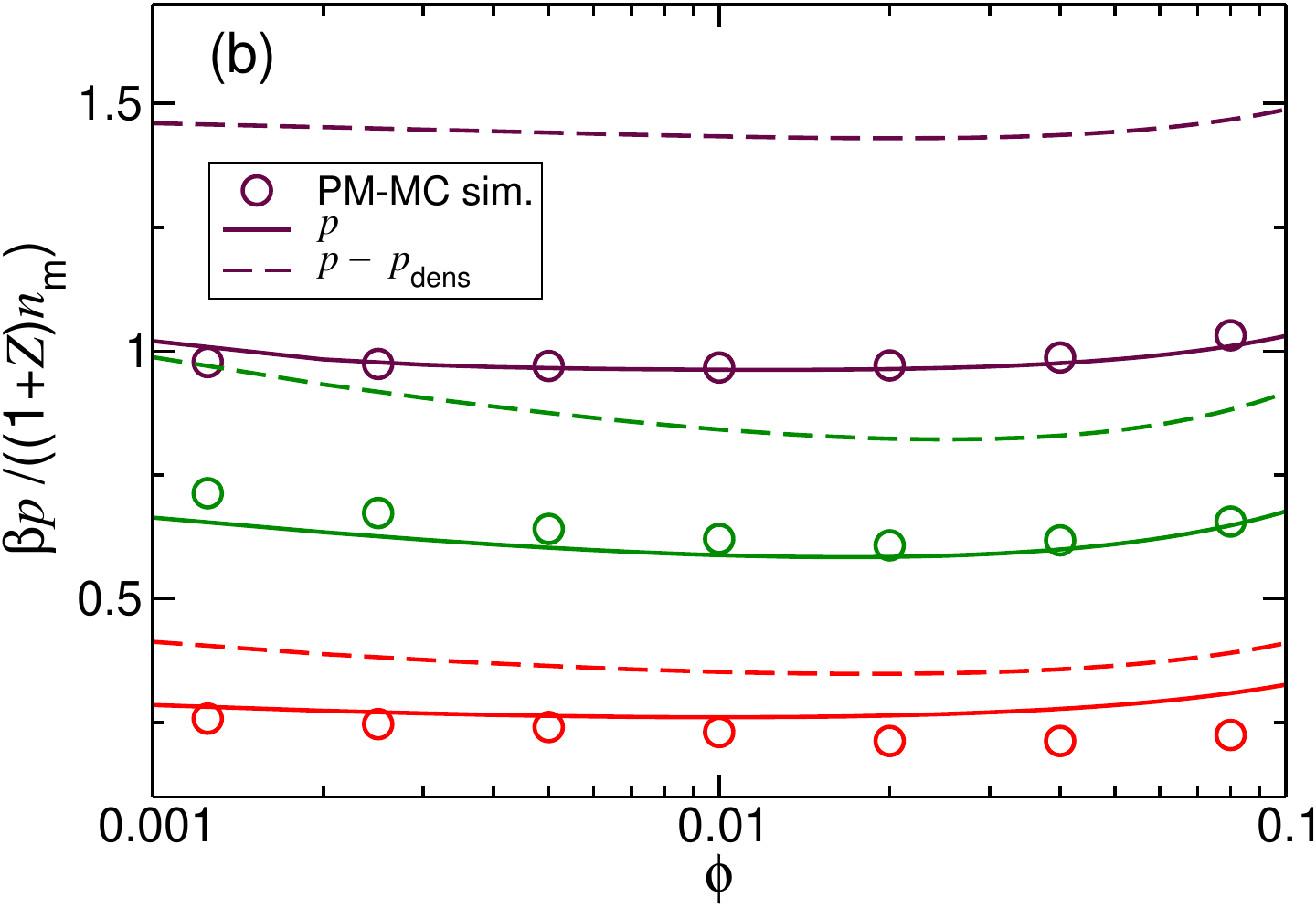}
\caption{(a) Reduced pressure, $p/p_\text{id}$, and its additive contributions as functions of $\phi$, with $Z\lambda_{\text{B}}/a=7.12$ for salt-free suspensions. (b) Reduced suspension pressure and pressure difference, $p-p_\text{den}$, versus $\phi$ for coupling strengths $Z\lambda_{\text{B}}/a=0.89,\,7.12$, and 28.46 (top to bottom), distinguished by different colors. In both (a) and (b), the generalized virial theorem in Eq.~(\ref{eq:PressureTwoBodyII}) is used in combination with the RLRT renormalization method. Symbols are PM-MC simulation data \cite{Linse2000}. Other system parameters are as in Fig.~\ref{pressure_linse}.}
\label{Diffpresscontrib_DentonVirial}
\end{figure}

It is interesting to analyze the different pressure contributions in the generalized one-component virial equation [Eq.~(\ref{eq:PressureTwoBodyII})] for the total suspension pressure. For the RLRT, in particular, we focus on the contribution
$p_\text{den}$, defined in Eq.~(\ref{p-den}),
resulting from the macroion density dependence of $u_\text{eff}$. Recall that this term is absent from the calculation of $p$ when using the SDHA method. For this reason, we decompose the suspension pressure, according to 
\begin{equation}
p=p_\text{vol}+p_\text{vir}+p_\text{den},
\end{equation}
where $p_\text{vol}$ is the contribution associated with the volume energy with renormalization included [Eq.~(\ref{pfree_RLRT})] and $p_\text{vir}$ is the virial contribution for an effective one-component model system with the state-dependence of $u_\text{eff}(r)$ disregarded, as defined in Eq.~(\ref{p-OCM}).
Figure~\ref{Diffpresscontrib_DentonVirial}(a) shows the RLRT-calculated pressure contributions for the most structured suspension treated in the PM-MC simulations. We observe that $p_\text{vol}$ is the dominant contribution in the considered concentration range for the salt-free suspension. Note further that $p_\text{den}$ is negative and practically compensates the positive contribution $p_\text{vir}$ for most $\phi$ values. At large $\phi$, the macroion-induced pressure contribution, $p_\text{m}=p_\text{vir}+p_\text{den}$, becomes positive and non-negligible. In fact, for $\phi>0.1$, $p_\text{m}$ contributes up to $20\%$ of the total pressure. 
		
Figure~\ref{Diffpresscontrib_DentonVirial}(b) quantifies the contribution of $p_\text{den}$ to $p$ for different indicated couplings. The RLRT method predicts a non-negligible (negative) contribution of $p_\text{den}$ with a larger relative contribution for weaker coupling. Neglecting $p_\text{den}$ would result in an unphysically large contribution of the macroion-induced pressure $p_\text{m}$ at low $\phi$.
\begin{figure}
\centering
\includegraphics[width=8cm]{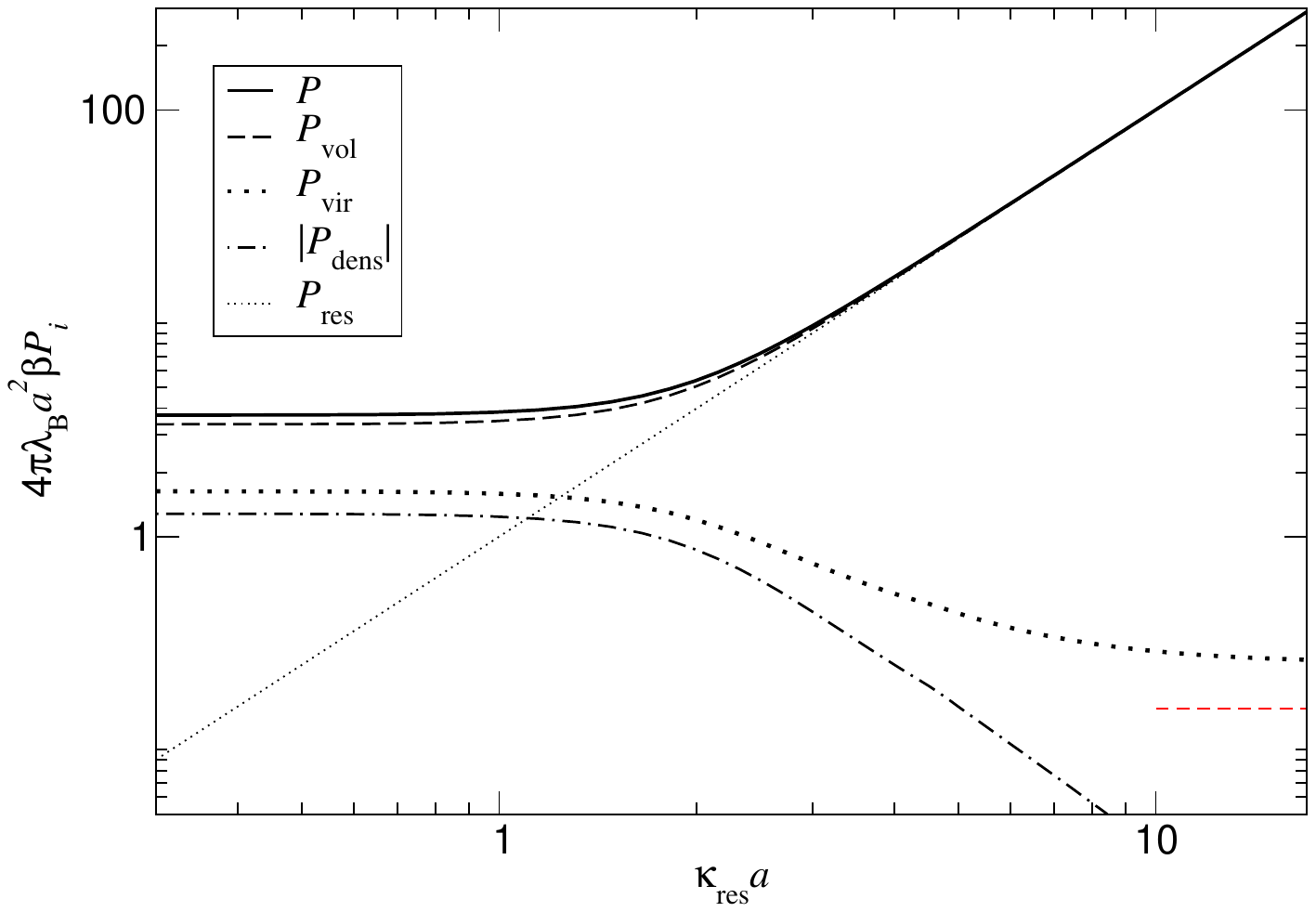}
\caption{Reduced suspension pressure, $4\pi\lambda_{\text{B}}a^2\beta p$, and its additive constituents versus $\kappa_\text{res}a$ ($\kappa_\text{res}^2\sim n_\text{res}$) for $\phi=0.2$. The pressure is calculated using the RLRT method. The red dashed horizontal line segment represents the pressure of a hard-sphere system, computed using the Carnahan-Starling equation of state at the same $\phi$. Other system parameters: $Z\lambda_\text{B}/a= 7.12$, $Z=40$, $\lambda_\text{B}=0.714$ nm.}
\label{comparevirial_salt}
\end{figure}
		
The influence of adding salt on the pressure and its constituent contributions is analyzed in Fig.~\ref{comparevirial_salt}. For low salt concentrations in the counterion-dominated regime ($\kappa_\text{res}a\lesssim 1$), $p$ stays constant and is practically equal to the pressure of a salt-free system. With increasing reservoir salt concentration, $p$ grows monotonically, approaching the reservoir pressure $p_\text{res}=2k_\text{B}T n_\text{res}$ in the salt-dominated regime, where the Donnan effect is absent and $Z_\text{eff}\approx Z$ holds with $\kappa_\text{eff}\approx\kappa_\text{res}$. As noted above in the salt-free case, $p_\text{vol}$ is the main contributor to $p$, approaching $p_\text{res}$ in the high-salinity limit. Although $p_\text{vol}$ is dominant throughout in the counterion-dominated regime, $p_\text{m}=p_\text{vir}+p_\text{den}$ gives a non-negligible positive contribution to $p$, visible even on the depicted logarithmic scale. As seen in Fig.~\ref{comparevirial_salt}, the negatively valued $p_\text{den}$ tends to compensate $p_\text{vir}$ for $\kappa_\text{res}a\lesssim 1$. Similarly to $p$ and $p_\text{vol}$, $p_\text{vir}$ and $p_\text{den}$ are constant in the low-salt region. With further increasing reservoir salinity, $p_\text{vir}$ tends to the hard-sphere pressure value (red dashed horizontal line segment), while $p_\text{den}$ tends to zero.
\begin{figure}
\centering
\includegraphics[width=8.5cm]{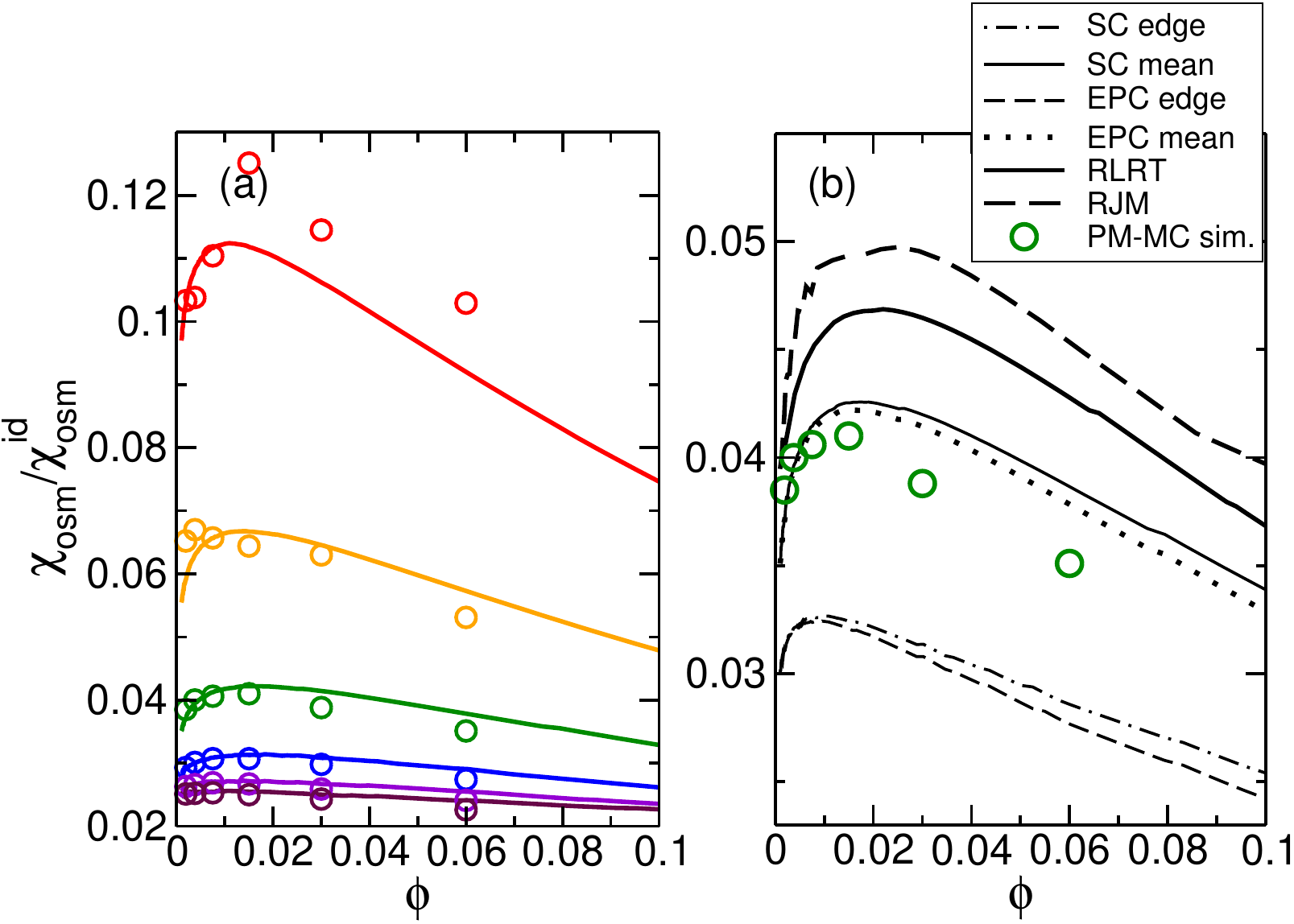}
\caption{(a) RY-calculated OCM zero-wavenumber limiting static structure factor $S(0)$ versus volume fraction $\phi$, viewed as an approximation for the reduced osmotic compressibility, $\chi_\text{osm}/\chi_\text{osm}^\text{id}$, and deduced from the EPC-mean renormalization method for different coupling parameters, $Z\lambda_\text{B}/a=0.89,\, 1.78,\, 3.56,\, 7.12,\, 14.23,$ and 28.46 (bottom to top), distinguished by different colors. (b) $S(0)$ versus $\phi$ from different renormalization methods as indicated and $Z\lambda_\text{B}/a= 7.12$. Open circles are PM-MC data for $\chi_\text{osm}/\chi_\text{osm}^\text{id}$, obtained from a numerical derivative of the PM-MC pressure data in Fig.~\ref{pressure_linse}. Other system parameters are as in Fig.~\ref{pressure_linse}.}
\label{compress}
\end{figure}
		
Another thermodynamic quantity of interest is the osmotic compressibility, $\chi_\text{osm}$. As discussed at the end of Sec. \ref{sec:StructureThermodynamics}, for a semi-open system in osmotic equilibrium with a salt reservoir, the (exact) reduced osmotic compressibility, $\chi_\text{osm}/\chi_\text{osm}^\text{id}$, is equal to the (exact) macroion-macroion static structure factor in the long-wavelength limit, $S_\text{mm}(q\to 0)$. Provided no approximations were involved in contracting out the microions and subsequently calculating the OCM $S(q)$, this structure factor would be identical to $S_\text{mm}(q)$. Hence, the accuracy of the considered renormalization methods in predicting thermodynamic properties is linked to their respective accuracy in describing the pair structure. In Fig.~\ref{compress}(a), the RY-calculated $S(0)$ obtained using $u_\text{eff}(r)$ with EPC-mean input for $Z_\text{eff}$ and $\kappa_\text{eff}$ is plotted versus $\phi$ for different coupling parameters. At fixed coupling, $S(0)$ has a non-monotonic $\phi$-dependence with a maximum at $\phi\approx 0.01$ for all considered cases. The position of this maximum coincides approximately with the volume concentration $\phi$ at which $Z_\text{eff}$ has its minimum [cf.~Fig.~\ref{pressure_linse}(b)]. For fixed $\phi$, $S(0)$ increases with increasing coupling, i.e., the suspension becomes more compressible as $Z_\text{eff}$ decreases. 
  
Figure~\ref{compress}(b) depicts the RY-calculated $S(0)$ versus $\phi$ for different renormalization methods and fixed $Z\lambda_\text{B}/a=7.12$. Evidently, $S(0)$ is larger for the methods predicting stronger charge renormalization, i.e., smaller $Z_\text{eff}$, except for the CM-based methods, for which this tendency is reversed using both edge and mean linearizations. The simulation data for $p$ from Fig.~\ref{pressure_linse} \cite{Linse2000} allow us to compute the compressibility factor from numerical differentiation according to Eq.~(\ref{OsmCompress_def}). While the non-monotonic shape of the PM-MC data for $\chi_\text{osm}/\chi_\text{osm}^\text{id}$ is qualitatively reproduced by the different methods, including the location of the maximum, there are significant quantitative differences. The figure shows that the methods using linearization around the mean potential give more accurate results for the osmotic compressibility factor than those using edge linearization.

\vspace*{-0.5cm}
\section{Conclusions}\label{conclusions}
Suspensions of highly charged colloidal particles in which nonlinear electrostatic screening is prevalent can be modeled using charge renormalization theories based on linear screening and mean-field approximations that predict effective interaction parameters. In this work, we analyzed several commonly used renormalization methods and identified conceptual differences between them. Furthermore, we numerically evaluated the considered methods and assessed their pros and cons by comparing the theoretical predictions for structural and thermodynamic properties of salt-free suspensions in the cell model and multi-center one-component models with corresponding data from MC simulations of suspensions in the primitive model \cite{Linse2000}. For simplicity, we have restricted our analysis to spherical hard colloids dispersed in solvents containing only monovalent microions, whose correlations are neglected. Our results and conclusions can help to guide the selection of methods for modeling charge-stabilized suspensions of impermeable colloids, and for interpreting experiments based on, e.g., light, x-ray, and neutron scattering.

The charge renormalization methods examined here differ from one another in how the Poisson-Boltzmann equation is linearized and whether they are based on the cell model or the one-component model. Methods invoking edge linearization include those based on cell models that linearize the electrostatic potential around its value at the cell edge and the RJM, which linearizes around the asymptotic value of the potential. Methods invoking mean linearization include those based on cell models that linearize the potential around its value averaged over the cell volume and the OCM-based RLRT, which linearizes around the potential averaged over the suspension volume. Regarding the resulting effective interaction parameters predicted by edge and mean linearization, we observe clear differences in the renormalized screening parameters, with $\kappa_{\text{eff}}(\text{mean})>\kappa_{\text{eff}}(\text{edge})$, but no clear differences for the renormalized valence $Z_\text{eff}$.

In analyzing the effect of added salt on the effective interaction parameters, the largest differences between the renormalization methods are observed at low salt concentrations in the counterion-dominated regime. While in the explored parameter space most of the methods predict similar effective interaction parameters, the RJM predicts notably lower values of $Z_\text{eff}$ and $\kappa_{\text{eff}}$ than the other methods. Consequently, the RJM predicts a distinctly stronger counterion quasi-condensation on the colloid surfaces.

Interestingly, the EPC and penetrating RJM methods predict an effective valence, $Z_\text{eff}$, larger than the bare one at comparatively low coupling strengths and high macroion concentrations. According to the PM-MSA, this unusual effect can be attributed to a reduced electrostatic screening caused by the geometric exclusion of the microion clouds by nearby colloids. The possibility that $Z_\text{eff}$ can exceed $Z$ in EPC \cite{Boon_PNAS_2015} renders this method the most accurate in predicting the peak value of the radial distribution function, $g(r_\text{m})$, as we noticed from the comparison with PM-MC simulation data.

From comparing the macroion $g(r)$ and $S(q)$ calculated using the RY scheme with the PM-MC simulation data, one observes considerably improved agreement at high coupling values, $5\lesssim Z\lambda_{\text{B}}/a\lesssim30$, if charge renormalization is incorporated. Most of the considered renormalization methods predict radial distribution functions with principal peak heights, $g(r_\text{m})$, that deviate from the MC data by less than $5\%$ for $Z\lambda_{\text{B}}/a<15$, and less than $10\%$ for $Z\lambda_{\text{B}}/a<30$. While yielding significantly stronger counterion condensation, the RJM predicts peak heights with an error of less than $10\%$ at high coupling values $Z\lambda_{\text{B}}/a>5$. Consistent with predictions for the renormalized interaction parameters, significant differences in $S(q)$ are observed in the counterion-dominated regime. All renormalization methods yield the correct hard-sphere limit at high salt concentrations.

To assess the performance of the renormalization methods in predicting thermodynamic properties, we analyzed the suspension pressure in the OCM, $p$, along with its constituent contributions, and the osmotic compressibility, $\chi_\text{osm}$. Our analysis shows that nonlinear electrostatic effects become relevant for coupling values $Z\lambda_{\text{B}}/a>5$, in accordance with the findings for the structural properties. We further analyzed the contribution to $p$ from the density derivative of the pair potential, $p_\text{den}$, 
which appears in the RLRT method for computing $p$ using the generalized virial equation [Eq.~(\ref{eq:PressureTwoBodyII})]. We also analyzed the volume pressure contribution, $p_\text{vol}$, and the macroion virial pressure contribution, $p_\text{vir}$, which must be included in both the RLRT and the SDHA methods.

In calculating the suspension pressure in the OCM, one consistently needs to account for the dependence of the effective pair potential, $u_\text{eff}(r;n_\text{m})$, on macroion density $n_\text{m}$. We emphasized the importance of separating the macroion density dependence emerging from charge renormalization from the thermodynamic density dependence. In this context, we showed that $p_\text{den}$ contributes essentially to $p$ in the RLRT method, but needs to be excluded from the SDHA-based pressure calculation in order to obtain the correct ideal-gas suspension pressure at high macroion dilution.

While predictions of structural properties by the various renormalization methods deviate significantly from the PM-MC data for coupling strengths $Z\lambda_\text{B}/a\gtrsim15$, the corresponding pressure predictions remain in close agreement with simulation data (within 5\%) even up to $Z\lambda_\text{B}/a\approx30$. The higher accuracy for thermodynamic properties is explained by the observation that for the considered zero-salt systems in the PM-MC simulations, where $\phi \leq 0.1$, the counterion-related pressure contributions,  associated with the volume energy-derived $p_\text{vol}$ in the RLRT and SDHA methods, and $p_\text{jell}$ and $p_\text{CT}$ in the RJM and CM methods, respectively, give the dominant contribution to $p$, and are overall well approximated by the considered renormalization methods.

Regarding the overall performance of the various methods, the following comments are in order. The CM-SC method with edge linearization has been widely used for impermeable colloids due to its simple implementation. However, mean linearization is not only conceptually preferable \cite{Trizac_Langmuir:2003}, but also improves the predictions for $g(r)$ and $S(q)$. Implementing this alternative linearization requires merely an extra integration step to compute the mean potential in the numerical solution of the nonlinear PB equation. 

The SDHA combined with the EPC method provides the most accurate structural description at low coupling and high concentrations, where it predicts $Z_\text{eff}>Z$. This method is as easy to implement as the pure CM, but its validity extends to high concentrations and salinities by using, e.g., the virial theorem for the pressure calculation. In this parameter range, macroion-macroion correlations are relevant, and predictions for the pressure based on the contact theorem become accordingly poor. 

The originators of the RJM argue \cite{Trizac_PRE_2004} that the renormalized valence predicted by this method is more appropriate than that from CM-based methods, since a DLVO-type $u_\text{eff}(r)$ arises naturally within the JA by integrating the electrostatic stress tensor over the colloid surfaces \cite{Levin2002_Review,RusselBook}. However, we showed that the RJM significantly underestimates the pair structure in comparison with the PM-MC simulation data for $g(r)$. Notwithstanding this property, RJM accurately predicts the suspension pressure over a wide range of colloid concentrations. 

The RLRT is shown in our analysis to be overall quite accurate in predicting both structural and thermodynamic properties. The thermodynamic basis of its counterion association mechanism offers a clear conceptual framework for the phenomenology behind charge renormalization. While its implementation requires some effort, the RLRT has the advantage of predicting the suspension free energy, including the contribution from macroion-macroion interactions.

In summary, we have provided a detailed analysis and assessment of widely used charge renormalization methods for charge-stabilized colloidal suspensions. With this work, we have aimed to guide the selection of appropriate methods, document their benefits and limitations, and outline their practical implementation under conditions where nonlinear screening effects are important.

\vspace*{-0.5cm}
\section*{Acknowledgments}
\vspace*{-0.3cm}
We thank Niels Boon for helpful correspondence. M.E.B. and G.N. thank J.~Riest (Viega Technology GmbH, Germany) for assistance with codes for integral-equation calculations. M.E.B. and G.N. further acknowledge support by the Deutsche Forschungsgemeinschaft (SFB 985, Project B6, Grant No.~191948804). A.R.D. acknowledges support of the National Science Foundation (Grant No.~DMR-1928073).

\begin{center}
{\bf DATA AVAILABILITY}
\end{center}
\vspace*{-0.2cm}
The data that support the findings of this study are available from the corresponding author
upon reasonable request.


\appendix*
\begin{table}
\caption{Table of common acronyms used in the paper.}
\begin{center}
\begin{tabular}{ l l }
 CM & Cell model\\
 CT & Contact theorem\\
 DFT & Density-functional theory \\
 DH & Debye-H\"uckel \\
 DLVO &  Derjaguin-Landau-Verwey-Overbeek (potential) \\ 
 EPC & Extrapolated point charge with PBCM \\
 HS & Hard sphere \\
 JA & Jellium approximation \\
 JM & Jellium model \\
 LRT & Linear response theory \\
 MC & Monte Carlo \\
 MSA & Mean spherical approximation \\
 OCM & One-component model (of dressed macroions) \\
 PB & Poisson-Boltzmann \\
 PBCM &  Poisson-Boltzmann cell model\\
 PM & Primitive model \\
 RJM & Renormalized jellium model \\
 RLRT &  Renormalized linear response theory \\
 RY & Rogers-Young (integral equation scheme) \\
 SC & Surface charge with PBCM \\
 SDHA & Shifted Debye-H\"uckel approximation 
\vspace*{-0.7cm}
\end{tabular}
\end{center}
\label{table1}
\end{table}

\vspace*{-0.3cm}
\appendix
\section{Suspension Pressure from RLRT Method} \label{AppendixRLRT}
\vspace*{-0.2cm}
We explain in more detail here how the suspension pressure, $p=p_\text{vol}+p_\text{m}$, with $p_\text{m}=p_\text{vir}+p_\text{den}$, is calculated in the OCM using the RLRT method \cite{Denton_JPCM_2008,LuDenton2010,Denton_JPCM_2010} based on Eq.~(\ref{eq:PressureTwoBody}) with renormalized effective pair potential
\begin{equation}
\beta u_\text{eff}(r)=\begin{cases}
{\displaystyle \infty}, & 0\le r\le\sigma\\
{\displaystyle \lambda_\text{B}Z_\text{eff}^2\left(\frac{\exp(\kappa_\text{eff} a)}{1+\kappa_\text{eff} a}\right)^2\frac{\exp(-\kappa_\text{eff} r)}{r}}, & r>\sigma\,.
\end{cases}
\label{EffectPotLinRespColII}
\end{equation}
The effective interaction parameters, $\kappa_\text{eff}$ and $Z_\text{eff}$, are calculated as described in Sec.~\ref{Sec:RLRT}. 
The volume pressure contribution, $p_\text{vol}$, derived from the volume energy, is due to the free microions outside of an association shell of thickness $\delta$. On accounting for charge renormalization, $p_\text{vol}$ is given by
\begin{equation}
\beta p_\text{vol}=\tilde{n}_++\tilde{n}_--\frac{Z_\text{eff}(\tilde{n}_+-\tilde{n}_-)\kappa_\text{eff}\lambda_\text{B}}{4[1+\kappa_\text{eff}(a+\delta)]^2}\,,
\end{equation}
where $\tilde{n}_\pm=\tilde{N}_\pm/[V(1-\phi_\text{eff})]$ and $\tilde{N}_\pm$ are mean number densities and numbers of free microions, $\phi_\text{eff}=\phi(1+\delta/a)^3$ is an effective macroion volume fraction that accounts for the volume of the association shell, and $\kappa_\text{eff}=\sqrt{4\pi\lambda_\text{B}(\tilde{n}_+ + \tilde{n}_-)}.$

The remaining two terms in Eq.~(\ref{eq:PressureTwoBody}) add up to the macroion pressure contribution
[Eq.~(\ref{eq:PressureTwoBodyII})],
\begin{equation}
p_\text{m}=p_\text{vir}+p_\text{den}.
\label{macro_press_own}
\end{equation}
The first term is the virial contribution for an OCM system with density dependence of pair interactions disregarded,
\begin{eqnarray}
p_\text{vir}
&=&n_\text{m}k_\text{B}T-\frac{2\pi}{3}n_\text{m}^2\int_{0}^{\infty}dr r^3g(r)\frac{\partial u_\text{eff}(r)}{\partial r}\nonumber\\&=&\frac{2\pi}{3} n_\text{m}^2\int_{\sigma^+}^\infty dr\,r^2g(r) u_\text{eff}(r)(\kappa_\text{eff}\,r+1)\nonumber\\&+&4n_\text{m}k_\text{B}T\phi \,g(\sigma^+)\,, 
\label{p1contrib}
\end{eqnarray}
where, in the second equality, we have singled out the contribution from the macroion hard cores involving the contact value of $g(r)$. The second term on the right side of Eq.~(\ref{macro_press_own}), involving the macroion density derivative of $u_\text{eff}$, is calculated by accounting only for variations of $u_\text{eff}$ with $n_\text{m}$ that are thermodynamically relevant, disregarding the $n_\text{m}$-dependencies of $\kappa_\text{eff}$ and $Z_\text{eff}$ due to charge renormalization. Furthermore, in taking the derivative with respect to $n_\text{m}$, electroneutrality must be maintained. For simplicity, in the canonical ensemble, in which $Z_\text{eff}N_\text{m}=\tilde{N}_+ - \tilde{N}_-$, one obtains $\partial Z_\text{eff}/\partial \phi=0$. Thus, the effective pair potential in RLRT depends on concentration only via the effective screening parameter.

Assuming $\delta$ to be independent of $n_\text{m}$, it follows that
\begin{eqnarray}
p_\text{den}
&=&2\pi n_\text{m}^3\int_{\sigma^+}^\infty dr\,r^2g(r)\,\frac{\partial u_\text{eff}}{\partial n_\text{m}}\nonumber\\
&=&2\pi n_\text{m}^2\int_{\sigma^+}^\infty dr\,r^{2}g(r) u_\text{eff}(r)\gamma\left(\alpha- \frac{r}{a}\right),
\label{p2contrib}
\end{eqnarray}
where
\begin{equation}
\alpha=\frac{2\kappa_\text{eff} a}{1+\kappa_\text{eff} a}
\label{alpha}
\end{equation}
and
\begin{equation}
\gamma=\phi\frac{\partial \kappa_\text{eff} a}{\partial \phi}=\frac{\kappa_\text{eff} a}{2\left(1-\phi_\text{eff}\right)}\,.
\label{gamma}
\end{equation}
For a charge-renormalized suspension, the bulk pressure from Eq.~(\ref{macro_press_own}) agrees precisely with that obtained from the (Gibbs-Bogoliubov) first-order thermodynamic perturbation expansion and OCM-based MC simulations \cite{Denton_JPCM_2010}.


%

\end{document}